\documentclass[ps,prd,preprintnumbers,superscriptaddress,floatfix,twocolumn,notitlepage,nofootinbib]{revtex4-2} %,a4paper,11pt
\pdfoutput=1 % if your are submitting a pdflatex (i.e. if you have
\usepackage{dcolumn}% Align table columns on decimal point
\usepackage{slashed}
\usepackage{bm,amssymb,slashed,graphicx,multirow,soul,mathtools,xspace,array,tikz,amsmath,siunitx}
\usepackage[compat=1.1.0]{tikz-feynman}   
\usepackage{float}
\usepackage[colorlinks=true,citecolor=blue,urlcolor=blue,linktocpage=true,
linkcolor=blue]{hyperref} 
\allowdisplaybreaks
\usepackage{ bbold }
\usepackage{braket}
\usepackage{multirow}

\definecolor{nicered}{rgb}{0.7,0.1,0.1}
\definecolor{nicegreen}{rgb}{0.1,0.5,0.1}
\definecolor{violet}{rgb}{0.7,0.3,0.3}
\newcommand{\lp}{\left(}
\newcommand{\rp}{\right)}

\newcommand{\SM}{\text{SM}}

\newcommand{\beq}{\begin{equation} }
\newcommand{\eeq}{\end{equation}} 
\newcommand{\bi}{\begin{itemize} }
\newcommand{\ei}{\end{itemize} }

\definecolor{Red}{rgb}{1.,0.,0.}
\definecolor{Grn}{rgb}{0.,0.75,0.}
\definecolor{Blu}{rgb}{0.,0.,1.}
\definecolor{Pink}{rgb}{1,0.08,0.58}

\DeclareMathOperator{\Binom}{Binom}

\usepackage[T1]{fontenc} % if needed

\usepackage{amsmath,amssymb,color,slashed}
\allowdisplaybreaks  

\begin{document} 

\title{Flavor violating Higgs and $Z$ decays at FCC-ee}

\author{Jernej F. Kamenik}
 \email{jernej.kamenik@cern.ch}
\affiliation{Jo\v{z}ef Stefan Institute, Jamova 39, 1000 Ljubljana, Slovenia}
\affiliation{Faculty of Mathematics and Physics, University of Ljubljana, Jadranska 19, 1000 Ljubljana, Slovenia}%

\author{Arman Korajac}
 \email{arman.korajac@ijs.si}
\affiliation{Jo\v{z}ef Stefan Institute, Jamova 39, 1000 Ljubljana, Slovenia}

\author{Manuel Szewc}
 \email{szewcml@ucmail.uc.edu}
\affiliation{Department of Physics, University of Cincinnati, Cincinnati, Ohio 45221, USA}

\author{Michele Tammaro}
 \email{michele.tammaro@ijs.si}
\affiliation{Jo\v{z}ef Stefan Institute, Jamova 39, 1000 Ljubljana, Slovenia}%

\author{Jure Zupan}
 \email{zupanje@ucmail.uc.edu}
 \affiliation{Department of Physics, University of Cincinnati, Cincinnati, Ohio 45221, USA}

\date{\today}% It is always \today, today,

\begin{abstract}
Recent advances in $b$, $c$, and $s$ quark tagging coupled with novel statistical analysis techniques will allow future high energy and high statistics electron-positron colliders, such as the FCC-ee, to place phenomenologically relevant bounds on flavor violating Higgs and $Z$ decays to quarks. We assess the FCC-ee reach for $Z/h\to bs, cu$ decays as a function of jet tagging performance. We also update the SM predictions for the corresponding branching ratios, as well as the indirect constraints on the flavor violating Higgs and $Z$ couplings to quarks. Using type III two Higgs doublet model as an example of beyond the standard model physics, we show that the searches for $h\to bs, cu$ decays at FCC-ee can probe new parameter space not excluded by indirect searches. We also reinterpret the FCC-ee reach for $Z\to bs , cu$ in terms of the constraints on models with vectorlike quarks. 
\end{abstract}

\maketitle

{\bf Introduction.} Flavor Changing Neutral Currents (FCNCs) are forbidden at tree level in the Standard Model (SM), and are as such ideal to search for effects of beyond the SM (BSM) physics. Most of the FCNC observables are accessible at experiments that are done at relatively low energies, but with large statistics. The list of such observables is very long, and involves both quarks and leptons. The classic examples are ${\mathcal{B}}(\mu\to e\gamma)$, $\mu \to e$ conversion rate, $B_{(s)}-\bar B_{(s)}$, or $K-\bar K$ mixing, ${\mathcal{B}}(B_s\to \mu^+\mu^-)$, and many more (for reviews see, e.g., \cite{Zupan:2019uoi,Gori:2019ybw,Kamenik:2016ygc,Altmannshofer:2022aml,Grossman:2017thq}). 

The situation is different for high energy FCNC observables, where the list is rather short and almost always involves leptons. Examples are ${\mathcal{B}}(h\to \ell\ell')$, ${\mathcal{B}}(Z\to  \ell\ell')$ and $\sigma(pp\to \ell\ell')$. The exception to this rule are the decays of top quarks, where $t\to ch, cg,\dots,$ can also be probed in high energy collisions, see, e.g., \cite{Kamenik:2020fgu,Greljo:2022jac,Fuentes-Martin:2020lea,Greljo:2018tzh,Greljo:2017vvb,Faroughy:2016osc,TopQuarkWorkingGroup:2013hxj,Cepeda:2019klc}. 

In this Letter we show that, somewhat surprisingly, the on-shell FCNC decays of the Higgs, ${\mathcal{B}}(h\to bs)\equiv {\mathcal{B}}(h\to \bar bs +b \bar s)$ and ${\mathcal{B}}(h\to cu)\equiv {\mathcal{B}}(h\to \bar cu +c \bar u)$, can be added to the list of high energy FCNC observables, since they can be probed at a phenomenologically interesting level at a future lepton collider, such as the FCC-ee \cite{Agapov:2022bhm}. 
Over the full running period of FCC-ee, the collider is expected to produce $N_h = 6.7\times 10^5$ $h$'s~\cite{deBlas:2019rxi} and $N_Z = 5\times 10^{12}$ $Z$'s~\cite{Mangano:2651294,FCC:2018evy}.
As we show in the following, FCC-ee is projected to have a sensitivity to ${\mathcal{B}}(h\to bs)$ and ${\mathcal{B}}(h\to cu)$ below the indirect bounds from $B_s-\bar B_s$ and $D-\bar D$ mixing, cf.~Table \ref{tab:SMpredict}, and we expect similar sensitivities to apply also to CEPC \cite{CEPCPhysicsStudyGroup:2022uwl}. For a recent analysis of the $h\to b s$ reach at ILC, but using $b$- and $c-$taggers, see \cite{Barducci:2017ioq}, where the leptonic channel reach is consistent with our results\footnote{Ref.~\cite{Barducci:2017ioq} also considered the $bj + \text{MET}$ final state where the authors find a slight improvement by going to higher energies, where the relative importance of the $Zh$ contribution decreases, and $W$-fusion production of the $h$ is the main signal channel. Because the backgrounds are lower, the power of the analysis therefore increases. We focus instead exclusively on the leptonic channel were $Zh$ can be more easily separated from the backgrounds and our Monte-Carlo-free analysis is more trustworthy. We do note that the increase in performance is still qualitatively consistent with our results. A quantitative comparison would require a more detailed study of the systematic uncertainties, which is beyond the scope of present work.}. The main reasons for these significant improvements are: {\it i)} the recent advances in $b$-, $c-$ and $s$-jet tagging, {\it ii)} the analysis technique that we advocate for below, which results in excellent sensitivity to these FCNC transitions,  and {\it iii)} the relatively clean environment of $e^+e^-$ collisions. The same approach can also be applied to ${\mathcal{B}}(Z\to bs)$ and ${\mathcal{B}}(Z\to cu)$, however, the phenomenologically interesting branching ratios are still below the floor set by the systematic uncertainties of taggers.

{\bf Accessing flavor violating transitions.}
An analysis strategy that has been successfully applied to $h\to c \bar c$ decays \cite{ATLAS:2022ers}, as well as to suppressed $t\to (s,d) W$ transitions~\cite{Faroughy:2022dyq,CMS:2020vac}, is to distribute events into different event types according to how many flavor tagged (and anti-tagged) jets they contain. In particular, the inclusion of information about events with light jets was shown in Ref.~\cite{Faroughy:2022dyq} to lead to significant improvement in sensitivity to $V_{ts,td}$. 

Here, we modify the approach of Ref.~\cite{Faroughy:2022dyq}  and apply it to the case of $h\to bs,cu$ and $Z\to bs,cu$ decays. For notational expediency we focus first on just the $bs$ final state, and then extend these results to the analysis of $cu$ decays.
In both  $h\to bs$ and $Z\to bs$ decays  there are two jets in the final state; in $e^+e^-\to hZ( h\to bs, Z\to ee, \mu\mu)$ there are also two isolated leptons, while the $e^+e^-\to Z\to bs$ events only have two jets. Applying the $b$- and $s$-taggers to the two jets, the events are distributed in  $(n_{b},n_{s})\in\{(0,0),(1,0),(0,1),(2,0), (1,1), (0,2)\}$ bins, where $n_{b(s)}$ denotes the number of $b(s)$-tagged jets in the event. The $b$- and $s$-taggers need to be orthogonal to ensure no event populates two different $(n_{b},n_{s})$ bins and is double-counted \footnote{Orthogonality can be achieved with the help of anti-tagging. For example, and as detailed in Ref.~\cite{Faroughy:2022dyq}, an $s$-tagger can be made orthogonal to a $b$-tagger simply by combining it with an anti-$b$-tagger. That is, for a jet to be $s$-tagged it has to be previously rejected by the $b$-tagger operating at a looser working point with a higher true and false positive rates than the one used for $b$-tagging itself.}. 
We denote the tagger efficiencies as $\epsilon^{b}_{\beta}$ and $\epsilon^{s}_{\beta}$, where $\beta=\{l,s,c,b\}$ denotes the flavor of the initial parton ($l=g$ for $h$ and $l=u,d$ for $Z$). 

The expected number of events in the bin $(n_b, n_s)$  is given by
\beq\label{eq:Nbar:nb:ns}
\bar N_{(n_{b},n_{s})} = \sum_f p(n_{b},n_{s}|f,\nu)\bar N_f(\nu) \,,
\eeq
where the summation is over the relevant (signal and background) decay channels, $f=\{gg,s \bar s,c\bar c,b\bar b,bs\}$ for the $h$ and $f=\{u \bar u+d \bar d,s \bar s,c \bar c,b \bar b,bs\}$ for the $Z$. The expect number of events in each decay channel is given by 
\beq
\label{eq:bar:Nf}
\bar N_f = {\cal B}(Z/h\to f)N_{Z/h}{\cal A}\,,
\eeq
where ${\cal B}(Z/h\to f)$ are the corresponding branching fractions,  $N_{Z/h}$ are the number of $Z$ and $h$ bosons expected to be produced during the FCC-ee run, while ${\cal A}$ is the detector acceptance including reconstruction efficiency, which we assume for simplicity to be the same for all the relevant decay channels.  

In writing down Eq.~\eqref{eq:Nbar:nb:ns} we have neglected the backgrounds: the  $\tau^{+}\tau^{-}$ for $Z\to bs$ and the Drell-Yan, $WW, ZZ$ for $h\to bs$. We expect that the inclusion of these backgrounds will not qualitatively change our results, since for most part they are small enough to  constitute only a subleading effect. Perhaps the most worrisome is the $ZZ$ background for $h\to bs$. Even this we expect in the actual experimental analysis to be either reduced enough through optimized selection to be ignored (e.g., through use of a multivariate classifier trained on other kinematic observables such as the invariant masses and angular correlations), or alternatively it can, in the proposed analysis strategy, be treated as an appropriate small re-scaling of the predicted $\bar N_{f}$.

\begin{figure}[t]
  \vspace{-0.3cm}
\begin{center}
\includegraphics[width=0.75\linewidth]{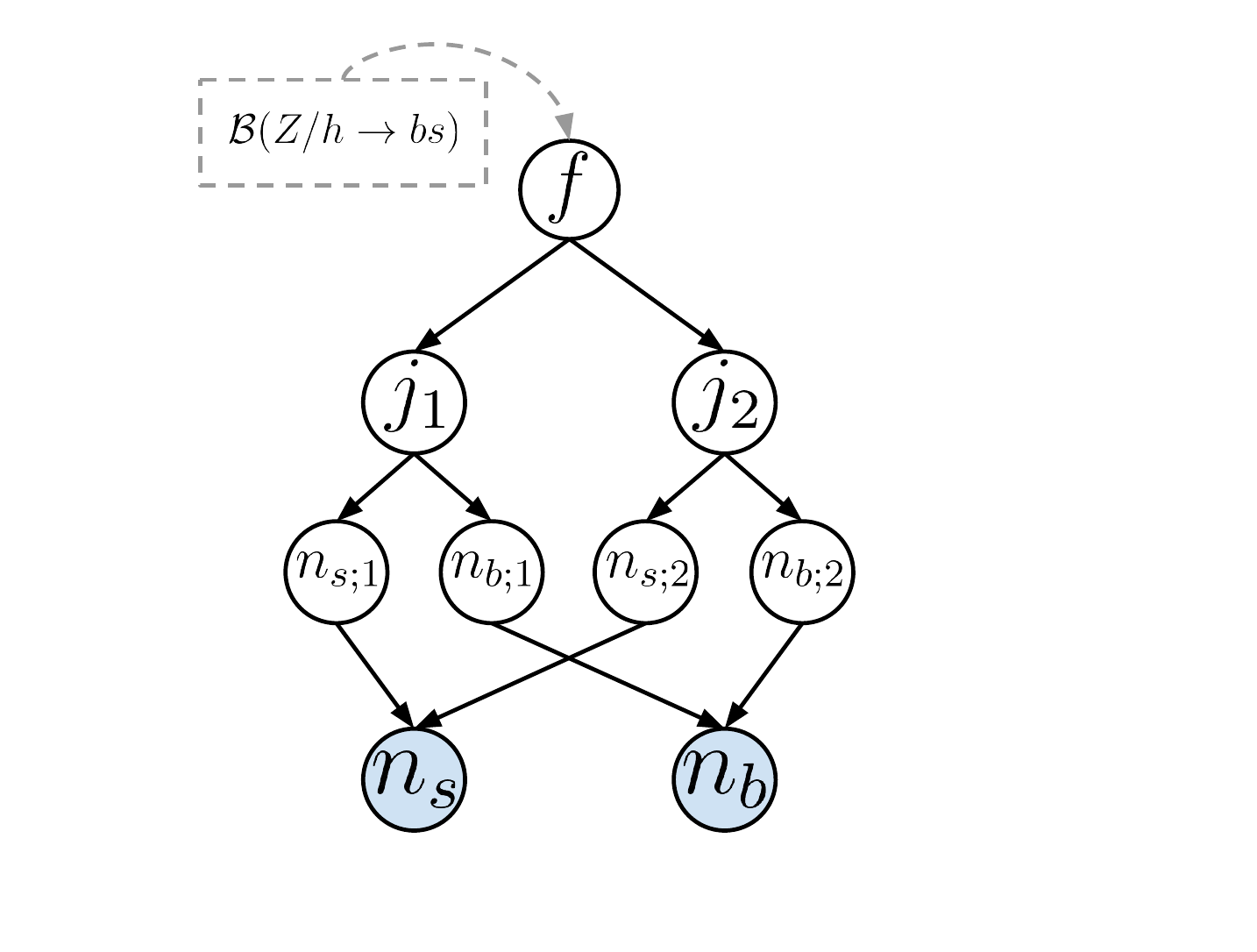}
  \vspace{-0.3cm}
\end{center}
\caption{Graphical representation of the probabilistic model for determining $\mathcal{B}(Z/h\to bs)$. Starting with the $Z/h\to f$ partonic decay, where 
$f=\{u \bar u+d \bar d (gg),s \bar s,c \bar c,b \bar b,bs\}$ for $Z(h)$,  the tagged flavours of the two final state jets, $Z/h \to j_{1} j_{2}$, are determined by the corresponding $s-$ and $b-$tagger efficiencies, $\epsilon_\beta^\alpha$. The arrows denote the probabilities for each event to end up in the $(n_b,n_s)$ bin.
}\label{fig:graphical_model}
\end{figure}

The probability distribution $p(n_{b},n_{s}|f,\nu)$ for a given event to end up in the $(n_b, n_s)$ bin depends on a number of nuisance parameters, $\nu=\{{\mathcal B}(h\to f), B(Z\to f'), \epsilon_\beta^\alpha, N_{Z/h}, {\mathcal A} \}$, which are varied within the uncertainties in the numerical analysis \footnote{The nominal values and uncertainties on the nuisance parameters used in our analysis are listed in the supplementary material, in Tables \ref{tab:syst_higgs}, \ref{tab:syst}.}.
We build a probabilistic model for $p(n_{b},n_{s}|f,\nu)$, with a graphical representation given in Fig.~\ref{fig:graphical_model} \footnote{The details of the model are relegated to the supplementary material, Sec. \ref{sec:supp:prob:model}. In particular, the rather lengthy explicit expression for $p(n_{b},n_{s}|f,\nu)$ is given in Eq.~\eqref{eq:pnbns}.}. The probability $p(n_{b},n_{s}|f,\nu)$ depends on the flavor of the initial $Z/h\to f$ parton decay, where 
$f=\{u \bar u+d \bar d (gg),s \bar s,c \bar c,b \bar b,bs\}$ for $Z(h)$, since the tagging efficiencies $\epsilon_\beta^\alpha$, $\alpha=b,s$, depend on the flavor of the initial parton. 

Experimentally, the value of ${\cal B}(Z/h\to bs)$ would be determined by comparing the measured number of events in each $(n_b,n_s)$ bin, $N_{(n_b,n_s)}$, with the expected value $\bar N_{(n_b,n_s)}$. The highest sensitivity to ${\cal B}(Z/h\to bs)$ is expected from the $(n_b,n_s)=(1,1)$ bin, however, keeping also the $(2,0)$ and $(0,2)$ bins increases the overall statistical power. In order to estimate the sensitivity of FCC-ee to ${\cal B}(Z/h\to bs)$, as a proof of concept, we can bypass the need for Monte Carlo simulations and work within the Asimov approximation~\cite{Cowan:2010js}\footnote{A recent preliminary analysis which employs Monte Carlo simulations and full tagger kinematic dependency shows similar results, see Ref.~\cite{HiggsFCNCstudy}}, both because of the simplicity of the study and especially due to the high statistics environment.  That is, we consider an ideal dataset where the observed number of events equals $N^A_{(n_b, n_s)}=\bar{N}_{(n_b, n_s)}({\mathcal B}(Z/h\to bs)_0,\nu=\nu_0)$, that is, it equals to the expected number of events for the nominal values of nuisance parameters and the input value of ${\mathcal B}(Z/h\to bs)_0$. The expected upper bound on ${\mathcal B}(Z/h\to bs)_0$ is then obtained from a maximum likelihood, allowing nuisance parameters to float \footnote{See Sec. \ref{sec:statistics} in the supplementary material for further details.}.

\begin{table}[t]
\renewcommand{\arraystretch}{1}
\centering
\begin{tabular}{ccccc}
\hline\hline
 ~Decay~ & ~~SM prediction~~  & ~exp. bound~ &~indir. constr.~ \\ \hline 
$\mathcal{B}(h\to bs)$ & $(8.9\pm 1.5)\cdot 10^{-8}$ & 0.16 & $2 \times 10^{-3}$  \\ 
$\mathcal{B}(h\to bd)$ & $(3.8\pm 0.6)\cdot 10^{-9}$ & 0.16  & $ 10^{-3}$\\ 
$\mathcal{B}(h\to cu)$ & $(2.7 \pm 0.5)\cdot 10^{-20}$ & 0.16  & $2 \times 10^{-2}$ \\ 
$\mathcal{B}(Z\to bs)$ & $(4.2\pm 0.7)\cdot 10^{-8}$ & $2.9 \times 10^{-3}$& $6 \times 10^{-8}$ \\ 
$\mathcal{B}(Z\to bd)$ & $(1.8\pm 0.3)\cdot 10^{-9}$ & $2.9 \times 10^{-3}$ & $6 \times 10^{-8}$ \\
$\mathcal{B}(Z\to cu)$ & $(1.4 \pm 0.2)\cdot 10^{-18}$ & $2.9 \times 10^{-3}$ & $4 \times 10^{-7}$ \\\hline\hline
\end{tabular}
\caption{The SM predictions and current experimental upper bounds on hadronic FCNC decays of $h$ and $Z$, either from direct searches (3rd column) or indirect constraints (4th column), where the indirect bounds on $\mathcal{B}(h\to qq')$ assume no large cancellations, see main text for details. For details on the SM calculations see supplementary material, Sec.~\ref{sec:UpdateSMpredictions}. 
} 
\label{tab:SMpredict}
\end{table}

{\bf Expected reach at FCC-ee.}
\begin{figure}[t]
  \vspace{-0.3cm}
\begin{center}
\includegraphics[width=0.75\linewidth]{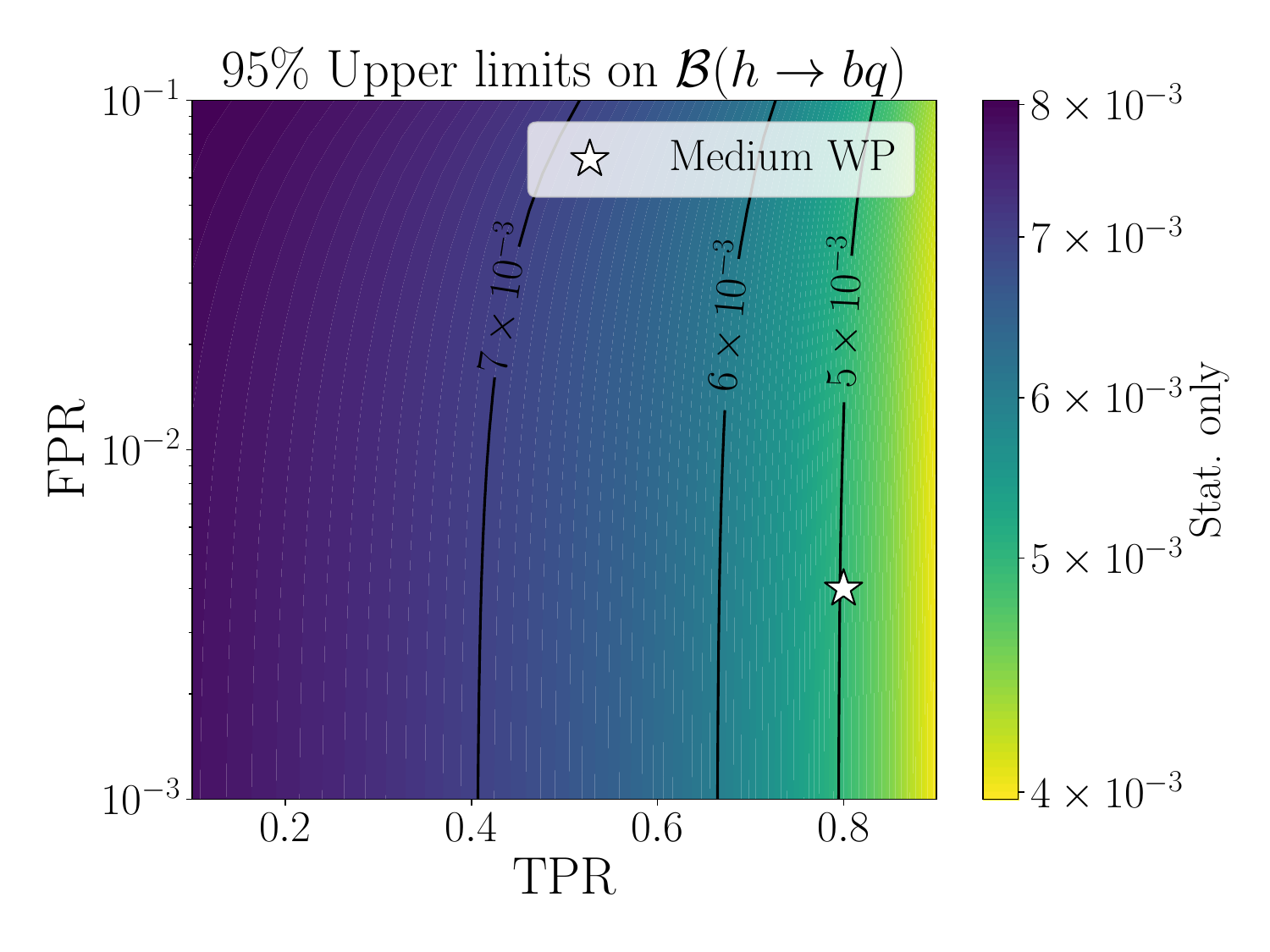}
\\
\includegraphics[width=0.75\linewidth]{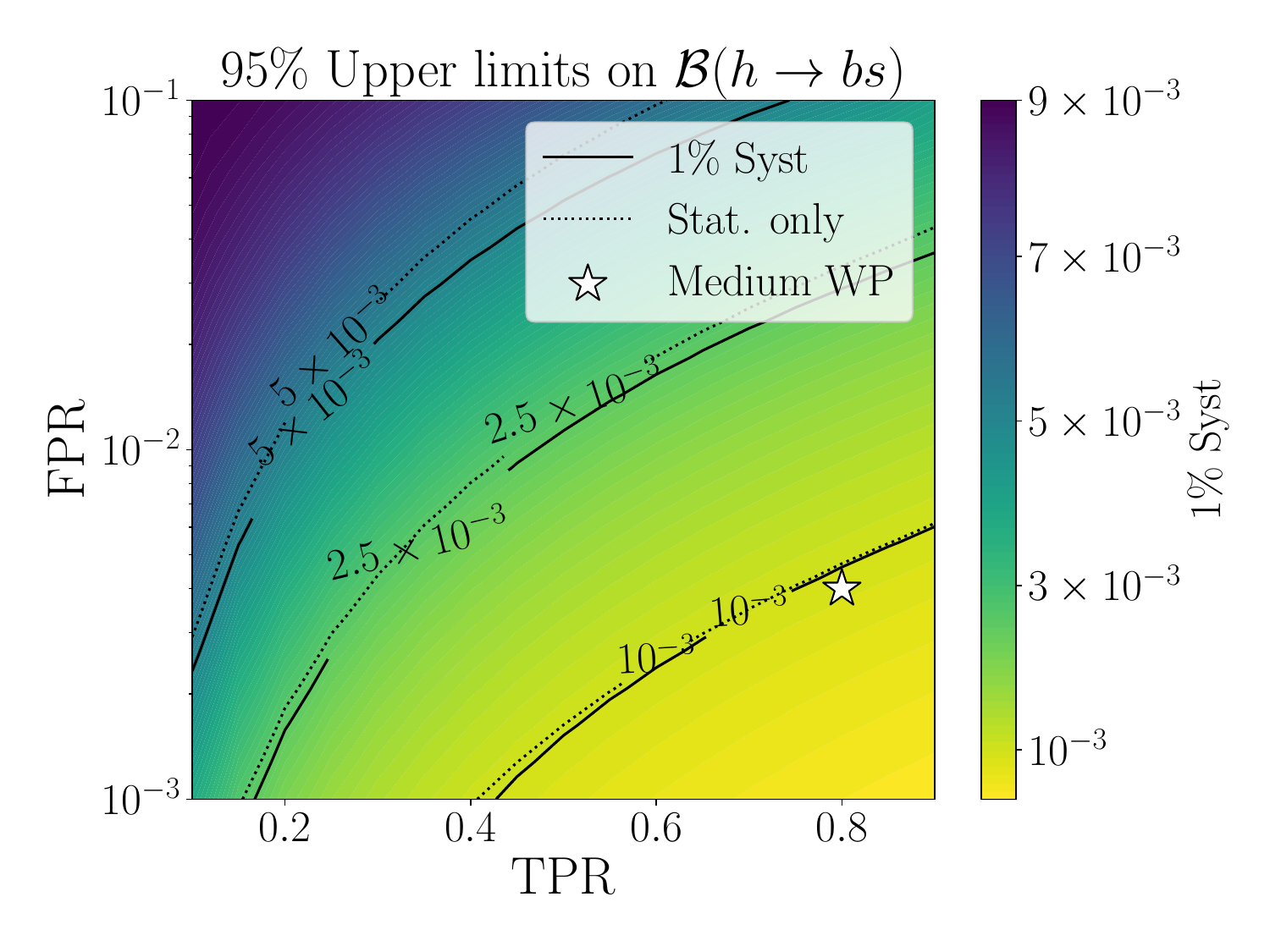}
  \vspace{-0.3cm}
  \end{center}
\caption{{\bf Top:} Expected $95\%$ CL upper bounds on $\mathcal{B}(h\to bq)$ as a function of the $b$-tagger efficiencies, neglecting systematic uncertainties. {\bf Bottom:} Expected $95\%$ CL upper bounds on $\mathcal{B}(h\to bs)$ as a function of TPR and FPR.  Solid (dashed) lines and colors are with default (no) systematic uncertainties. The Medium Working Point is based on the taggers introduced in Refs.~\cite{Bedeschi:2022rnj,tagger}. See main text for details.}\label{fig:result_Hbq}
\end{figure}
We first focus on the simplified case where only the $b$-tagger is used, and obtain the expected exclusion limits on FCNC decays summed over light quark flavors, $\mathcal{B}(h\to bq)=\mathcal{B}(h\to bd)+\mathcal{B}(h\to bs)$. The exclusions are derived from the observed yields in the $n_{b}=0,1,2$ bins. For simplicity, we parameterize the $b$-tagger as a function of two parameters: the true positive rate (TPR) $\epsilon^{b}_{b}$ and the overall effective false positive rate (FPR) for all the other initial parton flavors, $\epsilon^{b}_{gsc}$. 

\begin{figure}[t]
\begin{center}
\includegraphics[width=0.625\linewidth]{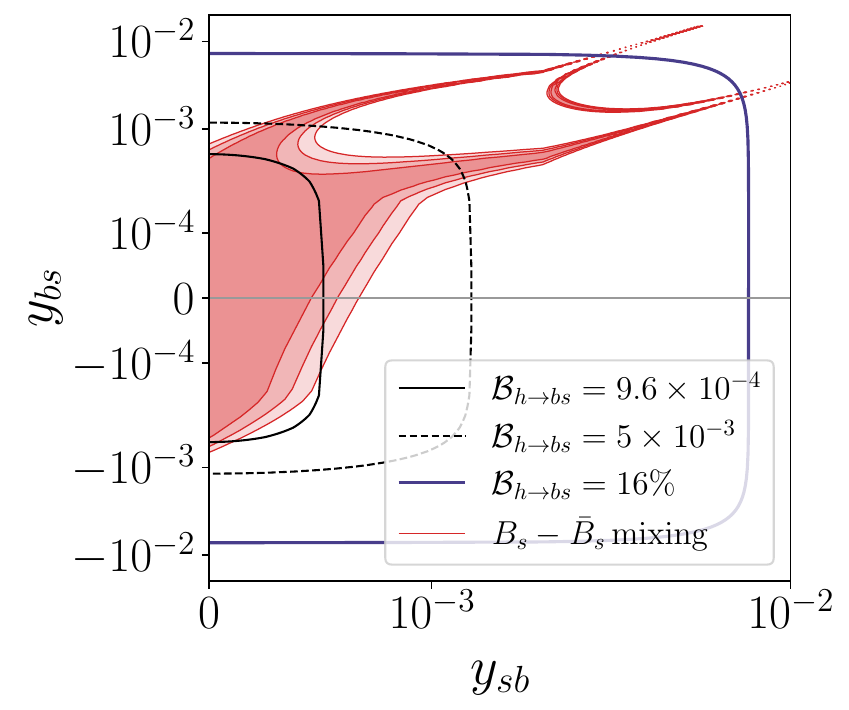}
\\
\includegraphics[width=0.625\linewidth]{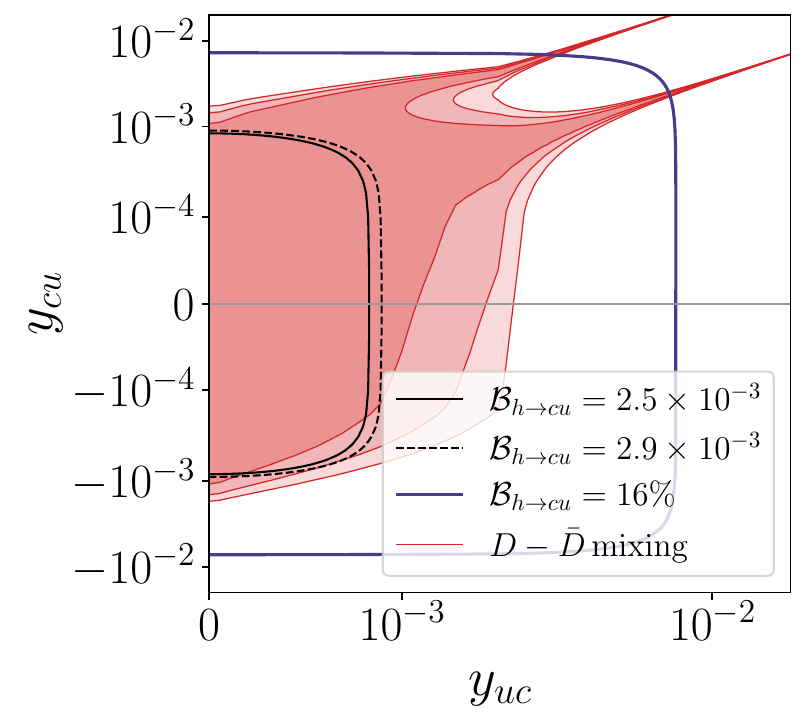}
  \vspace{-0.3cm}
  \end{center}
\caption{
{\bf Top:} Current and projected limits on $y_{sb}$ and $y_{bs}$.
{\bf Bottom:} Current and projected limits on $y_{uc}$ and $y_{cu}$. 
The  $1\sigma, 2\sigma, 3\sigma$ regions are depicted from darker to lighter red.
}
\label{fig:newPhysics_bottom_charm}
\end{figure}

The expected 95$\%$ CL upper limits on $\mathcal{B}(h\to bq)$, assuming only statistical uncertainties, are shown in Fig.~\ref{fig:result_Hbq} (top). We observe a saturation: for low enough FPR $\epsilon^{b}_{gsc}$ the upper limits become independent of  $\epsilon^{b}_{gsc}$ and depend only on $\epsilon^{b}_{b}$. With relatively modest TPR $\epsilon_b^b\in[0.4,0.8]$ and easily achievable FPR $\epsilon_{gsc}^b\lesssim 10^{-2}$ the projected bounds are $\mathcal{B}(h\to bq)\lesssim (5-7) \times 10^{-3}$. This is already in the regime that is interesting for the BSM physics searches, cf. Fig.~\ref{fig:newPhysics_bottom_charm} (top).  

However, the inclusion of strangeness tagging can result in further appreciable improvements in the expected sensitivity. Fig.~\ref{fig:result_Hbq} (bottom) shows the expected $95\%$ CL bounds on $\mathcal{B}(h\to bs)$ obtained from the comparison of all possible $(n_b,n_s)$ bins with the predictions. Here, the possible bins are $(n_b,n_s)=\{ (0,0), (0,1),(1,0), (1,1), (2,0),(0,2)\}$, where the signal mostly populates the $(n_b,n_s)= (1,1)$ bin, while the remaining bins constrain the backgrounds. To scan over possible taggers we assume in Fig.~\ref{fig:result_Hbq} (bottom) for the purpose of presentation a common TPR for $b$- and $s-$tagging, $\epsilon^{b}_{b}=\epsilon^{s}_{s}$, and similarly a common FPR, $\epsilon^{b}_{lsc}=\epsilon^{s}_{lcb}$. This assumption is not crucial, and is for instance relaxed in the analysis in Sec.~\ref{sec:details:sensitivity} of the supplementary material. 
Nevertheless, we anticipate it to give a reasonable guidance on the expected reach at FCC-ee, if the common FPR is identified as FPR=${\rm max}(\epsilon^{b}_{s},\epsilon^{s}_{b})$, where $\epsilon^{b}_{s},\epsilon^{s}_{b}$ are the actual tagger working point mis-identification rates. 
The reason is that the backgrounds with two misidentified jets are highly suppressed relative to the backgrounds with one misidentified jet, and this is more often than not dominated by the larger mis-identification rate. For instance, the performance of the common medium working point (TPR, FPR) = $(0.80,0.004)$, denoted with a star in Fig.~\ref{fig:result_Hbq} (bottom), is very close to the expected 95\% upper-limit $\mathcal{B}(h\to bs)<9.6\times 10^{-4}$, obtained when considering all the different efficiencies in the medium working point of the $b$- and $s$-taggers introduced in Refs.~\cite{Bedeschi:2022rnj,tagger}, and assuming a $1\%$ systematic uncertainty (the taggers still need to be calibrated). This limit, which does not consider other backgrounds such as Drell-Yan, $WW, ZZ, q\bar{q}$, which we expect to not affect significantly the projected reach, is competitive with indirect measurements and represents a complementary direct probe. We use this as a benchmark expected exclusion in our exploration of the impact on new physics (NP) searches. Note that the SM prediction is orders of magnitude smaller, see Table~\ref{tab:SMpredict}, so that any positive signal would mean discovery of NP.

In Fig.~\ref{fig:result_Hbq} (bottom) the relative uncertainties on the eight tagger parameters $\epsilon^\alpha_\beta$ are taken to be $1\%$ (the uncertainties are treated as independent, while the central values are common TPR, FPR). The $1\%$ uncertainty is currently below the calibrated scale factors in the LHC analyses~\cite{CMS:2017wtu,ATLAS:2019bwq}. However, given the high statistics environment at the FCC-ee, it is reasonable to expect that a dedicated calibration for high precision taggers could reach such relatively low uncertainties.  For $1\%$ systematic uncertainties the expected upper bounds on $\mathcal{B}(h\to bs)$ are statistics limited, except for very large FPR. Incidentally, this also justifies the neglect of systematics in Fig.~\ref{fig:result_Hbq} (top). 

A similar analysis can be performed to arrive at the expected FCC-ee sensitivity to $\mathcal{B}(h\to cu)$. The main difference is that the sensitivity is determined just by the performance of the $c$-tagger (there is currently no well established ``$u$-tagger''). Using the loose (medium) working point for the $c$-tagger~\cite{Bedeschi:2022rnj,tagger} leads to the $95\%$ CL expected bound for ${\cal B}(h\to cu)<2.9(2.5)\times 10^{-3}$.~\footnote{Further details can be found in the supplementary material, Sec. \ref{eq:sec:hcu}.}

\begin{figure}[t]
  \vspace{-0.3cm}
\begin{center}
\includegraphics[width=0.75\linewidth]{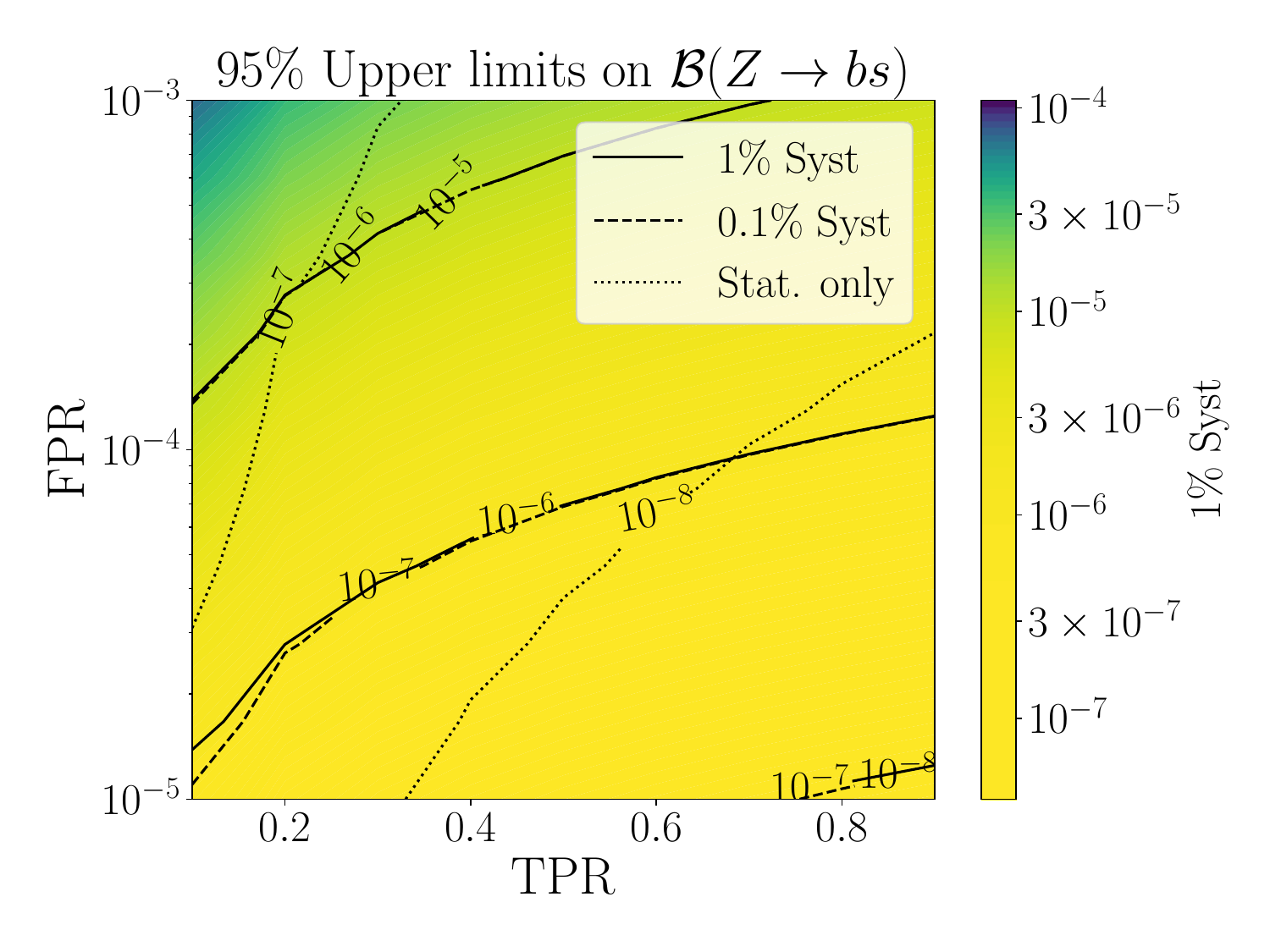}
  \vspace{-0.3cm}
  \end{center}
\caption{Expected $95\%$ CL upper bound on $\mathcal{B}(Z\to bs)$ as a function of  TPR and FPR. Solid (dashed, dotted) lines and colors are with default $1\%$ ($0.1\%$, no) systematic uncertainties.}
\label{fig:result_Zbq}
\end{figure}

We move next to the case of $Z\to bs$ decays. As before, we perform a scan over tagger efficiencies, taking the same TPR for $b$- and $s$-taggers, $\epsilon^{b}_{b}=\epsilon^{s}_{s}$, and similarly for the FPR, $\epsilon^{b}_{udsc}=\epsilon^{s}_{udcb}$. The resulting expected 95\% CL upper limits are shown in Fig. \ref{fig:result_Zbq}, where the solid (dashed, dotted) lines correspond to the default $1\%$ ($0.1\%$, no) systematic uncertainties. The FPR of $10^{-4}$ for $\epsilon_s^b$ and few$\times 10^{-3}$ for $\epsilon_b^s$ were estimated to be achievable at FCC-ee in Ref.~\cite{Bedeschi:2022rnj,tagger}. Obtaining the  $\epsilon_b^s$ well below $10^{-3}$ level will be hard, since this is roughly the fraction of $b$-quarks that decay effectively promptly, within the projected vertexing resolution of FCC-ee detectors~\cite{Barchetta:2021ibt}. To further improve on $\epsilon_b^s$ one would thus need to rely on jet shape variables to distinguish between $s$- and $b$-jets. For rather optimistic FPR of $10^{-4}$ the expected reach on $\mathcal{B}(Z\to bs)$ is $\mathcal O( 10^{-6})$ ($\mathcal O (10^{-7})$) when assuming systematics of $1\%$ (rather aggressive $0.1\%$), which is still well above the SM value (see Table~\ref{tab:SMpredict}). Given existing indirect constraints on effective $Zbs$ couplings coming from $b\to s \ell^+ \ell^-$ transitions, which have already been determined at SM rates, we conclude that it will be challenging to reach bounds on $\mathcal{B}(Z\to bs)$ that probe parameter space sensitive to NP. Similarly, the expected reach for $Z\to cu$ is $\mathcal{B}(Z\to cu)\sim 2\times10^{-3}$~\footnote{See section~\ref{eq:sec:Zcu} for further details.}, and thus well above the sensitivity of indirect probes, e.g., $\mathcal{B}(D^0 \to \mu^+\mu^-)$. We further quantify these statements below.

{\bf Sensitivity to NP.} 
We define the effective FCNC couplings of the $h$ and $Z$ bosons to $b$ and $s$ quarks as
\beq\label{eq:ZbsHbsLagrangianPheno}
\begin{split}
    \mathcal{L} \supset \, &g_{sb}^L(\bar{s}_L \gamma_\mu b_L)Z^\mu + g_{sb}^R(\bar{s}_R \gamma_\mu b_R)Z^\mu \\ 
    +& y_{sb}(\bar{s}_L b_R)h + y_{bs}(\bar{b}_L s_R)h + \mathrm{h.c.} \,,
\end{split}
\eeq
and similarly for couplings to $c$ and $u$ (or $b$ and $d$) quarks, with obvious changes in the notation.
Eq.~\eqref{eq:ZbsHbsLagrangianPheno} can be obtained as the effective low energy realization of various extensions of the SM, e.g., the addition of vector-like quarks~\cite{Fajfer:2013wca,Barducci:2017ioq}, or in the Two-Higgs-Doublet Model (2HDM)~\cite{Branco:2011iw,Crivellin:2013wna}. We provide details on these models in Sec.~\ref{sec:details:BSM} of the supplementary material, while here we focus on the relevant phenomenology.

Existing direct limits on the non-standard hadronic decays of the $Z$ follow from the agreement of the measurement and the SM prediction for the $Z$ hadronic width~\cite{OPAL:2000ufp}, giving $\mathcal{B} (Z \to qq') < 2.9 \times 10^{-3}$ at $95 \, \% \,\mathrm{CL}$, cf.~Table \ref{tab:SMpredict}. Similarly, existing Higgs boson studies at the LHC already impose limits on its undetermined decays $\mathcal{B} (h \to \mathrm{undet.}) < 0.16$ at $95 \, \% \,\mathrm{CL}$ \cite{ATLAS:2021vrm,CMS:2022dwd}. Assuming this bound is saturated by $h \to bs$ or $h \to cu$ decays, we obtain $|y_{ij},y_{ji}| \lesssim 7\times 10^{-3}$, where $ij = \{cu, bs\}$ (shown as  purple contours in Fig.~\ref{fig:newPhysics_bottom_charm}).  

At energies below the $h$ and $Z$ masses, the effective couplings in Eq.~\eqref{eq:ZbsHbsLagrangianPheno} give rise to additional contributions in numerous observables, such as the $B_s - \bar{B}_s$ mass splitting and the branching ratio for leptonic decay $B_s \to \mu^+ \mu^-$. Starting from Eq.~\eqref{eq:ZbsHbsLagrangianPheno}, we perform the matching to the Weak Effective Theory (WET) operators and employ the package \texttt{wilson}~\cite{Aebischer:2018bkb} to compute the RGE running down to the scale $\mu \sim m_b$, where we use \texttt{flavio}~\cite{Straub:2018kue} and \texttt{smelli}~\cite{Aebischer:2018iyb} to compute contributions to the relevant flavour observables and construct the resulting likelihoods. 

The $Z-bs$ couplings generate the effective $C_{9, \ell \ell}^{(\prime)}, C_{10, \ell \ell}^{(\prime)}$ coefficients in WET. The most stringent constraints on $g_{sb}^L, g_{sb}^R$ therefore come from the $b \to s \ell^+ \ell^-$ transitions. 
From the global fit we obtain $|g_{sb}^{L,R}|\lesssim10^{-5}$ with negative values of $g_{sb}^L$ slightly preferred by the current experimental results\footnote{For details see Sec.~\ref{sec:indirect:Z} and in particular Fig.~\ref{fig:SUPP:newphysicszbs}.}  (implying $\mathcal{B} (Z \to bs)$ is essentially constrained to the SM value, within uncertainties). The projected FCC-ee reach, ${\cal B}(Z\to bs)\lesssim10^{-6}$ (assuming $1\%$ systematics), can probe couplings of ${\cal O}(10^{-3})$ and is thus unable to put competitive constraints on NP. The analogous cases of $Z\to cu, bd$ are discussed in Sec.~\ref{sec:indirect:Z} with similar conclusions; the indirect bounds $|g_{uc}^{L,R}| \lesssim 3 \times 10^{-4}$ ($|g_{bd}^{L,R}| \lesssim 1 \times 10^{-4}$)  imply ${\cal B}(Z\to uc)< 4 \times 10^{-7}$ (${\cal B}(Z\to bd)< 6 \times 10^{-8}$), which are at least three orders of magnitude below the projected FCC-ee reach. 

The situation is very different for $h\to bs,cu $. The $h-bs$ effective couplings in Eq.~\eqref{eq:ZbsHbsLagrangianPheno} generate dominant contributions to scalar $(\bar b s)^2$ operators in WET, namely $C_{2,bs}^{(\prime)}$ and $C_{4,bs}$~\cite{Crivellin:2017upt}, which are probed by the $B_s$ meson mixing observables. The resulting bounds on flavor changing couplings read $|y_{bs},y_{sb}|\lesssim10^{-3}$ (baring large cancellations), as shown by the red regions in the upper panel in Fig.~\ref{fig:newPhysics_bottom_charm}. Similarly, the $D-\bar D$ mixing constraints lead to the indirect constraints on $|y_{cu},y_{uc}|\lesssim \text{few}\times10^{-3}$, shown in the lower panel in Fig.~\ref{fig:newPhysics_bottom_charm}. Excluding the regions with large cancellations, this leads to the approximate indirect bounds on $\mathcal{B} (h \to q_i q_j)$ quoted in Tab.~\ref{tab:SMpredict}\footnote{We quote present bounds on meson mixing, since  no large improvements on such indirect bounds are expected by the FCC-ee era~\cite{Charles:2020dfl}.}.
This is to be compared with the projected upper limits of FCC-ee on ${\cal B}(h\to bs)$ and ${\cal B}(h\to cu)$ shown with black lines in Fig.~\ref{fig:newPhysics_bottom_charm}. Taking the medium working point for jet-flavor taggers, the expected reach ${\cal B}(h\to bs)<9.6\times10^{-4}$ translates to the bound $|y_{bs},y_{sb}|\lesssim5\times10^{-4}$, whereas ${\cal B}(h\to cu)<2.5\times10^{-3}$ translates to $|y_{cu},y_{uc}|\lesssim 8\times10^{-4}$, as shown by the black solid lines. The latter thus improves the strongest indirect constraints on flavor-changing Higgs couplings by a factor of a few. For completeness, we show with lighter lines the expected bounds obtained employing less performative taggers. Details about $h \to bd$ can be found in Sec.~\ref{sec:details:BSM}, as well as more examples of constraints on 2HDM parameter space away from the limit of  light Higgs being the dominant contribution. 

{\bf Conclusions.}
The FCC-ee, running at the center of mass energies between the $Z$ boson mass and the $t\bar t$ threshold, will allow to measure flavor, electroweak and Higgs processes with an unprecedented level of precision. 
In this Letter we demonstrated the potential of FCC-ee  to explore flavor changing decays of the Higgs and $Z$ bosons (with similar expectations for CEPC). The projected sensitivities to ${\cal B}(h\to bs, cu)$, in particular, go well beyond the current constraints from indirect probes, such as the $B_s$ and $D$ meson oscillations. The expected reach does strongly depend on the performance of the flavor taggers, for which we explored a range of achievable efficiencies and uncertainties, based on existing measurements and ongoing studies. Auspiciously, even with rather conservative assumptions, where only the $b$-tagger is used in the analysis, the projected reach is already such that it will be able to probe significant portions of unconstrained NP parameter  space as demonstrated in Fig. \ref{fig:newPhysics_bottom_charm} (and on the example of a type III 2HDM in~\ref{sec:2HDM}). Finally, as a side-result we have also updated the SM predictions for the $h\to bs, cu$, and $Z\to bs, cu$ branching ratios. These are orders of magnitude smaller, so that any signal in these channels would unambiguously imply existence of New Physics.

{\bf Acknowledgments.} The authors would like to thank José Zurita for the updated references on the LHC upper limits on non-standard Higgs boson decays.
AK thanks Aleks Smolkovič for clarifications regarding \texttt{flavio}. JZ and MS acknowledge support in part by the DOE grant de-sc0011784 and NSF OAC-2103889. JFK, AK and MT acknowledge the financial support from the Slovenian Research Agency (grant No. J1-3013 and research core funding No. P1-0035). This work was performed in part at the Aspen Center for Physics, which is supported by National Science Foundation grant PHY-2210452.

\bibliography{references}

\begin{thebibliography}{71}%
\makeatletter
\providecommand \@ifxundefined [1]{%
 \@ifx{#1\undefined}
}%
\providecommand \@ifnum [1]{%
 \ifnum #1\expandafter \@firstoftwo
 \else \expandafter \@secondoftwo
 \fi
}%
\providecommand \@ifx [1]{%
 \ifx #1\expandafter \@firstoftwo
 \else \expandafter \@secondoftwo
 \fi
}%
\providecommand \natexlab [1]{#1}%
\providecommand \enquote  [1]{``#1''}%
\providecommand \bibnamefont  [1]{#1}%
\providecommand \bibfnamefont [1]{#1}%
\providecommand \citenamefont [1]{#1}%
\providecommand \href@noop [0]{\@secondoftwo}%
\providecommand \href [0]{\begingroup \@sanitize@url \@href}%
\providecommand \@href[1]{\@@startlink{#1}\@@href}%
\providecommand \@@href[1]{\endgroup#1\@@endlink}%
\providecommand \@sanitize@url [0]{\catcode `\\12\catcode `\$12\catcode
  `\&12\catcode `\#12\catcode `\^12\catcode `\_12\catcode `\%12\relax}%
\providecommand \@@startlink[1]{}%
\providecommand \@@endlink[0]{}%
\providecommand \url  [0]{\begingroup\@sanitize@url \@url }%
\providecommand \@url [1]{\endgroup\@href {#1}{\urlprefix }}%
\providecommand \urlprefix  [0]{URL }%
\providecommand \Eprint [0]{\href }%
\providecommand \doibase [0]{https://doi.org/}%
\providecommand \selectlanguage [0]{\@gobble}%
\providecommand \bibinfo  [0]{\@secondoftwo}%
\providecommand \bibfield  [0]{\@secondoftwo}%
\providecommand \translation [1]{[#1]}%
\providecommand \BibitemOpen [0]{}%
\providecommand \bibitemStop [0]{}%
\providecommand \bibitemNoStop [0]{.\EOS\space}%
\providecommand \EOS [0]{\spacefactor3000\relax}%
\providecommand \BibitemShut  [1]{\csname bibitem#1\endcsname}%
\let\auto@bib@innerbib\@empty
%</preamble>
\bibitem [{\citenamefont {Zupan}(2019)}]{Zupan:2019uoi}%
  \BibitemOpen
  \bibfield  {author} {\bibinfo {author} {\bibfnamefont {J.}~\bibnamefont
  {Zupan}},\ }\bibfield  {title} {\bibinfo {title} {{Introduction to flavour
  physics}},\ }\href {https://doi.org/10.23730/CYRSP-2019-006.181} {\bibfield
  {journal} {\bibinfo  {journal} {CERN Yellow Rep. School Proc.}\ }\textbf
  {\bibinfo {volume} {6}},\ \bibinfo {pages} {181} (\bibinfo {year} {2019})},\
  \Eprint {https://arxiv.org/abs/1903.05062} {arXiv:1903.05062 [hep-ph]}
  \BibitemShut {NoStop}%
\bibitem [{\citenamefont {Gori}(2019)}]{Gori:2019ybw}%
  \BibitemOpen
  \bibfield  {author} {\bibinfo {author} {\bibfnamefont {S.}~\bibnamefont
  {Gori}},\ }\bibfield  {title} {\bibinfo {title} {{TASI lectures on flavor
  physics}},\ }\href@noop {} {\bibfield  {journal} {\bibinfo  {journal} {PoS}\
  }\textbf {\bibinfo {volume} {TASI2018}},\ \bibinfo {pages} {013} (\bibinfo
  {year} {2019})}\BibitemShut {NoStop}%
\bibitem [{\citenamefont {Kamenik}(2016)}]{Kamenik:2016ygc}%
  \BibitemOpen
  \bibfield  {author} {\bibinfo {author} {\bibfnamefont {J.~F.}\ \bibnamefont
  {Kamenik}},\ }\bibfield  {title} {\bibinfo {title} {{Flavour Physics and CP
  Violation}},\ }in\ \href {https://doi.org/10.5170/CERN-2016-003.79} {\emph
  {\bibinfo {booktitle} {{2014 European School of High-Energy Physics}}}}\
  (\bibinfo {year} {2016})\ pp.\ \bibinfo {pages} {79--94},\ \Eprint
  {https://arxiv.org/abs/1708.00771} {arXiv:1708.00771 [hep-ph]} \BibitemShut
  {NoStop}%
\bibitem [{\citenamefont {Altmannshofer}\ and\ \citenamefont
  {Zupan}(2022)}]{Altmannshofer:2022aml}%
  \BibitemOpen
  \bibfield  {author} {\bibinfo {author} {\bibfnamefont {W.}~\bibnamefont
  {Altmannshofer}}\ and\ \bibinfo {author} {\bibfnamefont {J.}~\bibnamefont
  {Zupan}},\ }\bibfield  {title} {\bibinfo {title} {{Snowmass White Paper:
  Flavor Model Building}},\ }in\ \href@noop {} {\emph {\bibinfo {booktitle}
  {{Snowmass 2021}}}}\ (\bibinfo {year} {2022})\ \Eprint
  {https://arxiv.org/abs/2203.07726} {arXiv:2203.07726 [hep-ph]} \BibitemShut
  {NoStop}%
\bibitem [{\citenamefont {Grossman}\ and\ \citenamefont
  {Tanedo}(2018)}]{Grossman:2017thq}%
  \BibitemOpen
  \bibfield  {author} {\bibinfo {author} {\bibfnamefont {Y.}~\bibnamefont
  {Grossman}}\ and\ \bibinfo {author} {\bibfnamefont {P.}~\bibnamefont
  {Tanedo}},\ }\bibfield  {title} {\bibinfo {title} {{Just a Taste: Lectures on
  Flavor Physics}},\ }in\ \href {https://doi.org/10.1142/9789813233348_0004}
  {\emph {\bibinfo {booktitle} {{Theoretical Advanced Study Institute in
  Elementary Particle Physics}: {Anticipating the Next Discoveries in Particle
  Physics}}}}\ (\bibinfo {year} {2018})\ pp.\ \bibinfo {pages} {109--295},\
  \Eprint {https://arxiv.org/abs/1711.03624} {arXiv:1711.03624 [hep-ph]}
  \BibitemShut {NoStop}%
\bibitem [{\citenamefont {Kamenik}(2020)}]{Kamenik:2020fgu}%
  \BibitemOpen
  \bibfield  {author} {\bibinfo {author} {\bibfnamefont {J.~F.}\ \bibnamefont
  {Kamenik}},\ }\bibfield  {title} {\bibinfo {title} {{Flavor at low and high
  $p_T$}},\ }\href {https://doi.org/10.22323/1.377.0046} {\bibfield  {journal}
  {\bibinfo  {journal} {PoS}\ }\textbf {\bibinfo {volume} {Beauty2019}},\
  \bibinfo {pages} {046} (\bibinfo {year} {2020})}\BibitemShut {NoStop}%
\bibitem [{\citenamefont {Greljo}\ \emph {et~al.}(2023)\citenamefont {Greljo},
  \citenamefont {Salko}, \citenamefont {Smolkovi\v{c}},\ and\ \citenamefont
  {Stangl}}]{Greljo:2022jac}%
  \BibitemOpen
  \bibfield  {author} {\bibinfo {author} {\bibfnamefont {A.}~\bibnamefont
  {Greljo}}, \bibinfo {author} {\bibfnamefont {J.}~\bibnamefont {Salko}},
  \bibinfo {author} {\bibfnamefont {A.}~\bibnamefont {Smolkovi\v{c}}},\ and\
  \bibinfo {author} {\bibfnamefont {P.}~\bibnamefont {Stangl}},\ }\bibfield
  {title} {\bibinfo {title} {{Rare b decays meet high-mass Drell-Yan}},\ }\href
  {https://doi.org/10.1007/JHEP05(2023)087} {\bibfield  {journal} {\bibinfo
  {journal} {JHEP}\ }\textbf {\bibinfo {volume} {05}},\ \bibinfo {pages}
  {087}},\ \Eprint {https://arxiv.org/abs/2212.10497} {arXiv:2212.10497
  [hep-ph]} \BibitemShut {NoStop}%
\bibitem [{\citenamefont {Fuentes-Martin}\ \emph {et~al.}(2020)\citenamefont
  {Fuentes-Martin}, \citenamefont {Greljo}, \citenamefont {Martin~Camalich},\
  and\ \citenamefont {Ruiz-Alvarez}}]{Fuentes-Martin:2020lea}%
  \BibitemOpen
  \bibfield  {author} {\bibinfo {author} {\bibfnamefont {J.}~\bibnamefont
  {Fuentes-Martin}}, \bibinfo {author} {\bibfnamefont {A.}~\bibnamefont
  {Greljo}}, \bibinfo {author} {\bibfnamefont {J.}~\bibnamefont
  {Martin~Camalich}},\ and\ \bibinfo {author} {\bibfnamefont {J.~D.}\
  \bibnamefont {Ruiz-Alvarez}},\ }\bibfield  {title} {\bibinfo {title} {{Charm
  physics confronts high-p$_{T}$ lepton tails}},\ }\href
  {https://doi.org/10.1007/JHEP11(2020)080} {\bibfield  {journal} {\bibinfo
  {journal} {JHEP}\ }\textbf {\bibinfo {volume} {11}},\ \bibinfo {pages}
  {080}},\ \Eprint {https://arxiv.org/abs/2003.12421} {arXiv:2003.12421
  [hep-ph]} \BibitemShut {NoStop}%
\bibitem [{\citenamefont {Greljo}\ \emph {et~al.}(2019)\citenamefont {Greljo},
  \citenamefont {Martin~Camalich},\ and\ \citenamefont
  {Ruiz-\'Alvarez}}]{Greljo:2018tzh}%
  \BibitemOpen
  \bibfield  {author} {\bibinfo {author} {\bibfnamefont {A.}~\bibnamefont
  {Greljo}}, \bibinfo {author} {\bibfnamefont {J.}~\bibnamefont
  {Martin~Camalich}},\ and\ \bibinfo {author} {\bibfnamefont {J.~D.}\
  \bibnamefont {Ruiz-\'Alvarez}},\ }\bibfield  {title} {\bibinfo {title}
  {{Mono-$\tau$ Signatures at the LHC Constrain Explanations of $B$-decay
  Anomalies}},\ }\href {https://doi.org/10.1103/PhysRevLett.122.131803}
  {\bibfield  {journal} {\bibinfo  {journal} {Phys. Rev. Lett.}\ }\textbf
  {\bibinfo {volume} {122}},\ \bibinfo {pages} {131803} (\bibinfo {year}
  {2019})},\ \Eprint {https://arxiv.org/abs/1811.07920} {arXiv:1811.07920
  [hep-ph]} \BibitemShut {NoStop}%
\bibitem [{\citenamefont {Greljo}\ and\ \citenamefont
  {Marzocca}(2017)}]{Greljo:2017vvb}%
  \BibitemOpen
  \bibfield  {author} {\bibinfo {author} {\bibfnamefont {A.}~\bibnamefont
  {Greljo}}\ and\ \bibinfo {author} {\bibfnamefont {D.}~\bibnamefont
  {Marzocca}},\ }\bibfield  {title} {\bibinfo {title} {{High-$p_T$ dilepton
  tails and flavor physics}},\ }\href
  {https://doi.org/10.1140/epjc/s10052-017-5119-8} {\bibfield  {journal}
  {\bibinfo  {journal} {Eur. Phys. J. C}\ }\textbf {\bibinfo {volume} {77}},\
  \bibinfo {pages} {548} (\bibinfo {year} {2017})},\ \Eprint
  {https://arxiv.org/abs/1704.09015} {arXiv:1704.09015 [hep-ph]} \BibitemShut
  {NoStop}%
\bibitem [{\citenamefont {Faroughy}\ \emph {et~al.}(2017)\citenamefont
  {Faroughy}, \citenamefont {Greljo},\ and\ \citenamefont
  {Kamenik}}]{Faroughy:2016osc}%
  \BibitemOpen
  \bibfield  {author} {\bibinfo {author} {\bibfnamefont {D.~A.}\ \bibnamefont
  {Faroughy}}, \bibinfo {author} {\bibfnamefont {A.}~\bibnamefont {Greljo}},\
  and\ \bibinfo {author} {\bibfnamefont {J.~F.}\ \bibnamefont {Kamenik}},\
  }\bibfield  {title} {\bibinfo {title} {{Confronting lepton flavor
  universality violation in B decays with high-$p_T$ tau lepton searches at
  LHC}},\ }\href {https://doi.org/10.1016/j.physletb.2016.11.011} {\bibfield
  {journal} {\bibinfo  {journal} {Phys. Lett. B}\ }\textbf {\bibinfo {volume}
  {764}},\ \bibinfo {pages} {126} (\bibinfo {year} {2017})},\ \Eprint
  {https://arxiv.org/abs/1609.07138} {arXiv:1609.07138 [hep-ph]} \BibitemShut
  {NoStop}%
\bibitem [{\citenamefont {Agashe}\ \emph {et~al.}(2013)\citenamefont {Agashe}
  \emph {et~al.}}]{TopQuarkWorkingGroup:2013hxj}%
  \BibitemOpen
  \bibfield  {author} {\bibinfo {author} {\bibfnamefont {K.}~\bibnamefont
  {Agashe}} \emph {et~al.} (\bibinfo {collaboration} {Top Quark Working
  Group}),\ }\bibfield  {title} {\bibinfo {title} {{Working Group Report: Top
  Quark}},\ }in\ \href@noop {} {\emph {\bibinfo {booktitle} {{Snowmass 2013}:
  {Snowmass on the Mississippi}}}}\ (\bibinfo {year} {2013})\ \Eprint
  {https://arxiv.org/abs/1311.2028} {arXiv:1311.2028 [hep-ph]} \BibitemShut
  {NoStop}%
\bibitem [{\citenamefont {Cepeda}\ \emph {et~al.}(2019)\citenamefont {Cepeda}
  \emph {et~al.}}]{Cepeda:2019klc}%
  \BibitemOpen
  \bibfield  {author} {\bibinfo {author} {\bibfnamefont {M.}~\bibnamefont
  {Cepeda}} \emph {et~al.},\ }\bibfield  {title} {\bibinfo {title} {{Report
  from Working Group 2}: {Higgs Physics at the HL-LHC and HE-LHC}},\ }\href
  {https://doi.org/10.23731/CYRM-2019-007.221} {\bibfield  {journal} {\bibinfo
  {journal} {CERN Yellow Rep. Monogr.}\ }\textbf {\bibinfo {volume} {7}},\
  \bibinfo {pages} {221} (\bibinfo {year} {2019})},\ \Eprint
  {https://arxiv.org/abs/1902.00134} {arXiv:1902.00134 [hep-ph]} \BibitemShut
  {NoStop}%
\bibitem [{\citenamefont {Agapov}\ \emph {et~al.}(2022)\citenamefont {Agapov}
  \emph {et~al.}}]{Agapov:2022bhm}%
  \BibitemOpen
  \bibfield  {author} {\bibinfo {author} {\bibfnamefont {I.}~\bibnamefont
  {Agapov}} \emph {et~al.},\ }\bibfield  {title} {\bibinfo {title} {{Future
  Circular Lepton Collider FCC-ee: Overview and Status}},\ }in\ \href@noop {}
  {\emph {\bibinfo {booktitle} {{Snowmass 2021}}}}\ (\bibinfo {year} {2022})\
  \Eprint {https://arxiv.org/abs/2203.08310} {arXiv:2203.08310
  [physics.acc-ph]} \BibitemShut {NoStop}%
\bibitem [{\citenamefont {de~Blas}\ \emph {et~al.}(2020)\citenamefont {de~Blas}
  \emph {et~al.}}]{deBlas:2019rxi}%
  \BibitemOpen
  \bibfield  {author} {\bibinfo {author} {\bibfnamefont {J.}~\bibnamefont
  {de~Blas}} \emph {et~al.},\ }\bibfield  {title} {\bibinfo {title} {{Higgs
  Boson Studies at Future Particle Colliders}},\ }\href
  {https://doi.org/10.1007/JHEP01(2020)139} {\bibfield  {journal} {\bibinfo
  {journal} {JHEP}\ }\textbf {\bibinfo {volume} {01}},\ \bibinfo {pages}
  {139}},\ \Eprint {https://arxiv.org/abs/1905.03764} {arXiv:1905.03764
  [hep-ph]} \BibitemShut {NoStop}%
\bibitem [{\citenamefont {Mangano}\ \emph {et~al.}(2019)\citenamefont
  {Mangano}, \citenamefont {Azzi}, \citenamefont {Benedikt}, \citenamefont
  {Blondel}, \citenamefont {Britzger}, \citenamefont {Dainese}, \citenamefont
  {Dam}, \citenamefont {de~Blas}, \citenamefont {Enterria}, \citenamefont
  {Fischer}, \citenamefont {Grojean}, \citenamefont {Gutleber}, \citenamefont
  {Gwenlan}, \citenamefont {Helsens}, \citenamefont {Janot}, \citenamefont
  {Klein}, \citenamefont {Klein}, \citenamefont {Mccullough}, \citenamefont
  {Monteil}, \citenamefont {Poole}, \citenamefont {Ramsey-Musolf},
  \citenamefont {Schwanenberger}, \citenamefont {Selvaggi}, \citenamefont
  {Zimmermann},\ and\ \citenamefont {You}}]{Mangano:2651294}%
  \BibitemOpen
  \bibfield  {author} {\bibinfo {author} {\bibfnamefont {M.}~\bibnamefont
  {Mangano}}, \bibinfo {author} {\bibfnamefont {P.}~\bibnamefont {Azzi}},
  \bibinfo {author} {\bibfnamefont {M.}~\bibnamefont {Benedikt}}, \bibinfo
  {author} {\bibfnamefont {A.}~\bibnamefont {Blondel}}, \bibinfo {author}
  {\bibfnamefont {D.~A.}\ \bibnamefont {Britzger}}, \bibinfo {author}
  {\bibfnamefont {A.}~\bibnamefont {Dainese}}, \bibinfo {author} {\bibfnamefont
  {M.}~\bibnamefont {Dam}}, \bibinfo {author} {\bibfnamefont {J.}~\bibnamefont
  {de~Blas}}, \bibinfo {author} {\bibfnamefont {D.}~\bibnamefont {Enterria}},
  \bibinfo {author} {\bibfnamefont {O.}~\bibnamefont {Fischer}}, \bibinfo
  {author} {\bibfnamefont {C.}~\bibnamefont {Grojean}}, \bibinfo {author}
  {\bibfnamefont {J.}~\bibnamefont {Gutleber}}, \bibinfo {author}
  {\bibfnamefont {C.}~\bibnamefont {Gwenlan}}, \bibinfo {author} {\bibfnamefont
  {C.}~\bibnamefont {Helsens}}, \bibinfo {author} {\bibfnamefont
  {P.}~\bibnamefont {Janot}}, \bibinfo {author} {\bibfnamefont
  {M.}~\bibnamefont {Klein}}, \bibinfo {author} {\bibfnamefont
  {U.}~\bibnamefont {Klein}}, \bibinfo {author} {\bibfnamefont {M.~P.}\
  \bibnamefont {Mccullough}}, \bibinfo {author} {\bibfnamefont
  {S.}~\bibnamefont {Monteil}}, \bibinfo {author} {\bibfnamefont
  {J.}~\bibnamefont {Poole}}, \bibinfo {author} {\bibfnamefont
  {M.}~\bibnamefont {Ramsey-Musolf}}, \bibinfo {author} {\bibfnamefont
  {C.}~\bibnamefont {Schwanenberger}}, \bibinfo {author} {\bibfnamefont
  {M.}~\bibnamefont {Selvaggi}}, \bibinfo {author} {\bibfnamefont
  {F.}~\bibnamefont {Zimmermann}},\ and\ \bibinfo {author} {\bibfnamefont
  {T.}~\bibnamefont {You}},\ }\href
  {https://doi.org/10.1140/epjc/s10052-019-6904-3} {\emph {\bibinfo {title}
  {{FCC Physics Opportunities: Future Circular Collider Conceptual Design
  Report Volume 1. Future Circular Collider}}}},\ \bibinfo {type} {Tech. Rep.}\
  \bibinfo {number} {6}\ (\bibinfo  {institution} {CERN},\ \bibinfo {address}
  {Geneva},\ \bibinfo {year} {2019})\BibitemShut {NoStop}%
\bibitem [{\citenamefont {Abada}\ \emph {et~al.}(2019)\citenamefont {Abada}
  \emph {et~al.}}]{FCC:2018evy}%
  \BibitemOpen
  \bibfield  {author} {\bibinfo {author} {\bibfnamefont {A.}~\bibnamefont
  {Abada}} \emph {et~al.} (\bibinfo {collaboration} {FCC}),\ }\bibfield
  {title} {\bibinfo {title} {{FCC-ee: The Lepton Collider}: {Future Circular
  Collider Conceptual Design Report Volume 2}},\ }\href
  {https://doi.org/10.1140/epjst/e2019-900045-4} {\bibfield  {journal}
  {\bibinfo  {journal} {Eur. Phys. J. ST}\ }\textbf {\bibinfo {volume} {228}},\
  \bibinfo {pages} {261} (\bibinfo {year} {2019})}\BibitemShut {NoStop}%
\bibitem [{\citenamefont {Cheng}\ \emph {et~al.}(2022)\citenamefont {Cheng}
  \emph {et~al.}}]{CEPCPhysicsStudyGroup:2022uwl}%
  \BibitemOpen
  \bibfield  {author} {\bibinfo {author} {\bibfnamefont {H.}~\bibnamefont
  {Cheng}} \emph {et~al.} (\bibinfo {collaboration} {CEPC Physics Study
  Group}),\ }\bibfield  {title} {\bibinfo {title} {{The Physics potential of
  the CEPC. Prepared for the US Snowmass Community Planning Exercise (Snowmass
  2021)}},\ }in\ \href@noop {} {\emph {\bibinfo {booktitle} {{Snowmass
  2021}}}}\ (\bibinfo {year} {2022})\ \Eprint
  {https://arxiv.org/abs/2205.08553} {arXiv:2205.08553 [hep-ph]} \BibitemShut
  {NoStop}%
\bibitem [{\citenamefont {Barducci}\ and\ \citenamefont
  {Helmboldt}(2017)}]{Barducci:2017ioq}%
  \BibitemOpen
  \bibfield  {author} {\bibinfo {author} {\bibfnamefont {D.}~\bibnamefont
  {Barducci}}\ and\ \bibinfo {author} {\bibfnamefont {A.~J.}\ \bibnamefont
  {Helmboldt}},\ }\bibfield  {title} {\bibinfo {title} {{Quark
  flavour-violating Higgs decays at the ILC}},\ }\href
  {https://doi.org/10.1007/JHEP12(2017)105} {\bibfield  {journal} {\bibinfo
  {journal} {JHEP}\ }\textbf {\bibinfo {volume} {12}},\ \bibinfo {pages}
  {105}},\ \Eprint {https://arxiv.org/abs/1710.06657} {arXiv:1710.06657
  [hep-ph]} \BibitemShut {NoStop}%
\bibitem [{\citenamefont {Aad}\ \emph {et~al.}(2022)\citenamefont {Aad} \emph
  {et~al.}}]{ATLAS:2022ers}%
  \BibitemOpen
  \bibfield  {author} {\bibinfo {author} {\bibfnamefont {G.}~\bibnamefont
  {Aad}} \emph {et~al.} (\bibinfo {collaboration} {ATLAS}),\ }\bibfield
  {title} {\bibinfo {title} {{Direct constraint on the Higgs-charm coupling
  from a search for Higgs boson decays into charm quarks with the ATLAS
  detector}},\ }\href {https://doi.org/10.1140/epjc/s10052-022-10588-3}
  {\bibfield  {journal} {\bibinfo  {journal} {Eur. Phys. J. C}\ }\textbf
  {\bibinfo {volume} {82}},\ \bibinfo {pages} {717} (\bibinfo {year} {2022})},\
  \Eprint {https://arxiv.org/abs/2201.11428} {arXiv:2201.11428 [hep-ex]}
  \BibitemShut {NoStop}%
\bibitem [{\citenamefont {Faroughy}\ \emph {et~al.}(2022)\citenamefont
  {Faroughy}, \citenamefont {Kamenik}, \citenamefont {Szewc},\ and\
  \citenamefont {Zupan}}]{Faroughy:2022dyq}%
  \BibitemOpen
  \bibfield  {author} {\bibinfo {author} {\bibfnamefont {D.~A.}\ \bibnamefont
  {Faroughy}}, \bibinfo {author} {\bibfnamefont {J.~F.}\ \bibnamefont
  {Kamenik}}, \bibinfo {author} {\bibfnamefont {M.}~\bibnamefont {Szewc}},\
  and\ \bibinfo {author} {\bibfnamefont {J.}~\bibnamefont {Zupan}},\ }\bibfield
   {title} {\bibinfo {title} {{Accessing CKM suppressed top decays at the
  LHC}},\ }\href@noop {} {\  (\bibinfo {year} {2022})},\ \Eprint
  {https://arxiv.org/abs/2209.01222} {arXiv:2209.01222 [hep-ph]} \BibitemShut
  {NoStop}%
\bibitem [{\citenamefont {Sirunyan}\ \emph {et~al.}(2020)\citenamefont
  {Sirunyan} \emph {et~al.}}]{CMS:2020vac}%
  \BibitemOpen
  \bibfield  {author} {\bibinfo {author} {\bibfnamefont {A.~M.}\ \bibnamefont
  {Sirunyan}} \emph {et~al.} (\bibinfo {collaboration} {CMS}),\ }\bibfield
  {title} {\bibinfo {title} {{Measurement of CKM matrix elements in single top
  quark $t$-channel production in proton-proton collisions at $\sqrt{s} = $ 13
  TeV}},\ }\href {https://doi.org/10.1016/j.physletb.2020.135609} {\bibfield
  {journal} {\bibinfo  {journal} {Phys. Lett. B}\ }\textbf {\bibinfo {volume}
  {808}},\ \bibinfo {pages} {135609} (\bibinfo {year} {2020})},\ \Eprint
  {https://arxiv.org/abs/2004.12181} {arXiv:2004.12181 [hep-ex]} \BibitemShut
  {NoStop}%
\bibitem [{\citenamefont {Cowan}\ \emph {et~al.}(2011)\citenamefont {Cowan},
  \citenamefont {Cranmer}, \citenamefont {Gross},\ and\ \citenamefont
  {Vitells}}]{Cowan:2010js}%
  \BibitemOpen
  \bibfield  {author} {\bibinfo {author} {\bibfnamefont {G.}~\bibnamefont
  {Cowan}}, \bibinfo {author} {\bibfnamefont {K.}~\bibnamefont {Cranmer}},
  \bibinfo {author} {\bibfnamefont {E.}~\bibnamefont {Gross}},\ and\ \bibinfo
  {author} {\bibfnamefont {O.}~\bibnamefont {Vitells}},\ }\bibfield  {title}
  {\bibinfo {title} {{Asymptotic formulae for likelihood-based tests of new
  physics}},\ }\href {https://doi.org/10.1140/epjc/s10052-011-1554-0}
  {\bibfield  {journal} {\bibinfo  {journal} {Eur. Phys. J. C}\ }\textbf
  {\bibinfo {volume} {71}},\ \bibinfo {pages} {1554} (\bibinfo {year}
  {2011})},\ \bibinfo {note} {[Erratum: Eur.Phys.J.C 73, 2501 (2013)]},\
  \Eprint {https://arxiv.org/abs/1007.1727} {arXiv:1007.1727 [physics.data-an]}
  \BibitemShut {NoStop}%
\bibitem [{\citenamefont {ee~Higgs Working~Group}()}]{HiggsFCNCstudy}%
  \BibitemOpen
  \bibfield  {author} {\bibinfo {author} {\bibfnamefont {F.}~\bibnamefont
  {ee~Higgs Working~Group}} (\bibinfo {collaboration} {FCC}),\ }\bibfield
  {title} {\bibinfo {title} {Fcc-ee higgs working group},\ }\bibinfo {note}
  {\url{https://indico.cern.ch/event/1304164/contributions/5513419/attachments/2689281/4666483/ZHvvjj_24072023.pdf}}\BibitemShut
  {NoStop}%
\bibitem [{\citenamefont {Bedeschi}\ \emph {et~al.}(2022)\citenamefont
  {Bedeschi}, \citenamefont {Gouskos},\ and\ \citenamefont
  {Selvaggi}}]{Bedeschi:2022rnj}%
  \BibitemOpen
  \bibfield  {author} {\bibinfo {author} {\bibfnamefont {F.}~\bibnamefont
  {Bedeschi}}, \bibinfo {author} {\bibfnamefont {L.}~\bibnamefont {Gouskos}},\
  and\ \bibinfo {author} {\bibfnamefont {M.}~\bibnamefont {Selvaggi}},\
  }\bibfield  {title} {\bibinfo {title} {{Jet flavour tagging for future
  colliders with fast simulation}},\ }\href
  {https://doi.org/10.1140/epjc/s10052-022-10609-1} {\bibfield  {journal}
  {\bibinfo  {journal} {Eur. Phys. J. C}\ }\textbf {\bibinfo {volume} {82}},\
  \bibinfo {pages} {646} (\bibinfo {year} {2022})},\ \Eprint
  {https://arxiv.org/abs/2202.03285} {arXiv:2202.03285 [hep-ex]} \BibitemShut
  {NoStop}%
\bibitem [{\citenamefont {Gouskos}(2023)}]{tagger}%
  \BibitemOpen
  \bibfield  {author} {\bibinfo {author} {\bibfnamefont {L.}~\bibnamefont
  {Gouskos}},\ }\href
  {https://indico.cern.ch/event/1176398/contributions/5207197/attachments/2582238/4453976/lg-jettagging-fccee-krakow2023.pdf}
  {\bibinfo {title} {{Jet flavor identification for FCCee}}} (\bibinfo {year}
  {2023})\BibitemShut {NoStop}%
\bibitem [{\citenamefont {Sirunyan}\ \emph {et~al.}(2018)\citenamefont
  {Sirunyan} \emph {et~al.}}]{CMS:2017wtu}%
  \BibitemOpen
  \bibfield  {author} {\bibinfo {author} {\bibfnamefont {A.~M.}\ \bibnamefont
  {Sirunyan}} \emph {et~al.} (\bibinfo {collaboration} {CMS}),\ }\bibfield
  {title} {\bibinfo {title} {{Identification of heavy-flavour jets with the CMS
  detector in pp collisions at 13 TeV}},\ }\href
  {https://doi.org/10.1088/1748-0221/13/05/P05011} {\bibfield  {journal}
  {\bibinfo  {journal} {JINST}\ }\textbf {\bibinfo {volume} {13}}\bibfield
  {number} {\bibinfo  {number} { (05)},\ \bibinfo {pages} {P05011}},\ }\Eprint
  {https://arxiv.org/abs/1712.07158} {arXiv:1712.07158 [physics.ins-det]}
  \BibitemShut {NoStop}%
\bibitem [{\citenamefont {Aad}\ \emph {et~al.}(2019)\citenamefont {Aad} \emph
  {et~al.}}]{ATLAS:2019bwq}%
  \BibitemOpen
  \bibfield  {author} {\bibinfo {author} {\bibfnamefont {G.}~\bibnamefont
  {Aad}} \emph {et~al.} (\bibinfo {collaboration} {ATLAS}),\ }\bibfield
  {title} {\bibinfo {title} {{ATLAS b-jet identification performance and
  efficiency measurement with $t{\bar{t}}$ events in pp collisions at
  $\sqrt{s}=13$ TeV}},\ }\href {https://doi.org/10.1140/epjc/s10052-019-7450-8}
  {\bibfield  {journal} {\bibinfo  {journal} {Eur. Phys. J. C}\ }\textbf
  {\bibinfo {volume} {79}},\ \bibinfo {pages} {970} (\bibinfo {year} {2019})},\
  \Eprint {https://arxiv.org/abs/1907.05120} {arXiv:1907.05120 [hep-ex]}
  \BibitemShut {NoStop}%
\bibitem [{\citenamefont {Barchetta}\ \emph {et~al.}(2022)\citenamefont
  {Barchetta}, \citenamefont {Collins},\ and\ \citenamefont
  {Riedler}}]{Barchetta:2021ibt}%
  \BibitemOpen
  \bibfield  {author} {\bibinfo {author} {\bibfnamefont {N.}~\bibnamefont
  {Barchetta}}, \bibinfo {author} {\bibfnamefont {P.}~\bibnamefont {Collins}},\
  and\ \bibinfo {author} {\bibfnamefont {P.}~\bibnamefont {Riedler}},\
  }\bibfield  {title} {\bibinfo {title} {{Tracking and vertex detectors at
  FCC-ee}},\ }\href {https://doi.org/10.1140/epjp/s13360-021-02323-w}
  {\bibfield  {journal} {\bibinfo  {journal} {Eur. Phys. J. Plus}\ }\textbf
  {\bibinfo {volume} {137}},\ \bibinfo {pages} {231} (\bibinfo {year}
  {2022})},\ \Eprint {https://arxiv.org/abs/2112.13019} {arXiv:2112.13019
  [physics.ins-det]} \BibitemShut {NoStop}%
\bibitem [{\citenamefont {Fajfer}\ \emph {et~al.}(2013)\citenamefont {Fajfer},
  \citenamefont {Greljo}, \citenamefont {Kamenik},\ and\ \citenamefont
  {Mustac}}]{Fajfer:2013wca}%
  \BibitemOpen
  \bibfield  {author} {\bibinfo {author} {\bibfnamefont {S.}~\bibnamefont
  {Fajfer}}, \bibinfo {author} {\bibfnamefont {A.}~\bibnamefont {Greljo}},
  \bibinfo {author} {\bibfnamefont {J.~F.}\ \bibnamefont {Kamenik}},\ and\
  \bibinfo {author} {\bibfnamefont {I.}~\bibnamefont {Mustac}},\ }\bibfield
  {title} {\bibinfo {title} {{Light Higgs and Vector-like Quarks without
  Prejudice}},\ }\href {https://doi.org/10.1007/JHEP07(2013)155} {\bibfield
  {journal} {\bibinfo  {journal} {JHEP}\ }\textbf {\bibinfo {volume} {07}},\
  \bibinfo {pages} {155}},\ \Eprint {https://arxiv.org/abs/1304.4219}
  {arXiv:1304.4219 [hep-ph]} \BibitemShut {NoStop}%
\bibitem [{\citenamefont {Branco}\ \emph {et~al.}(2012)\citenamefont {Branco},
  \citenamefont {Ferreira}, \citenamefont {Lavoura}, \citenamefont {Rebelo},
  \citenamefont {Sher},\ and\ \citenamefont {Silva}}]{Branco:2011iw}%
  \BibitemOpen
  \bibfield  {author} {\bibinfo {author} {\bibfnamefont {G.~C.}\ \bibnamefont
  {Branco}}, \bibinfo {author} {\bibfnamefont {P.~M.}\ \bibnamefont
  {Ferreira}}, \bibinfo {author} {\bibfnamefont {L.}~\bibnamefont {Lavoura}},
  \bibinfo {author} {\bibfnamefont {M.~N.}\ \bibnamefont {Rebelo}}, \bibinfo
  {author} {\bibfnamefont {M.}~\bibnamefont {Sher}},\ and\ \bibinfo {author}
  {\bibfnamefont {J.~P.}\ \bibnamefont {Silva}},\ }\bibfield  {title} {\bibinfo
  {title} {{Theory and phenomenology of two-Higgs-doublet models}},\ }\href
  {https://doi.org/10.1016/j.physrep.2012.02.002} {\bibfield  {journal}
  {\bibinfo  {journal} {Phys. Rept.}\ }\textbf {\bibinfo {volume} {516}},\
  \bibinfo {pages} {1} (\bibinfo {year} {2012})},\ \Eprint
  {https://arxiv.org/abs/1106.0034} {arXiv:1106.0034 [hep-ph]} \BibitemShut
  {NoStop}%
\bibitem [{\citenamefont {Crivellin}\ \emph {et~al.}(2013)\citenamefont
  {Crivellin}, \citenamefont {Kokulu},\ and\ \citenamefont
  {Greub}}]{Crivellin:2013wna}%
  \BibitemOpen
  \bibfield  {author} {\bibinfo {author} {\bibfnamefont {A.}~\bibnamefont
  {Crivellin}}, \bibinfo {author} {\bibfnamefont {A.}~\bibnamefont {Kokulu}},\
  and\ \bibinfo {author} {\bibfnamefont {C.}~\bibnamefont {Greub}},\ }\bibfield
   {title} {\bibinfo {title} {{Flavor-phenomenology of two-Higgs-doublet models
  with generic Yukawa structure}},\ }\href
  {https://doi.org/10.1103/PhysRevD.87.094031} {\bibfield  {journal} {\bibinfo
  {journal} {Phys. Rev. D}\ }\textbf {\bibinfo {volume} {87}},\ \bibinfo
  {pages} {094031} (\bibinfo {year} {2013})},\ \Eprint
  {https://arxiv.org/abs/1303.5877} {arXiv:1303.5877 [hep-ph]} \BibitemShut
  {NoStop}%
\bibitem [{\citenamefont {Abbiendi}\ \emph {et~al.}(2001)\citenamefont
  {Abbiendi} \emph {et~al.}}]{OPAL:2000ufp}%
  \BibitemOpen
  \bibfield  {author} {\bibinfo {author} {\bibfnamefont {G.}~\bibnamefont
  {Abbiendi}} \emph {et~al.} (\bibinfo {collaboration} {OPAL}),\ }\bibfield
  {title} {\bibinfo {title} {{Precise determination of the Z resonance
  parameters at LEP: 'Zedometry'}},\ }\href
  {https://doi.org/10.1007/s100520100627} {\bibfield  {journal} {\bibinfo
  {journal} {Eur. Phys. J. C}\ }\textbf {\bibinfo {volume} {19}},\ \bibinfo
  {pages} {587} (\bibinfo {year} {2001})},\ \Eprint
  {https://arxiv.org/abs/hep-ex/0012018} {arXiv:hep-ex/0012018} \BibitemShut
  {NoStop}%
\bibitem [{ATL(2021)}]{ATLAS:2021vrm}%
  \BibitemOpen
  \bibfield  {title} {\bibinfo {title} {{Combined measurements of Higgs boson
  production and decay using up to $139$ fb$^{-1}$ of proton-proton collision
  data at $\sqrt{s}= 13$ TeV collected with the ATLAS experiment}},\
  }\href@noop {} {\  (\bibinfo {year} {2021})}\BibitemShut {NoStop}%
\bibitem [{\citenamefont {Tumasyan}\ \emph {et~al.}(2022)\citenamefont
  {Tumasyan} \emph {et~al.}}]{CMS:2022dwd}%
  \BibitemOpen
  \bibfield  {author} {\bibinfo {author} {\bibfnamefont {A.}~\bibnamefont
  {Tumasyan}} \emph {et~al.} (\bibinfo {collaboration} {CMS}),\ }\bibfield
  {title} {\bibinfo {title} {{A portrait of the Higgs boson by the CMS
  experiment ten years after the discovery}},\ }\href
  {https://doi.org/10.1038/s41586-022-04892-x} {\bibfield  {journal} {\bibinfo
  {journal} {Nature}\ }\textbf {\bibinfo {volume} {607}},\ \bibinfo {pages}
  {60} (\bibinfo {year} {2022})},\ \Eprint {https://arxiv.org/abs/2207.00043}
  {arXiv:2207.00043 [hep-ex]} \BibitemShut {NoStop}%
\bibitem [{\citenamefont {Aebischer}\ \emph {et~al.}(2018)\citenamefont
  {Aebischer}, \citenamefont {Kumar},\ and\ \citenamefont
  {Straub}}]{Aebischer:2018bkb}%
  \BibitemOpen
  \bibfield  {author} {\bibinfo {author} {\bibfnamefont {J.}~\bibnamefont
  {Aebischer}}, \bibinfo {author} {\bibfnamefont {J.}~\bibnamefont {Kumar}},\
  and\ \bibinfo {author} {\bibfnamefont {D.~M.}\ \bibnamefont {Straub}},\
  }\bibfield  {title} {\bibinfo {title} {{Wilson: a Python package for the
  running and matching of Wilson coefficients above and below the electroweak
  scale}},\ }\href {https://doi.org/10.1140/epjc/s10052-018-6492-7} {\bibfield
  {journal} {\bibinfo  {journal} {Eur. Phys. J. C}\ }\textbf {\bibinfo {volume}
  {78}},\ \bibinfo {pages} {1026} (\bibinfo {year} {2018})},\ \Eprint
  {https://arxiv.org/abs/1804.05033} {arXiv:1804.05033 [hep-ph]} \BibitemShut
  {NoStop}%
\bibitem [{\citenamefont {Straub}(2018)}]{Straub:2018kue}%
  \BibitemOpen
  \bibfield  {author} {\bibinfo {author} {\bibfnamefont {D.~M.}\ \bibnamefont
  {Straub}},\ }\bibfield  {title} {\bibinfo {title} {{flavio: a Python package
  for flavour and precision phenomenology in the Standard Model and beyond}},\
  }\href@noop {} {\  (\bibinfo {year} {2018})},\ \Eprint
  {https://arxiv.org/abs/1810.08132} {arXiv:1810.08132 [hep-ph]} \BibitemShut
  {NoStop}%
\bibitem [{\citenamefont {Aebischer}\ \emph {et~al.}(2019)\citenamefont
  {Aebischer}, \citenamefont {Kumar}, \citenamefont {Stangl},\ and\
  \citenamefont {Straub}}]{Aebischer:2018iyb}%
  \BibitemOpen
  \bibfield  {author} {\bibinfo {author} {\bibfnamefont {J.}~\bibnamefont
  {Aebischer}}, \bibinfo {author} {\bibfnamefont {J.}~\bibnamefont {Kumar}},
  \bibinfo {author} {\bibfnamefont {P.}~\bibnamefont {Stangl}},\ and\ \bibinfo
  {author} {\bibfnamefont {D.~M.}\ \bibnamefont {Straub}},\ }\bibfield  {title}
  {\bibinfo {title} {{A Global Likelihood for Precision Constraints and Flavour
  Anomalies}},\ }\href {https://doi.org/10.1140/epjc/s10052-019-6977-z}
  {\bibfield  {journal} {\bibinfo  {journal} {Eur. Phys. J. C}\ }\textbf
  {\bibinfo {volume} {79}},\ \bibinfo {pages} {509} (\bibinfo {year} {2019})},\
  \Eprint {https://arxiv.org/abs/1810.07698} {arXiv:1810.07698 [hep-ph]}
  \BibitemShut {NoStop}%
\bibitem [{\citenamefont {Crivellin}\ \emph {et~al.}(2018)\citenamefont
  {Crivellin}, \citenamefont {Heeck},\ and\ \citenamefont
  {M\"uller}}]{Crivellin:2017upt}%
  \BibitemOpen
  \bibfield  {author} {\bibinfo {author} {\bibfnamefont {A.}~\bibnamefont
  {Crivellin}}, \bibinfo {author} {\bibfnamefont {J.}~\bibnamefont {Heeck}},\
  and\ \bibinfo {author} {\bibfnamefont {D.}~\bibnamefont {M\"uller}},\
  }\bibfield  {title} {\bibinfo {title} {{Large $h\to b s$ in generic
  two-Higgs-doublet models}},\ }\href
  {https://doi.org/10.1103/PhysRevD.97.035008} {\bibfield  {journal} {\bibinfo
  {journal} {Phys. Rev. D}\ }\textbf {\bibinfo {volume} {97}},\ \bibinfo
  {pages} {035008} (\bibinfo {year} {2018})},\ \Eprint
  {https://arxiv.org/abs/1710.04663} {arXiv:1710.04663 [hep-ph]} \BibitemShut
  {NoStop}%
\bibitem [{\citenamefont {Charles}\ \emph {et~al.}(2020)\citenamefont
  {Charles}, \citenamefont {Descotes-Genon}, \citenamefont {Ligeti},
  \citenamefont {Monteil}, \citenamefont {Papucci}, \citenamefont {Trabelsi},\
  and\ \citenamefont {Vale~Silva}}]{Charles:2020dfl}%
  \BibitemOpen
  \bibfield  {author} {\bibinfo {author} {\bibfnamefont {J.}~\bibnamefont
  {Charles}}, \bibinfo {author} {\bibfnamefont {S.}~\bibnamefont
  {Descotes-Genon}}, \bibinfo {author} {\bibfnamefont {Z.}~\bibnamefont
  {Ligeti}}, \bibinfo {author} {\bibfnamefont {S.}~\bibnamefont {Monteil}},
  \bibinfo {author} {\bibfnamefont {M.}~\bibnamefont {Papucci}}, \bibinfo
  {author} {\bibfnamefont {K.}~\bibnamefont {Trabelsi}},\ and\ \bibinfo
  {author} {\bibfnamefont {L.}~\bibnamefont {Vale~Silva}},\ }\bibfield  {title}
  {\bibinfo {title} {{New physics in $B$ meson mixing: future sensitivity and
  limitations}},\ }\href {https://doi.org/10.1103/PhysRevD.102.056023}
  {\bibfield  {journal} {\bibinfo  {journal} {Phys. Rev. D}\ }\textbf {\bibinfo
  {volume} {102}},\ \bibinfo {pages} {056023} (\bibinfo {year} {2020})},\
  \Eprint {https://arxiv.org/abs/2006.04824} {arXiv:2006.04824 [hep-ph]}
  \BibitemShut {NoStop}%
\bibitem [{\citenamefont {Dembinski}\ \emph {et~al.}(2022)\citenamefont
  {Dembinski}, \citenamefont {Ongmongkolkul}, \citenamefont {Deil},
  \citenamefont {Schreiner}, \citenamefont {Feickert}, \citenamefont {Andrew},
  \citenamefont {Burr}, \citenamefont {Watson}, \citenamefont {Rost},
  \citenamefont {Pearce}, \citenamefont {Geiger}, \citenamefont {Wiedemann},
  \citenamefont {Gohlke}, \citenamefont {Gonzalo}, \citenamefont {Drotleff},
  \citenamefont {Eschle}, \citenamefont {Neste}, \citenamefont {Gorelli},
  \citenamefont {Baak}, \citenamefont {Zapata},\ and\ \citenamefont
  {odidev}}]{hans_dembinski_2022_6389982}%
  \BibitemOpen
  \bibfield  {author} {\bibinfo {author} {\bibfnamefont {H.}~\bibnamefont
  {Dembinski}}, \bibinfo {author} {\bibfnamefont {P.}~\bibnamefont
  {Ongmongkolkul}}, \bibinfo {author} {\bibfnamefont {C.}~\bibnamefont {Deil}},
  \bibinfo {author} {\bibfnamefont {H.}~\bibnamefont {Schreiner}}, \bibinfo
  {author} {\bibfnamefont {M.}~\bibnamefont {Feickert}}, \bibinfo {author}
  {\bibnamefont {Andrew}}, \bibinfo {author} {\bibfnamefont {C.}~\bibnamefont
  {Burr}}, \bibinfo {author} {\bibfnamefont {J.}~\bibnamefont {Watson}},
  \bibinfo {author} {\bibfnamefont {F.}~\bibnamefont {Rost}}, \bibinfo {author}
  {\bibfnamefont {A.}~\bibnamefont {Pearce}}, \bibinfo {author} {\bibfnamefont
  {L.}~\bibnamefont {Geiger}}, \bibinfo {author} {\bibfnamefont {B.~M.}\
  \bibnamefont {Wiedemann}}, \bibinfo {author} {\bibfnamefont {C.}~\bibnamefont
  {Gohlke}}, \bibinfo {author} {\bibnamefont {Gonzalo}}, \bibinfo {author}
  {\bibfnamefont {J.}~\bibnamefont {Drotleff}}, \bibinfo {author}
  {\bibfnamefont {J.}~\bibnamefont {Eschle}}, \bibinfo {author} {\bibfnamefont
  {L.}~\bibnamefont {Neste}}, \bibinfo {author} {\bibfnamefont {M.~E.}\
  \bibnamefont {Gorelli}}, \bibinfo {author} {\bibfnamefont {M.}~\bibnamefont
  {Baak}}, \bibinfo {author} {\bibfnamefont {O.}~\bibnamefont {Zapata}},\ and\
  \bibinfo {author} {\bibnamefont {odidev}},\ }\href
  {https://doi.org/10.5281/zenodo.6389982} {\bibinfo {title}
  {scikit-hep/iminuit: v2.11.2}} (\bibinfo {year} {2022})\BibitemShut {NoStop}%
\bibitem [{\citenamefont {Marchiori}(2023)}]{citekey}%
  \BibitemOpen
  \bibfield  {author} {\bibinfo {author} {\bibfnamefont {G.}~\bibnamefont
  {Marchiori}},\ }\href
  {https://indico.cern.ch/event/1176398/contributions/5208222/attachments/2582981/4456976/2023_01_27%20-%20Higgs%20hadronic%20branching%20ratios%20with%20ZH%20at%20FCCee.pdf}
  {\bibinfo {title} {{Higgs $\to$ bb/cc/gg/ss with $Z(\ell\ell,\nu\nu)H$ at
  $\sqrt{s}$=240 GeV}}} (\bibinfo {year} {2023})\BibitemShut {NoStop}%
\bibitem [{\citenamefont {Selvaggi}(2023)}]{utagger}%
  \BibitemOpen
  \bibfield  {author} {\bibinfo {author} {\bibfnamefont {M.}~\bibnamefont
  {Selvaggi}},\ }\href
  {https://indico.cern.ch/event/1278845/contributions/5461471/attachments/2679899/4648628/higgs_physics_exp_pheno.pdf}
  {\bibinfo {title} {{Higgs Physics at the FCC: experimental overview}}}
  (\bibinfo {year} {2023})\BibitemShut {NoStop}%
\bibitem [{\citenamefont {Workman}\ \emph
  {et~al.}(2022{\natexlab{a}})\citenamefont {Workman} \emph
  {et~al.}}]{ParticleDataGroup:2022pth}%
  \BibitemOpen
  \bibfield  {author} {\bibinfo {author} {\bibfnamefont {R.~L.}\ \bibnamefont
  {Workman}} \emph {et~al.} (\bibinfo {collaboration} {Particle Data Group}),\
  }\bibfield  {title} {\bibinfo {title} {{Review of Particle Physics}},\ }\href
  {https://doi.org/10.1093/ptep/ptac097} {\bibfield  {journal} {\bibinfo
  {journal} {PTEP}\ }\textbf {\bibinfo {volume} {2022}},\ \bibinfo {pages}
  {083C01} (\bibinfo {year} {2022}{\natexlab{a}})}\BibitemShut {NoStop}%
\bibitem [{\citenamefont {Benitez-Guzm\'an}\ \emph {et~al.}(2015)\citenamefont
  {Benitez-Guzm\'an}, \citenamefont {Garc\'\i{}a-Jim\'enez}, \citenamefont
  {L\'opez-Osorio}, \citenamefont {Mart\'\i{}nez-Pascual},\ and\ \citenamefont
  {Toscano}}]{Benitez-Guzman:2015ana}%
  \BibitemOpen
  \bibfield  {author} {\bibinfo {author} {\bibfnamefont {L.~G.}\ \bibnamefont
  {Benitez-Guzm\'an}}, \bibinfo {author} {\bibfnamefont {I.}~\bibnamefont
  {Garc\'\i{}a-Jim\'enez}}, \bibinfo {author} {\bibfnamefont {M.~A.}\
  \bibnamefont {L\'opez-Osorio}}, \bibinfo {author} {\bibfnamefont
  {E.}~\bibnamefont {Mart\'\i{}nez-Pascual}},\ and\ \bibinfo {author}
  {\bibfnamefont {J.~J.}\ \bibnamefont {Toscano}},\ }\bibfield  {title}
  {\bibinfo {title} {{Revisiting the flavor changing neutral current Higgs
  decays $H\;\to \;{q}_{i}{q}_{j}$ in the Standard Model}},\ }\href
  {https://doi.org/10.1088/0954-3899/42/8/085002} {\bibfield  {journal}
  {\bibinfo  {journal} {J. Phys. G}\ }\textbf {\bibinfo {volume} {42}},\
  \bibinfo {pages} {085002} (\bibinfo {year} {2015})},\ \Eprint
  {https://arxiv.org/abs/1506.02718} {arXiv:1506.02718 [hep-ph]} \BibitemShut
  {NoStop}%
\bibitem [{\citenamefont {Aranda}\ \emph {et~al.}(2020)\citenamefont {Aranda},
  \citenamefont {Gonz\'alez-Estrada}, \citenamefont {Monta\~no}, \citenamefont
  {Ram\'\i{}rez-Zavaleta},\ and\ \citenamefont {Tututi}}]{Aranda:2020tqw}%
  \BibitemOpen
  \bibfield  {author} {\bibinfo {author} {\bibfnamefont {J.~I.}\ \bibnamefont
  {Aranda}}, \bibinfo {author} {\bibfnamefont {G.}~\bibnamefont
  {Gonz\'alez-Estrada}}, \bibinfo {author} {\bibfnamefont {J.}~\bibnamefont
  {Monta\~no}}, \bibinfo {author} {\bibfnamefont {F.}~\bibnamefont
  {Ram\'\i{}rez-Zavaleta}},\ and\ \bibinfo {author} {\bibfnamefont {E.~S.}\
  \bibnamefont {Tututi}},\ }\bibfield  {title} {\bibinfo {title} {{Revisiting
  the rare $H\to q_iq_j$ decays in the standard model}},\ }\href
  {https://doi.org/10.1088/1361-6471/abb44d} {\bibfield  {journal} {\bibinfo
  {journal} {J. Phys. G}\ }\textbf {\bibinfo {volume} {47}},\ \bibinfo {pages}
  {125001} (\bibinfo {year} {2020})},\ \Eprint
  {https://arxiv.org/abs/2009.07166} {arXiv:2009.07166 [hep-ph]} \BibitemShut
  {NoStop}%
\bibitem [{\citenamefont {Ma}\ and\ \citenamefont
  {Pramudita}(1980)}]{PhysRevD.22.214}%
  \BibitemOpen
  \bibfield  {author} {\bibinfo {author} {\bibfnamefont {E.}~\bibnamefont
  {Ma}}\ and\ \bibinfo {author} {\bibfnamefont {A.}~\bibnamefont {Pramudita}},\
  }\bibfield  {title} {\bibinfo {title} {Flavor-changing effective
  neutral-current couplings in the weinberg-salam model},\ }\href
  {https://doi.org/10.1103/PhysRevD.22.214} {\bibfield  {journal} {\bibinfo
  {journal} {Phys. Rev. D}\ }\textbf {\bibinfo {volume} {22}},\ \bibinfo
  {pages} {214} (\bibinfo {year} {1980})}\BibitemShut {NoStop}%
\bibitem [{\citenamefont {Clements}\ \emph {et~al.}(1983)\citenamefont
  {Clements}, \citenamefont {Footman}, \citenamefont {Kronfeld}, \citenamefont
  {Narasimhan},\ and\ \citenamefont {Photiadis}}]{PhysRevD.27.570}%
  \BibitemOpen
  \bibfield  {author} {\bibinfo {author} {\bibfnamefont {M.}~\bibnamefont
  {Clements}}, \bibinfo {author} {\bibfnamefont {C.}~\bibnamefont {Footman}},
  \bibinfo {author} {\bibfnamefont {A.}~\bibnamefont {Kronfeld}}, \bibinfo
  {author} {\bibfnamefont {S.}~\bibnamefont {Narasimhan}},\ and\ \bibinfo
  {author} {\bibfnamefont {D.}~\bibnamefont {Photiadis}},\ }\bibfield  {title}
  {\bibinfo {title} {Flavor-changing decays of the ${Z}^{0}$},\ }\href
  {https://doi.org/10.1103/PhysRevD.27.570} {\bibfield  {journal} {\bibinfo
  {journal} {Phys. Rev. D}\ }\textbf {\bibinfo {volume} {27}},\ \bibinfo
  {pages} {570} (\bibinfo {year} {1983})}\BibitemShut {NoStop}%
\bibitem [{\citenamefont {Ganapathi}\ \emph {et~al.}(1983)\citenamefont
  {Ganapathi}, \citenamefont {Weiler}, \citenamefont {Laermann}, \citenamefont
  {Schmitt},\ and\ \citenamefont {Zerwas}}]{PhysRevD.27.579}%
  \BibitemOpen
  \bibfield  {author} {\bibinfo {author} {\bibfnamefont {V.}~\bibnamefont
  {Ganapathi}}, \bibinfo {author} {\bibfnamefont {T.}~\bibnamefont {Weiler}},
  \bibinfo {author} {\bibfnamefont {E.}~\bibnamefont {Laermann}}, \bibinfo
  {author} {\bibfnamefont {I.}~\bibnamefont {Schmitt}},\ and\ \bibinfo {author}
  {\bibfnamefont {P.~M.}\ \bibnamefont {Zerwas}},\ }\bibfield  {title}
  {\bibinfo {title} {Flavor-changing $z$ decays: A window to ultraheavy
  quarks?},\ }\href {https://doi.org/10.1103/PhysRevD.27.579} {\bibfield
  {journal} {\bibinfo  {journal} {Phys. Rev. D}\ }\textbf {\bibinfo {volume}
  {27}},\ \bibinfo {pages} {579} (\bibinfo {year} {1983})}\BibitemShut
  {NoStop}%
\bibitem [{\citenamefont {Mihaila}\ \emph {et~al.}(2012)\citenamefont
  {Mihaila}, \citenamefont {Salomon},\ and\ \citenamefont
  {Steinhauser}}]{Mihaila:2012fm}%
  \BibitemOpen
  \bibfield  {author} {\bibinfo {author} {\bibfnamefont {L.~N.}\ \bibnamefont
  {Mihaila}}, \bibinfo {author} {\bibfnamefont {J.}~\bibnamefont {Salomon}},\
  and\ \bibinfo {author} {\bibfnamefont {M.}~\bibnamefont {Steinhauser}},\
  }\bibfield  {title} {\bibinfo {title} {{Gauge Coupling Beta Functions in the
  Standard Model to Three Loops}},\ }\href
  {https://doi.org/10.1103/PhysRevLett.108.151602} {\bibfield  {journal}
  {\bibinfo  {journal} {Phys. Rev. Lett.}\ }\textbf {\bibinfo {volume} {108}},\
  \bibinfo {pages} {151602} (\bibinfo {year} {2012})},\ \Eprint
  {https://arxiv.org/abs/1201.5868} {arXiv:1201.5868 [hep-ph]} \BibitemShut
  {NoStop}%
\bibitem [{\citenamefont {Chetyrkin}\ and\ \citenamefont
  {Zoller}(2012)}]{Chetyrkin:2012rz}%
  \BibitemOpen
  \bibfield  {author} {\bibinfo {author} {\bibfnamefont {K.~G.}\ \bibnamefont
  {Chetyrkin}}\ and\ \bibinfo {author} {\bibfnamefont {M.~F.}\ \bibnamefont
  {Zoller}},\ }\bibfield  {title} {\bibinfo {title} {{Three-loop
  \textbackslash{}beta-functions for top-Yukawa and the Higgs self-interaction
  in the Standard Model}},\ }\href {https://doi.org/10.1007/JHEP06(2012)033}
  {\bibfield  {journal} {\bibinfo  {journal} {JHEP}\ }\textbf {\bibinfo
  {volume} {06}},\ \bibinfo {pages} {033}},\ \Eprint
  {https://arxiv.org/abs/1205.2892} {arXiv:1205.2892 [hep-ph]} \BibitemShut
  {NoStop}%
\bibitem [{\citenamefont {Charles}\ \emph {et~al.}(2005)\citenamefont
  {Charles}, \citenamefont {Hocker}, \citenamefont {Lacker}, \citenamefont
  {Laplace}, \citenamefont {Le~Diberder}, \citenamefont {Malcles},
  \citenamefont {Ocariz}, \citenamefont {Pivk},\ and\ \citenamefont
  {Roos}}]{Charles:2004jd}%
  \BibitemOpen
  \bibfield  {author} {\bibinfo {author} {\bibfnamefont {J.}~\bibnamefont
  {Charles}}, \bibinfo {author} {\bibfnamefont {A.}~\bibnamefont {Hocker}},
  \bibinfo {author} {\bibfnamefont {H.}~\bibnamefont {Lacker}}, \bibinfo
  {author} {\bibfnamefont {S.}~\bibnamefont {Laplace}}, \bibinfo {author}
  {\bibfnamefont {F.~R.}\ \bibnamefont {Le~Diberder}}, \bibinfo {author}
  {\bibfnamefont {J.}~\bibnamefont {Malcles}}, \bibinfo {author} {\bibfnamefont
  {J.}~\bibnamefont {Ocariz}}, \bibinfo {author} {\bibfnamefont
  {M.}~\bibnamefont {Pivk}},\ and\ \bibinfo {author} {\bibfnamefont
  {L.}~\bibnamefont {Roos}} (\bibinfo {collaboration} {CKMfitter Group}),\
  }\bibfield  {title} {\bibinfo {title} {{CP violation and the CKM matrix:
  Assessing the impact of the asymmetric $B$ factories}},\ }\href
  {https://doi.org/10.1140/epjc/s2005-02169-1} {\bibfield  {journal} {\bibinfo
  {journal} {Eur. Phys. J. C}\ }\textbf {\bibinfo {volume} {41}},\ \bibinfo
  {pages} {1} (\bibinfo {year} {2005})},\ \Eprint
  {https://arxiv.org/abs/hep-ph/0406184} {arXiv:hep-ph/0406184} \BibitemShut
  {NoStop}%
\bibitem [{\citenamefont {Hahn}(2001)}]{Hahn:2000kx}%
  \BibitemOpen
  \bibfield  {author} {\bibinfo {author} {\bibfnamefont {T.}~\bibnamefont
  {Hahn}},\ }\bibfield  {title} {\bibinfo {title} {{Generating Feynman diagrams
  and amplitudes with FeynArts 3}},\ }\href
  {https://doi.org/10.1016/S0010-4655(01)00290-9} {\bibfield  {journal}
  {\bibinfo  {journal} {Comput. Phys. Commun.}\ }\textbf {\bibinfo {volume}
  {140}},\ \bibinfo {pages} {418} (\bibinfo {year} {2001})},\ \Eprint
  {https://arxiv.org/abs/hep-ph/0012260} {arXiv:hep-ph/0012260} \BibitemShut
  {NoStop}%
\bibitem [{\citenamefont {Mertig}\ \emph {et~al.}(1991)\citenamefont {Mertig},
  \citenamefont {Bohm},\ and\ \citenamefont {Denner}}]{Mertig:1990an}%
  \BibitemOpen
  \bibfield  {author} {\bibinfo {author} {\bibfnamefont {R.}~\bibnamefont
  {Mertig}}, \bibinfo {author} {\bibfnamefont {M.}~\bibnamefont {Bohm}},\ and\
  \bibinfo {author} {\bibfnamefont {A.}~\bibnamefont {Denner}},\ }\bibfield
  {title} {\bibinfo {title} {{FEYN CALC: Computer algebraic calculation of
  Feynman amplitudes}},\ }\href {https://doi.org/10.1016/0010-4655(91)90130-D}
  {\bibfield  {journal} {\bibinfo  {journal} {Comput. Phys. Commun.}\ }\textbf
  {\bibinfo {volume} {64}},\ \bibinfo {pages} {345} (\bibinfo {year}
  {1991})}\BibitemShut {NoStop}%
\bibitem [{\citenamefont {Shtabovenko}\ \emph {et~al.}(2016)\citenamefont
  {Shtabovenko}, \citenamefont {Mertig},\ and\ \citenamefont
  {Orellana}}]{Shtabovenko:2016sxi}%
  \BibitemOpen
  \bibfield  {author} {\bibinfo {author} {\bibfnamefont {V.}~\bibnamefont
  {Shtabovenko}}, \bibinfo {author} {\bibfnamefont {R.}~\bibnamefont
  {Mertig}},\ and\ \bibinfo {author} {\bibfnamefont {F.}~\bibnamefont
  {Orellana}},\ }\bibfield  {title} {\bibinfo {title} {{New Developments in
  FeynCalc 9.0}},\ }\href {https://doi.org/10.1016/j.cpc.2016.06.008}
  {\bibfield  {journal} {\bibinfo  {journal} {Comput. Phys. Commun.}\ }\textbf
  {\bibinfo {volume} {207}},\ \bibinfo {pages} {432} (\bibinfo {year}
  {2016})},\ \Eprint {https://arxiv.org/abs/1601.01167} {arXiv:1601.01167
  [hep-ph]} \BibitemShut {NoStop}%
\bibitem [{\citenamefont {Shtabovenko}\ \emph {et~al.}(2020)\citenamefont
  {Shtabovenko}, \citenamefont {Mertig},\ and\ \citenamefont
  {Orellana}}]{Shtabovenko:2020gxv}%
  \BibitemOpen
  \bibfield  {author} {\bibinfo {author} {\bibfnamefont {V.}~\bibnamefont
  {Shtabovenko}}, \bibinfo {author} {\bibfnamefont {R.}~\bibnamefont
  {Mertig}},\ and\ \bibinfo {author} {\bibfnamefont {F.}~\bibnamefont
  {Orellana}},\ }\bibfield  {title} {\bibinfo {title} {{FeynCalc 9.3: New
  features and improvements}},\ }\href
  {https://doi.org/10.1016/j.cpc.2020.107478} {\bibfield  {journal} {\bibinfo
  {journal} {Comput. Phys. Commun.}\ }\textbf {\bibinfo {volume} {256}},\
  \bibinfo {pages} {107478} (\bibinfo {year} {2020})},\ \Eprint
  {https://arxiv.org/abs/2001.04407} {arXiv:2001.04407 [hep-ph]} \BibitemShut
  {NoStop}%
\bibitem [{\citenamefont {Hahn}\ and\ \citenamefont
  {Perez-Victoria}(1999)}]{Hahn:1998yk}%
  \BibitemOpen
  \bibfield  {author} {\bibinfo {author} {\bibfnamefont {T.}~\bibnamefont
  {Hahn}}\ and\ \bibinfo {author} {\bibfnamefont {M.}~\bibnamefont
  {Perez-Victoria}},\ }\bibfield  {title} {\bibinfo {title} {{Automatized one
  loop calculations in four-dimensions and D-dimensions}},\ }\href
  {https://doi.org/10.1016/S0010-4655(98)00173-8} {\bibfield  {journal}
  {\bibinfo  {journal} {Comput. Phys. Commun.}\ }\textbf {\bibinfo {volume}
  {118}},\ \bibinfo {pages} {153} (\bibinfo {year} {1999})},\ \Eprint
  {https://arxiv.org/abs/hep-ph/9807565} {arXiv:hep-ph/9807565} \BibitemShut
  {NoStop}%
\bibitem [{\citenamefont {Patel}(2017)}]{Patel:2016fam}%
  \BibitemOpen
  \bibfield  {author} {\bibinfo {author} {\bibfnamefont {H.~H.}\ \bibnamefont
  {Patel}},\ }\bibfield  {title} {\bibinfo {title} {{Package-X 2.0: A
  Mathematica package for the analytic calculation of one-loop integrals}},\
  }\href {https://doi.org/10.1016/j.cpc.2017.04.015} {\bibfield  {journal}
  {\bibinfo  {journal} {Comput. Phys. Commun.}\ }\textbf {\bibinfo {volume}
  {218}},\ \bibinfo {pages} {66} (\bibinfo {year} {2017})},\ \Eprint
  {https://arxiv.org/abs/1612.00009} {arXiv:1612.00009 [hep-ph]} \BibitemShut
  {NoStop}%
\bibitem [{\citenamefont {Chetyrkin}\ and\ \citenamefont
  {Kuhn}(1990)}]{Chetyrkin:1990kr}%
  \BibitemOpen
  \bibfield  {author} {\bibinfo {author} {\bibfnamefont {K.~G.}\ \bibnamefont
  {Chetyrkin}}\ and\ \bibinfo {author} {\bibfnamefont {J.~H.}\ \bibnamefont
  {Kuhn}},\ }\bibfield  {title} {\bibinfo {title} {{Mass corrections to the Z
  decay rate}},\ }\href {https://doi.org/10.1016/0370-2693(90)90306-Q}
  {\bibfield  {journal} {\bibinfo  {journal} {Phys. Lett. B}\ }\textbf
  {\bibinfo {volume} {248}},\ \bibinfo {pages} {359} (\bibinfo {year}
  {1990})}\BibitemShut {NoStop}%
\bibitem [{\citenamefont {Novikov}\ \emph {et~al.}(1995)\citenamefont
  {Novikov}, \citenamefont {Okun}, \citenamefont {Rozanov},\ and\ \citenamefont
  {Vysotsky}}]{Novikov:1995vu}%
  \BibitemOpen
  \bibfield  {author} {\bibinfo {author} {\bibfnamefont {V.}~\bibnamefont
  {Novikov}}, \bibinfo {author} {\bibfnamefont {L.}~\bibnamefont {Okun}},
  \bibinfo {author} {\bibfnamefont {A.~N.}\ \bibnamefont {Rozanov}},\ and\
  \bibinfo {author} {\bibfnamefont {M.}~\bibnamefont {Vysotsky}},\ }\bibfield
  {title} {\bibinfo {title} {{LEPTOP}},\ }\href@noop {} {\  (\bibinfo {year}
  {1995})},\ \Eprint {https://arxiv.org/abs/hep-ph/9503308}
  {arXiv:hep-ph/9503308} \BibitemShut {NoStop}%
\bibitem [{\citenamefont {Novikov}\ \emph {et~al.}(1999)\citenamefont
  {Novikov}, \citenamefont {Okun}, \citenamefont {Rozanov},\ and\ \citenamefont
  {Vysotsky}}]{Novikov:1999af}%
  \BibitemOpen
  \bibfield  {author} {\bibinfo {author} {\bibfnamefont {V.~A.}\ \bibnamefont
  {Novikov}}, \bibinfo {author} {\bibfnamefont {L.~B.}\ \bibnamefont {Okun}},
  \bibinfo {author} {\bibfnamefont {A.~N.}\ \bibnamefont {Rozanov}},\ and\
  \bibinfo {author} {\bibfnamefont {M.~I.}\ \bibnamefont {Vysotsky}},\
  }\bibfield  {title} {\bibinfo {title} {{Theory of $Z$ boson decays}},\ }\href
  {https://doi.org/10.1088/0034-4885/62/9/201} {\bibfield  {journal} {\bibinfo
  {journal} {Rept. Prog. Phys.}\ }\textbf {\bibinfo {volume} {62}},\ \bibinfo
  {pages} {1275} (\bibinfo {year} {1999})},\ \Eprint
  {https://arxiv.org/abs/hep-ph/9906465} {arXiv:hep-ph/9906465} \BibitemShut
  {NoStop}%
\bibitem [{\citenamefont {Workman}\ \emph
  {et~al.}(2022{\natexlab{b}})\citenamefont {Workman} \emph
  {et~al.}}]{Workman:2022ynf}%
  \BibitemOpen
  \bibfield  {author} {\bibinfo {author} {\bibfnamefont {R.~L.}\ \bibnamefont
  {Workman}} \emph {et~al.} (\bibinfo {collaboration} {Particle Data Group}),\
  }\bibfield  {title} {\bibinfo {title} {{Review of Particle Physics}},\ }\href
  {https://doi.org/10.1093/ptep/ptac097} {\bibfield  {journal} {\bibinfo
  {journal} {PTEP}\ }\textbf {\bibinfo {volume} {2022}},\ \bibinfo {pages}
  {083C01} (\bibinfo {year} {2022}{\natexlab{b}})}\BibitemShut {NoStop}%
\bibitem [{\citenamefont {Andersen}\ \emph {et~al.}(2013)\citenamefont
  {Andersen} \emph {et~al.}}]{LHCHiggsCrossSectionWorkingGroup:2013rie}%
  \BibitemOpen
  \bibfield  {author} {\bibinfo {author} {\bibfnamefont {J.~R.}\ \bibnamefont
  {Andersen}} \emph {et~al.} (\bibinfo {collaboration} {LHC Higgs Cross Section
  Working Group}),\ }\bibfield  {title} {\bibinfo {title} {{Handbook of LHC
  Higgs Cross Sections: 3. Higgs Properties}}\ }\href
  {https://doi.org/10.5170/CERN-2013-004} {10.5170/CERN-2013-004} (\bibinfo
  {year} {2013}),\ \Eprint {https://arxiv.org/abs/1307.1347} {arXiv:1307.1347
  [hep-ph]} \BibitemShut {NoStop}%
\bibitem [{\citenamefont {Alguer\'o}\ \emph {et~al.}(2023)\citenamefont
  {Alguer\'o}, \citenamefont {Biswas}, \citenamefont {Capdevila}, \citenamefont
  {Descotes-Genon}, \citenamefont {Matias},\ and\ \citenamefont
  {Novoa-Brunet}}]{Alguero:2023jeh}%
  \BibitemOpen
  \bibfield  {author} {\bibinfo {author} {\bibfnamefont {M.}~\bibnamefont
  {Alguer\'o}}, \bibinfo {author} {\bibfnamefont {A.}~\bibnamefont {Biswas}},
  \bibinfo {author} {\bibfnamefont {B.}~\bibnamefont {Capdevila}}, \bibinfo
  {author} {\bibfnamefont {S.}~\bibnamefont {Descotes-Genon}}, \bibinfo
  {author} {\bibfnamefont {J.}~\bibnamefont {Matias}},\ and\ \bibinfo {author}
  {\bibfnamefont {M.}~\bibnamefont {Novoa-Brunet}},\ }\bibfield  {title}
  {\bibinfo {title} {{To (b)e or not to (b)e: No electrons at LHCb}},\
  }\href@noop {} {\  (\bibinfo {year} {2023})},\ \Eprint
  {https://arxiv.org/abs/2304.07330} {arXiv:2304.07330 [hep-ph]} \BibitemShut
  {NoStop}%
\bibitem [{LHC(2022)}]{LHCb:2022uzt}%
  \BibitemOpen
  \bibfield  {title} {\bibinfo {title} {{Search for rare decays of $D^0$ mesons
  into two muons}},\ }\href@noop {} {\  (\bibinfo {year} {2022})},\ \Eprint
  {https://arxiv.org/abs/2212.11203} {arXiv:2212.11203 [hep-ex]} \BibitemShut
  {NoStop}%
\bibitem [{\citenamefont {Petric}\ \emph {et~al.}(2010)\citenamefont {Petric}
  \emph {et~al.}}]{Belle:2010ouj}%
  \BibitemOpen
  \bibfield  {author} {\bibinfo {author} {\bibfnamefont {M.}~\bibnamefont
  {Petric}} \emph {et~al.} (\bibinfo {collaboration} {Belle}),\ }\bibfield
  {title} {\bibinfo {title} {{Search for leptonic decays of $D^0$ mesons}},\
  }\href {https://doi.org/10.1103/PhysRevD.81.091102} {\bibfield  {journal}
  {\bibinfo  {journal} {Phys. Rev. D}\ }\textbf {\bibinfo {volume} {81}},\
  \bibinfo {pages} {091102} (\bibinfo {year} {2010})},\ \Eprint
  {https://arxiv.org/abs/1003.2345} {arXiv:1003.2345 [hep-ex]} \BibitemShut
  {NoStop}%
\bibitem [{\citenamefont {Ablikim}\ \emph {et~al.}(2022)\citenamefont {Ablikim}
  \emph {et~al.}}]{BESIII:2021slf}%
  \BibitemOpen
  \bibfield  {author} {\bibinfo {author} {\bibfnamefont {M.}~\bibnamefont
  {Ablikim}} \emph {et~al.} (\bibinfo {collaboration} {BESIII}),\ }\bibfield
  {title} {\bibinfo {title} {{Search for the decay $D^{0} \to \pi^{0} \nu
  \bar{\nu}$}},\ }\href {https://doi.org/10.1103/PhysRevD.105.L071102}
  {\bibfield  {journal} {\bibinfo  {journal} {Phys. Rev. D}\ }\textbf {\bibinfo
  {volume} {105}},\ \bibinfo {pages} {L071102} (\bibinfo {year} {2022})},\
  \Eprint {https://arxiv.org/abs/2112.14236} {arXiv:2112.14236 [hep-ex]}
  \BibitemShut {NoStop}%
\bibitem [{\citenamefont {Bause}\ \emph {et~al.}(2020)\citenamefont {Bause},
  \citenamefont {Golz}, \citenamefont {Hiller},\ and\ \citenamefont
  {Tayduganov}}]{Bause:2019vpr}%
  \BibitemOpen
  \bibfield  {author} {\bibinfo {author} {\bibfnamefont {R.}~\bibnamefont
  {Bause}}, \bibinfo {author} {\bibfnamefont {M.}~\bibnamefont {Golz}},
  \bibinfo {author} {\bibfnamefont {G.}~\bibnamefont {Hiller}},\ and\ \bibinfo
  {author} {\bibfnamefont {A.}~\bibnamefont {Tayduganov}},\ }\bibfield  {title}
  {\bibinfo {title} {{The new physics reach of null tests with $D \rightarrow
  \pi \ell \ell $ and $D_s \rightarrow K \ell \ell $ decays}},\ }\href
  {https://doi.org/10.1140/epjc/s10052-020-7621-7} {\bibfield  {journal}
  {\bibinfo  {journal} {Eur. Phys. J. C}\ }\textbf {\bibinfo {volume} {80}},\
  \bibinfo {pages} {65} (\bibinfo {year} {2020})},\ \bibinfo {note} {[Erratum:
  Eur.Phys.J.C 81, 219 (2021)]},\ \Eprint {https://arxiv.org/abs/1909.11108}
  {arXiv:1909.11108 [hep-ph]} \BibitemShut {NoStop}%
\bibitem [{\citenamefont {Aaij}\ \emph {et~al.}(2013)\citenamefont {Aaij} \emph
  {et~al.}}]{LHCb:2013hxr}%
  \BibitemOpen
  \bibfield  {author} {\bibinfo {author} {\bibfnamefont {R.}~\bibnamefont
  {Aaij}} \emph {et~al.} (\bibinfo {collaboration} {LHCb}),\ }\bibfield
  {title} {\bibinfo {title} {{Search for D+(s) to pi+ mu+ mu- and D+(s) to pi-
  mu+ mu+ decays}},\ }\href {https://doi.org/10.1016/j.physletb.2013.06.010}
  {\bibfield  {journal} {\bibinfo  {journal} {Phys. Lett. B}\ }\textbf
  {\bibinfo {volume} {724}},\ \bibinfo {pages} {203} (\bibinfo {year}
  {2013})},\ \Eprint {https://arxiv.org/abs/1304.6365} {arXiv:1304.6365
  [hep-ex]} \BibitemShut {NoStop}%
\bibitem [{\citenamefont {Descotes-Genon}\ \emph {et~al.}(2023)\citenamefont
  {Descotes-Genon}, \citenamefont {Faroughy}, \citenamefont {Plakias},\ and\
  \citenamefont {Sumensari}}]{Descotes-Genon:2023pen}%
  \BibitemOpen
  \bibfield  {author} {\bibinfo {author} {\bibfnamefont {S.}~\bibnamefont
  {Descotes-Genon}}, \bibinfo {author} {\bibfnamefont {D.~A.}\ \bibnamefont
  {Faroughy}}, \bibinfo {author} {\bibfnamefont {I.}~\bibnamefont {Plakias}},\
  and\ \bibinfo {author} {\bibfnamefont {O.}~\bibnamefont {Sumensari}},\
  }\bibfield  {title} {\bibinfo {title} {{Probing Lepton Flavor Violation in
  Meson Decays with LHC Data}},\ }\href@noop {} {\  (\bibinfo {year} {2023})},\
  \Eprint {https://arxiv.org/abs/2303.07521} {arXiv:2303.07521 [hep-ph]}
  \BibitemShut {NoStop}%
\bibitem [{\citenamefont {Allwicher}\ \emph {et~al.}(2023)\citenamefont
  {Allwicher}, \citenamefont {Faroughy}, \citenamefont {Jaffredo},
  \citenamefont {Sumensari},\ and\ \citenamefont {Wilsch}}]{Allwicher:2022mcg}%
  \BibitemOpen
  \bibfield  {author} {\bibinfo {author} {\bibfnamefont {L.}~\bibnamefont
  {Allwicher}}, \bibinfo {author} {\bibfnamefont {D.~A.}\ \bibnamefont
  {Faroughy}}, \bibinfo {author} {\bibfnamefont {F.}~\bibnamefont {Jaffredo}},
  \bibinfo {author} {\bibfnamefont {O.}~\bibnamefont {Sumensari}},\ and\
  \bibinfo {author} {\bibfnamefont {F.}~\bibnamefont {Wilsch}},\ }\bibfield
  {title} {\bibinfo {title} {{HighPT: A tool for high-pT Drell-Yan tails beyond
  the standard model}},\ }\href {https://doi.org/10.1016/j.cpc.2023.108749}
  {\bibfield  {journal} {\bibinfo  {journal} {Comput. Phys. Commun.}\ }\textbf
  {\bibinfo {volume} {289}},\ \bibinfo {pages} {108749} (\bibinfo {year}
  {2023})},\ \Eprint {https://arxiv.org/abs/2207.10756} {arXiv:2207.10756
  [hep-ph]} \BibitemShut {NoStop}%
\end{thebibliography}%

\clearpage
\onecolumngrid

\setcounter{equation}{0}
\setcounter{figure}{0}
\setcounter{table}{0}
\setcounter{section}{0}
\makeatletter
\renewcommand{\theequation}{S\arabic{equation}}
\renewcommand{\thefigure}{S\arabic{figure}}
\renewcommand{\thetable}{S\arabic{table}}
\renewcommand{\thesection}{S\arabic{section}}
%%%%%%%%%%%%%%%%%%%%%%%%%%%%%%%%%%%%%%%%%%%%%%%%%%%%%%%%%%%%%%%%%%%%%%
\begin{center}
 \large{\bf 
 Flavor violating Higgs and $Z$ decays at the FCC-ee
 }\\
 Supplementary Material
\end{center}
%%%%%%%%%%%%%%%%%%%%%%%%%%%%%%%%%%%%%%%%%%%%%%%%%%%%%%%%%%%%%%%%%%%%%%
\begin{center}
Jernej F. Kamenik, Arman Korajac, Manuel Szewc, Michele Tammaro, and Jure Zupan
\end{center}

In this supplementary material we give further details on the probabilistic model implemented in the analysis of the projected FCC-ee reach, in Sec.~\ref{sec:supp:prob:model}, and define the relevant statistical estimates implemented in this work, in Sec. \ref{sec:statistics}. We expand on the results for $h\to bs$, $h\to cu$, $Z\to bs$ and $Z\to cu$ in Sec.~\ref{sec:details:sensitivity}. Updated theoretical calculations for the SM FCNC branching ratios  are given in Sec.~\ref{sec:UpdateSMpredictions}, while Sec.~\ref{sec:details:BSM} contains additional details about the two BSM examples: type III 2HDM and a model with vectorlike quarks.

\section{The probabilistic model}
\label{sec:supp:prob:model}

In this section we provide further details on the probabilistic model implemented to obtain the FCC-ee reach. The probability distribution functions $p(n_{b},n_{s}|f,\nu)$ give the probability for an event with the initial parton flavor configuration $f$ to end up in the $(n_b, n_s)$ bin. By convoluting these distributions with the expected number of events in each $f$ decay channel and summing over all possible $f$, we can obtain the total number of expected events in the $(n_b, n_s)$ bin, see  Eq.~\eqref{eq:Nbar:nb:ns}. 

Given the parton configuration of the two final state jets, indicated as  $j_1$ and $j_2$ respectively, we denote the flavor tagging of the two final state jets\footnote{The jet flavor tagging labels are not independent, but rather satisfy $n_{q;2}=n_{q}-n_{q;1}$. The $(n_b,n_s)=(0,2)$ bin thus has events with the $(n_{b;1},n_{b;2},n_{s;1},n_{s;2})=(0,0,1,1)$  individual jet flavor tag, while $(n_b,n_s)=(1,1)$ bin has events with $(n_{b;1},n_{b;2},n_{s;1},n_{s;2})=(1,0,0,1)$, or $(n_{b;1},n_{b;2},n_{s;1},n_{s;2})=(0,1,1,0)$. The $(n_b,n_s)=(2,0)$ bin contains events with the $(n_{b;1},n_{b;2},n_{s;1},n_{s;2})=(1,1,0,0)$ jet flavor tag. }
with $n_{q;i}= 0,1$, where $q=b,s$, and $i=1,2$. The probability density distribution then is (see also Fig.~\ref{fig:graphical_model} for a representation of the model)

\beq
p(n_{b},n_{s}|f,\nu)= 
\sum_{n_{b;1}=0}^{\text{min}(n_{b},1)}\sum_{n_{s;1}=0}^{\text{min}(n_s,1-n_{b;1})}
p(n_{b;1}|j_{1}) p(n_{s;1}|j_{1},n_{b;1})p(n_{b;2}|j_{2})p(n_{s;2}|j_{2},n_{b;2})\,,  
\eeq
where we sum over all the possible jet flavor tagging allowed in the $(n_b,n_s)$ bin, with the constraint $n_{b;i} + n_{s;i} = 1$.
Here $p(n_{b;1}|j_{1})$ is the (anti)-$b-$tagging probability of jet $j_1$ when $n_{b;1}=(0)1$. Assuming no kinematic dependence of the taggers, this is simply the probability to get $n_{b;1}$ $b$-tags, with tagging probability $\epsilon_1^b$, where $\epsilon_1^b\equiv \epsilon_{j_1}^b$ is the $b$-tagger efficiency for the initial parton configuration of jet $j_1$. Thus, we can model such probability as a Binomial distribution,
\beq
p(n_{b;1}|j_{1}) = \Binom(n_{b;1},1,\epsilon_1^b)\,,
\eeq
where 
\beq
\Binom(k,n,p) \equiv \binom{n}{k} p^k (1 - p)^{n-k}\,, \quad  \binom{n}{k} = \frac{n!}{k!(n-k)!}\,.
\eeq
Similarly, $p(n_{s;1}|j_{1},n_{b;1})$ is the (anti)-$s$-tagging probability of jet $j_1$, conditioned over the $n_{b;1}$ flavor tagging. Namely,
\beq
p(n_{s;1}|j_{1},n_{b;1}) = \Binom\lp n_{s;1},1-n_{b;1},\frac{\epsilon_1^s}{1-\epsilon_1^b} \rp\,,
\eeq
where the probability $\epsilon_1^s$ is weigthed by a factor $(1 - \epsilon_1^b)^{-1}$ to account for the fact that we are $s$-tagging the non $b$-tagged jets. 

The same reasoning applies to the flavor-tagging of jet $j_2$, thus the full probability distribution reads
\beq\label{eq:pnbns}
\begin{split}
p(n_b, n_s|f,\nu)=\sum_{n_{b;1}=0}^{\text{min}(n_{b},1)}~\sum_{n_{s;1}=0}^{\text{min}(n_s,1-n_{b;1})} &\Binom(n_{b;1},1,\epsilon^{b}_{1}) \Binom\lp n_{s;1},1-n_{b;1},\frac{\epsilon_1^s}{1-\epsilon_1^b} \rp \times \\
\times&\Binom(n_{b}-n_{b;1},1,\epsilon^{b}_{2}) \Binom\lp n_{s}-n_{s;1},1-(n_{b}-n_{b;1}),\frac{\epsilon_2^s}{1-\epsilon_2^b} \rp.
\end{split}
\eeq
For flavour-conserving decays, i.e., for all $f$ parton configurations apart from $bs$, this expression simplifies to
\beq
p(n_{b},n_{s}|f,\nu)=\Binom(n_{b},2,\epsilon^{b}_{1})\Binom\lp n_{s},2-n_{b},\frac{\epsilon_1^s}{1-\epsilon_1^b}\rp\,.
\eeq
Note that the ordering of the tagging that we assumed in the probabilistic model is irrelevant. By imposing the constraint $n_{b;i} + n_{s;i} = 1$, one can easily verify that
\beq
p(n_{b;i}|j_i) p(n_{s;i}|j_i,n_{b;i}) = p(n_{s;i}|j_i) p(n_{b;i}|j_i,n_{s;i})\,.
\eeq

Finally, note that all the efficiencies are implicit functions of the nuisance parameters. 

\section{The details on statistical estimates}
\label{sec:statistics}

Here we give further details on how we arrived at the statistical estimates of the FCC-ee sensitivities, quoted in the main text. As the parameter of interest we use the ratio of the observed FCNC branching ratio normalized to the  SM prediction 

\beq
\mu = \frac{\mathcal{B}(Z/h\to bs)}{\mathcal{B}(Z/h\to bs)_{\SM}}\,.
\eeq

To relate the observed number of events in the $(n_b,n_s)$ bin $N_{(n_b, n_s)}$ to the number of expected events in the same bin we can construct the extended likelihood
\beq
{\cal L}(\mu,\nu) = {\cal P}(N_{(n_b, n_s)} | \bar{N}_{(n_b, n_s)}(\mu,\nu))p(\nu)\,,
\eeq
where $\mathcal{P}$ is the Poisson likelihood and $p(\nu)$ is the appropriate distribution for the nuisance parameters. Note that  $\bar{N}_{(n_b, n_s)}(\mu,\nu)$ depends on the parameter of interest through the number of expected decays to the $bs$ partonic final state, $N_{bs}$, cf. Eq. \eqref{eq:Nbar:nb:ns}. The probability density functions $p(n_{b},n_{s}|f,\nu)$, which also enter Eq. \eqref{eq:Nbar:nb:ns}, are given explicitly in Eq. \eqref{eq:pnbns}.

Given this probabilistic model for $\bar{N}_{(n_b, n_s)}(\mu,\nu)$, we follow Ref.~\cite{Cowan:2010js} and define the profile likelihood ratio

\begin{equation}
    \lambda(\mu)=
    \frac{\mathcal{L}(\mu,\hat{\hat{\nu}}(\mu))}{\mathcal{L}(\hat{\mu},\hat{\nu})}\,,
    \label{eq:plr}
\end{equation}
with the associated test statistic
\begin{equation}
   t_{\mu}= -2\text{ Ln }\lambda(\mu).
    \label{eq:test_stat}
\end{equation} 
Here, $\hat{\hat{\nu}}(\mu)$ are the maximum likelihood estimates (MLE) of the nuisance parameters, obtained by  maximizing $\mathcal{L}({\mu},{\nu})$, varying $\nu$, but keeping $\mu$ fixed. The maximum likelihood estimates, $\hat{\mu}$, $\hat{\nu}$, are instead obtained by finding the global maximum of $\mathcal{L}(\mu,\nu)$, varying both $\nu$ and $\mu$. 

Since the expected statistics of events in each $(n_b,n_s)$ bin is large, we work in the Asimov approximation~\cite{Cowan:2010js} to obtain the expected confidence level interval. In the Asimov approximation, one sets the observed values of $N_{(n_b,n_s)}$ to $N^A_{(n_b, n_s)}=\bar{N}_{(n_b, n_s)}(\mu=\mu_{\rm true},\nu=0)$, where $\mu_{\rm true}$ is the input value of $\mu$, and we have defined the nuisance parameters such that $\nu$ is the difference to their nominal input values. For Asimov dataset the MLE are thus by definition $\hat{\mu}=\mu_{\rm true}$ and $\hat{\nu}=0$. Using the Asimov dataset in conjunction with the test statistic $t_{\mu}$, we can make three statements regarding $\mu$:
\begin{itemize}
    \item We set confidence intervals on $\mu$ assuming $\mu_{\rm true}=1$. In particular, the $68\%$ confidence interval $[\mu_{\rm low},\mu_{\rm up}]$ is obtained by solving for $t_{\mu} = 1$. For simplicity, we report as the result the upper relative uncertainty $(\mu_{\rm up}-\hat{\mu})/\hat{\mu}=(\mu_{\rm up}-1)$. If we modify Eq.~\eqref{eq:test_stat} to account for the fact that $\mu \geq 0$, the lower relative uncertainty will be the minimum between $(1-\mu_{\rm low})$ and $1$.
    \item We obtain the discovery significance as the significance of ruling out $\mu=0$, given that $\mu_{\rm true}=1$. This is done by computing $q_{0}=t_{0}$, with the median expected significance $Z_{0,A}\equiv\sqrt{q_{0}}$.
    \item We set upper limits on ${\cal B}(h/Z\to bs)$ assuming $\mu_{\rm true}=0$. In particular, the $95\%$ CL upper bound is obtained by solving for $t_{\mu} = (\Phi^{-1}(1-0.05))^{2}$, where $\Phi^{-1}$ is the inverse of the Gaussian error function. Note that here the limit is one-sided and thus it will not coincide with the $95\%$ $\mu_{\rm up}$ which is instead obtained from a two-sided test.
\end{itemize}

In all three instances, the minimization over the relevant nuisance parameters is assumed. This is achieved with the {\tt iminuit} python package~\cite{hans_dembinski_2022_6389982}. The nuisance parameters for the $h\to bs$ and $Z\to bs$ searches are listed in Tables~\ref{tab:syst_higgs} and \ref{tab:syst}, respectively. In our simplified model we know the functional dependency of $\bar N_{(n_{b},n_{s})}$ on all these parameters  are given explicitly by the probabilistic model, Eqs. \eqref{eq:Nbar:nb:ns},  \eqref{eq:bar:Nf}, \eqref{eq:pnbns}. The $\nu_i$ are taken to be constrained by additional measurements which have yielded the nominal values of nuisance parameters, which we subtract, so that $\nu_{i,0}=0$. The additional likelihood term will thus be
\beq
p(\nu) = \prod_{i}\mathcal{N}(\nu_{i,0};\nu_{i},\sigma_{i})\,,
\eeq
where $\sigma_{i}$ is the estimated uncertainty for the $i$-th nuisance parameter.

\section{Further details on the FCC-ee reach}
\label{sec:details:sensitivity}

In this section we collect further details regarding our estimates of the FCC-ee sensitivity to the $Z/h\to bs, bd, cu$ decays. 

\subsection{The $h\to bs$ decays}

The $h\to bs$ study has a relatively low statistic, cf. $N_h$  in Table~\ref{tab:syst_higgs}. Because of this, and the smallness of the SM $\mathcal{B}(h\to bs)$ branching ratio, the expected sensitivity is well above the SM values. We therefore focus mainly on setting the upper limits on $\mathcal{B}(h\to bs)$ assuming no $h\to bs$ decays in the data. The nominal values and uncertainties of several nuisance parameters can be found on Table~\ref{tab:syst_higgs}.

The $N_{h}$ and $\mathcal{A}$ are taken from Ref.~\cite{deBlas:2019rxi} (note that their relative uncertainties are higher than for the equivalent parameters for $Z\to bs$, discussed in the next subsection). The nominal values and the uncertainties on the flavor conserving branching ratios are taken to be the quoted central values and statistical uncertainties on the signal strength from the preliminary projections of the FCC-ee fits in the $Z(\to \nu\bar{\nu})h$ mode~\cite{citekey}, which are consistent with Ref.~\cite{deBlas:2019rxi}. We verified that increasing the uncertainties on the branching ratios by one order of magnitude will not impact significantly the performance of the analysis. The leading uncertainties are the systematic uncertainties on the tagger efficiencies. We set these  to  $1\%$ relative uncertainty across all the taggers, which we anticipate to be realistically achievable for the $b$- and $s$-taggers at the FCC-ee. 
 
In the main part of the manuscript we showed the results for the  true positive rate (TPR) $\epsilon^{b}_{b}=\epsilon_s^s$ and the false positive rate (FPR), $\epsilon^{b}_{gsc}=\epsilon_{gcb}^s$.  This parameterization assumes that the two sets of probabilities are the same for both taggers as well as, perhaps more importantly, that the mistag probabilities are the same for every type of jet. This is almost certainly not the case for the actual taggers that will be used, since most taggers have quite different behaviors in the case of heavy and light jets. However, we can interpret any (TPR, FPR) choice as a conservative choice where all mistags are the same as the least stringent of the ones achievable for the individual cases, $\epsilon^{b}_{g}, \epsilon_s^b,\epsilon_c^b,\epsilon_g^s,\epsilon_c^s,\epsilon_b^s$. Another factor that encourages us to take this approximation is that because $gg$ (or $uu+dd$ for $Z$) and $cc$ final states require two misidentifications in order to populate the signal dominated region, their contributions will be strongly suppressed both with respect to the signal and with respect to the two main backgrounds, $ss$ and $bb$. In this sense, FPR can be thought of as representing mostly the $\epsilon^{b}_{s}$ and $\epsilon^{s}_{b}$ values (taken in most of the analyses to be also the same).

 In Table~\ref{tab:comparison_higgs} we list instead the 95\% CL upper limits on $\mathcal{B}(h\to bs)$, which were obtained for two realistic working points for the  $b$- and $s$-taggers, introduced in Refs.~\cite{Bedeschi:2022rnj,tagger},
\begin{align}
& {\rm Loose:}& \epsilon^{b}_{\beta;{\rm Loose}}= \{0.02,\,0.001,\,0.02,\,0.90\},& &\epsilon^{s}_{\beta;{\rm Loose}}= \{0.20,\,0.90,\,0.10,\,0.01\},& 
  \\
  &{\rm Medium:}& \epsilon^{b}_{\beta;{\rm Med}}=\{0.007,\, 0.0001,\, 0.003,\,0.80\},& &\epsilon^{s}_{\beta;{\rm Med}}=\{0.09,\,0.80,\, 0.06,\, 0.004\},&
  \end{align}
where the label runs over $\beta=\{g,s,c,b\}$. The best performance is achieved by combining the Loose $b$-tagger and Medium $s$-tagger, although all performances are very similar. This projected limit on $\mathcal{B}(h\to bs)$ does not take into account other backgrounds such as Drell-Yan, $WW, ZZ, q\bar{q}$, which we expect to be subleading, and should not affect significantly the projected reach. The projected FCC-ee sensitivity to $h\to bs$ decays is competitive with indirect measurements and represents a complementary direct probe, as we discuss in the main text.

 \begin{table}[t]
\renewcommand{\arraystretch}{1}
\centering
\begin{tabular}{ccc}
\hline\hline
 ~Nuisance Param.~ & ~~Nominal Value~~  & ~Rel. uncert. (\%)~ \\ \hline 
$\mathcal{B}(h\to gg)$ & $1.4\%$ & 1.2 \\ 
$\mathcal{B}(h\to ss)$ & $0.024\%$ & 160 \\
$\mathcal{B}(h\to cc)$ & $2.9\%$ & 2.8 \\ 
$\mathcal{B}(h\to bb)$ & $56\%$ & 0.4 \\ 
$\epsilon^{\alpha}_{\beta}$ & See text &  1.0 \\ 
$N_{h}$ &  $6.7 \times 10^{5}$ & $0.5$ \\ 
$\mathcal{A}$ & 0.70 &  $0.1$ \\ \hline\hline
\end{tabular}
\caption{Nuisance parameters and their relative uncertainties, entering the $h\to bs$ sensitivity estimation.}
\label{tab:syst_higgs}
\end{table}

\begin{table*}[t]
\renewcommand{\arraystretch}{1}
\centering
\begin{tabular}{ccc}
\hline\hline
~~~$\epsilon^{b}_{\beta}$~~~ & ~~~$\epsilon^{s}_{\beta}$~~~ &  $\mathcal{B}(h\to bs)$ (95$\%$ CL)  \\ \hline 
$\epsilon^{b}_{\beta;{\rm Loose}}$ &  $\epsilon^{s}_{\beta;{\rm Loose}}$& $1.3\times 10^{-3} $ \\ 
$\epsilon^{b}_{\beta;{\rm Loose}}$ &  $\epsilon^{s}_{\beta;{\rm Med}}$ & $9.6\times 10^{-4} $\\ 
$\epsilon^{b}_{\beta;{\rm Med}}$  & $\epsilon^{s}_{\beta;{\rm Loose}}$ & $1.4\times 10^{-3} $\\ 
$\epsilon^{b}_{\beta;{\rm Med}}$&  $\epsilon^{s}_{\beta;{\rm Med}}$ & $1.0\times 10^{-3} $
 \\ \hline\hline
 \end{tabular}
\caption{Several examples of possible tagger working point choices, where $\epsilon^{b}_{\beta;{\rm Loose}}= \{0.02,\,0.001,\,0.02,\,0.90\}$, $\epsilon^{b}_{\beta;{\rm Med}}=\{0.007,\, 0.0001,\, 0.003,\,0.80\} $, $\epsilon^{s}_{\beta;{\rm Loose}}= \{0.20,\,0.90,\,0.10,\,0.01\}$, $\epsilon^{s}_{\beta;{\rm Med}}=\{0.09,\,0.80,\, 0.06,\, 0.004\}$, with the label running over $\beta=\{g,s,c,b\}$. The last column in the table gives the resulting $95\%$ CL upper limits on $\mathcal{B}(h\to bs)$. All tagging efficiencies are assumed to have a relative systematic uncertainty of $1\%$.}
\label{tab:comparison_higgs}
\end{table*}

\subsection{The $h\to cu$ decays}
\label{eq:sec:hcu}

A similar analysis can be performed for the FCC-ee sensitivity to $\mathcal{B}(h\to cu)$. Due to the difficulties in implementing a ``$u$-tagger'',  the main text considers just the case where only the $c$-tagger is applied (similar to the $\mathcal{B}(h\to bq)$ case shown in the top row in Fig.~\ref{fig:result_Hbq}). We parameterize the $c$-tagger in terms of the TPR, $\epsilon^{c}_{c}$, and a common FPR for all the other initial partons. We consider a $c$-tagger with four parameter $\epsilon^{c}_{g},\epsilon^{c}_{uds},\epsilon^{c}_{c},\epsilon^{c}_{b}$, so that the number of nuisance parameters is thus four, one for each efficiency. The systematic uncertainties considered are the same as those listed in Table~\ref{tab:syst_higgs}, with the caveat that the efficiencies now refer to the four $\epsilon^{c}_{\beta}$.  

\begin{figure}[t]
  \vspace{-0.3cm}
\begin{center}
\includegraphics[width=0.375\linewidth]{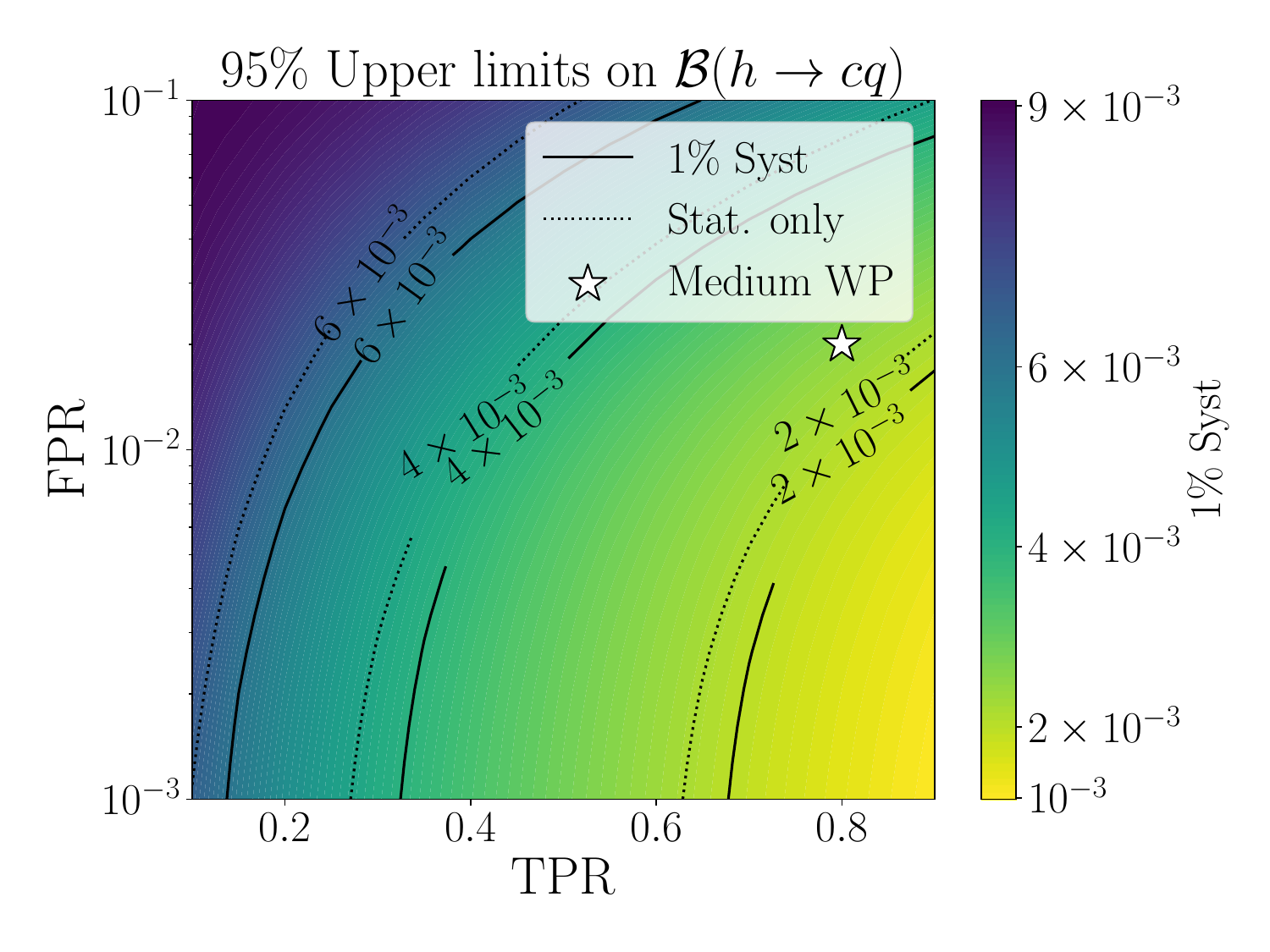}~
\includegraphics[width=0.375\linewidth]{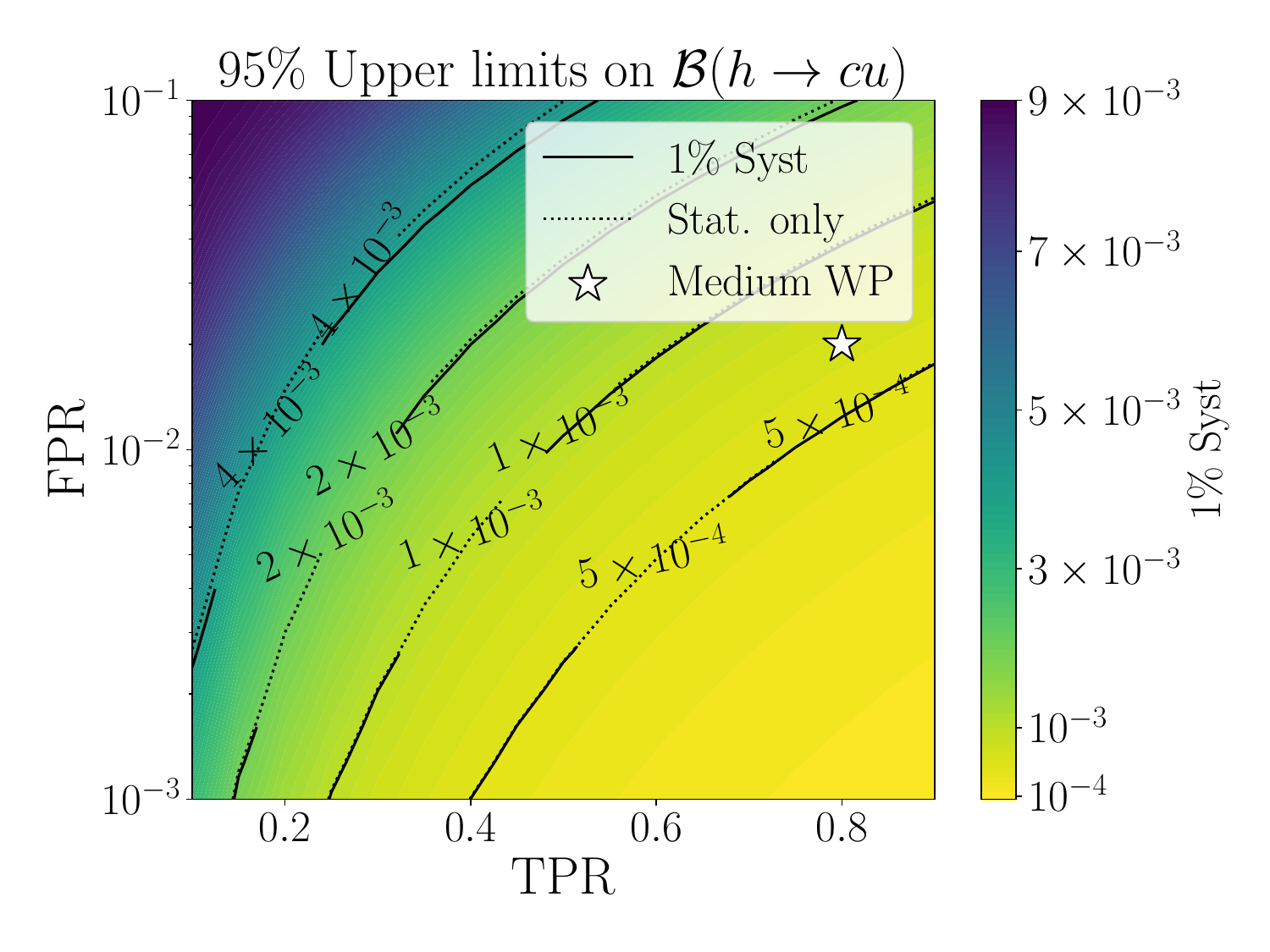}
  \vspace{-0.3cm}
  \end{center}
\caption{{\bf Left:} Expected $95\%$ CL upper bounds on $\mathcal{B}(h\to cu)$ as a function of the $c$-tagger efficiencies  {\bf Right:} Expected $95\%$ CL upper bounds on $\mathcal{B}(h\to cu)$ as a function of the TPR and FPR of the $c$- and $u$-taggers.  Solid (dashed) lines and colors are with default (no) systematic uncertainties. The Medium Working Point is based on the $c$-tagger introduced in Refs.~\cite{Bedeschi:2022rnj,tagger} and the $u$-tagger introduced in Ref.~\cite{utagger}.}\label{fig:result_Hcq}
\end{figure}

The resulting $95\%$ CL upper limits on $\mathcal{B}(h\to cu)$ as a function of FPR and TPR are shown in Fig.~\ref{fig:result_Hcq} (left). For a given FPR, TPR values the expected upper bounds on $\mathcal{B}(h\to cu)$, assuming just statistical errors, are stronger than the ones for $\mathcal{B}(h\to bq)$, Fig.~\ref{fig:result_Hbq} (top), because the dominant background for $h\to cu$, due to $h\to cc$, is smaller than the $h\to bb$ background for $h\to bq$.

Besides the two dimensional scan, we also list in Table~\ref{tab:comparison_higgs_cu} the 95\% CL upper limits on $\mathcal{B}(h\to cu)$ obtained for two realistic working points for the $c$-tagger introduced in Refs.~\cite{Bedeschi:2022rnj,tagger},
 \begin{align}
 & {\rm Loose:}& \epsilon^{c}_{\beta;{\rm Loose}}= \{0.07,\,0.07,\,0.90,\,0.04\},& 
  \\
  &{\rm Medium:}& \epsilon^{c}_{\beta;{\rm Med}}=\{0.02,\,0.008,\,0.80,\,0.02\},&
  \end{align}
  where the labels run over $\beta=\{g,uds,c,b\}$.
The best performance is achieved with the medium working point, although the two performances are again very similar. As for $\mathcal{B}(h\to bs)$, this projected limit on $\mathcal{B}(h\to cu)$ does not take into account other backgrounds such as Drell-Yan, $WW, ZZ, q\bar{q}$, which we expect to not affect significantly the projected reach. The projected FCC-ee sensitivity to $h\to cu$ decays is competitive with indirect measurements and represents a complementary direct probe, as we discuss in the main text.

%Regarding the possibility of a similar two tagger approach as for $h\to bs$ to increase the statistical power, there have been  
Regarding the $u$-tagger, there were very recently developments based on the jet charge~\cite{utagger}. Fig.~\ref{fig:result_Hcq} (right) shows the expected $95\%$ CL upper limits on $\mathcal{B}(h\to cu)$ as a function of FPR and TPR, obtained from the same two tagger approach that we used for $h\to bs$ in the main text, but now using the $c$ and the $u$-taggers. We use the same medium working point as we did for the case of just the $c$-tagger (TPR,FPR) = $(0.80,0.02)$, shown in Fig.~\ref{fig:result_Hcq} (left). Note that this is consistent with the tagger performance curves reported in Ref.~\cite{utagger}. Note that the $u/d$ separation is rather weak with the $u$-tagger, but it is not needed in our analysis, since none of the backgrounds contain a $d$ quark. Both the $c$-tagger and the $u$-tagger were assumed to distinguish between $u/d$ and $s$-quarks, $\epsilon^{c,u}_{g},\epsilon^{c,u}_{ud},\epsilon^{c,u}_{s},\epsilon^{u,c}_{c},\epsilon^{u,c}_{b}$ with a resulting total of ten nuisance parameters, one for each efficiency (we lump together $u$ and $d$ quarks, but this is mostly for completeness: neither the backgrounds nor the signal posses a $d$-quark). The systematic uncertainties considered are the same as those listed in Table~\ref{tab:syst_higgs}, with the caveat that the efficiencies now refer to the ten $\epsilon^{c,u}_{\beta}$.  

If such a $u$-tagger were to be implemented in practice, the power of the analysis would increase considerably, with the medium working point yielding a  $95\%$ CL expected bound $\mathcal{B}(h\to cu) < 6.6 \times 10^{-4}$, compared to $2.5 \times 10^{-3}$ for the $c$-tagger only analysis. The increase in the statistical power is due to both the additional observables, as well as due to the lack of a background with $u$-jets that could populate the $(n_{c},n_{u})=(1,1)$ bin (this is different from the $h\to bs$ case, where the $h\to ss$ decays can end up in the  $(n_{b},n_{s})=(1,1)$ with a still relatively high probability).
% as the  case does for . 
%We do not report a Table with more complete working points beyond the (TPR,FPR) parameterization because they are not numerically provided in Ref.~\cite{utagger}. However, we expect the results to be similar as with all other examples reported in this work.

Although very encouraging, one should keep in mind that  the $u$-taggers are still in the early stages of development. The performance one will be able to achieve in practice may thus well differ from the medium working point we assumed above. We also comment in passing that in principle the $s$-tagger could be applied as a light-jet tagger instead of the $u$-tagger. The hope would be that this would increase the sensitivity much in the same way as the $s$-tagger increases the sensitivity of the $h\to bs$ analysis compared to the one with the $b$-tagger alone. However, we find that the performance of the analysis does not noticeably increase, because the $s$-tagger acts mostly just as the background rejector. In conclusion, while the preliminary results from  a two-tagger analysis are encouraging, a more detailed analysis is called for.
%, with future work in this direction is warranted.
%, and we leave this study for future work.

\begin{table*}[t]
\renewcommand{\arraystretch}{1}
\centering
\begin{tabular}{cc}
\hline\hline
~~~$\epsilon^{c}_{\beta}$~~~ & $\mathcal{B}(h\to cu)$ (95$\%$ CL)  \\ \hline 
$\epsilon^{c}_{\beta;{\rm Loose}}$ & $2.9\times 10^{-3} $ \\ 
$\epsilon^{c}_{\beta;{\rm Med}}$ &  $2.5\times 10^{-3} $
 \\ \hline\hline
 \end{tabular}
\caption{Two examples of possible tagger working point choices, where $\epsilon^{c}_{\beta;{\rm Loose}}= \{0.07,\,0.07,\,0.90,\,0.04\}$, $\epsilon^{c}_{\beta;{\rm Med}}=\{0.02,\,0.008,\,0.80,\,0.02\} $, with the label running over $\beta=\{g,uds,c,b\}$. The last column in the table gives the resulting $95\%$ CL upper limits on $\mathcal{B}(h\to cu)$. All tagging efficiencies are assumed to have a relative systematic uncertainty of $1\%$.}
\label{tab:comparison_higgs_cu}
\end{table*}

% \begin{table*}[t]
% \renewcommand{\arraystretch}{1}
% \centering
% \begin{tabular}{ccc}
% \hline\hline
% ~~~$\epsilon^{c}_{\beta}$~~~ & ~~~$\epsilon^{u}_{\beta}$~~~ & $\mathcal{B}(h\to cu)$ (95$\%$ CL)  \\ \hline 
% $\epsilon^{c}_{\beta;{\rm Loose}}$ & $\epsilon^{u}_{\beta;{\rm Loose}}$ & \\ 
% $\epsilon^{c}_{\beta;{\rm Loose}}$ & $\epsilon^{u}_{\beta;{\rm Med}}$ & \\ 
% $\epsilon^{c}_{\beta;{\rm Med}}$ &  $\epsilon^{u}_{\beta;{\rm Loose}}$ & \\
% $\epsilon^{c}_{\beta;{\rm Med}}$ &  $\epsilon^{u}_{\beta;{\rm Med}}$ &
%  \\ \hline\hline
%  \end{tabular}
% \caption{Several examples of possible tagger working point choices, where $\epsilon^{c}_{\beta;{\rm Loose}}= \{0.07,\,0.07,\,0.07,\,0.90,\,0.04\}$, $\epsilon^{c}_{\beta;{\rm Med}}=\{0.02,\,0.008,\,0.008,\,0.80,\,0.02\} $, $\epsilon^{u}_{\beta;{\rm Loose}}= \{0.90\}$, $\epsilon^{u}_{\beta;{\rm Med}}=\{0.80\} $, with the label running over $\beta=\{g,u,s,c,b\}$. The last column in the table gives the resulting $95\%$ CL upper limits on $\mathcal{B}(h\to cu)$. All tagging efficiencies are assumed to have a relative systematic uncertainty of $1\%$.}
% \label{tab:comparison_higgs_cu}
% \end{table*}

\subsection{The $Z\to bs$ decays}

\begin{table}[t]
\renewcommand{\arraystretch}{1}
\centering
\begin{tabular}{ccc}
\hline\hline
 Nuisance param. & ~Nominal value~  & Rel. uncert. (in \%) \\ \hline 
$\mathcal{B}(Z\to uu + dd)$ & $27.01\%$ & 5.0 \\ 
$\mathcal{B}(Z\to ss)$ & $15.84\%$ & 3.8 \\ 
$\mathcal{B}(Z\to cc)$ & $12.03\%$ & 1.7 \\ 
$\mathcal{B}(Z\to bb)$ & $15.12\%$ & 0.33 \\ 
$\epsilon^{\alpha}_{\beta}$ & See text &  1.0 \\ 
$N_{Z}$ & $5\times 10^{12}$ &  $10^{-3}$ \\ 
$\mathcal{A}$ & 0.994 &  $10^{-3}$ \\ \hline\hline
\end{tabular}
\caption{Nuisance parameters and their relative uncertainties, entering the estimate of  sensitivity to ${\cal B}(Z\to bs)$.}
\label{tab:syst}
\end{table}

 \begin{figure}[t]
  \vspace{-0.3cm}
\begin{center}
\includegraphics[width=0.33\linewidth]{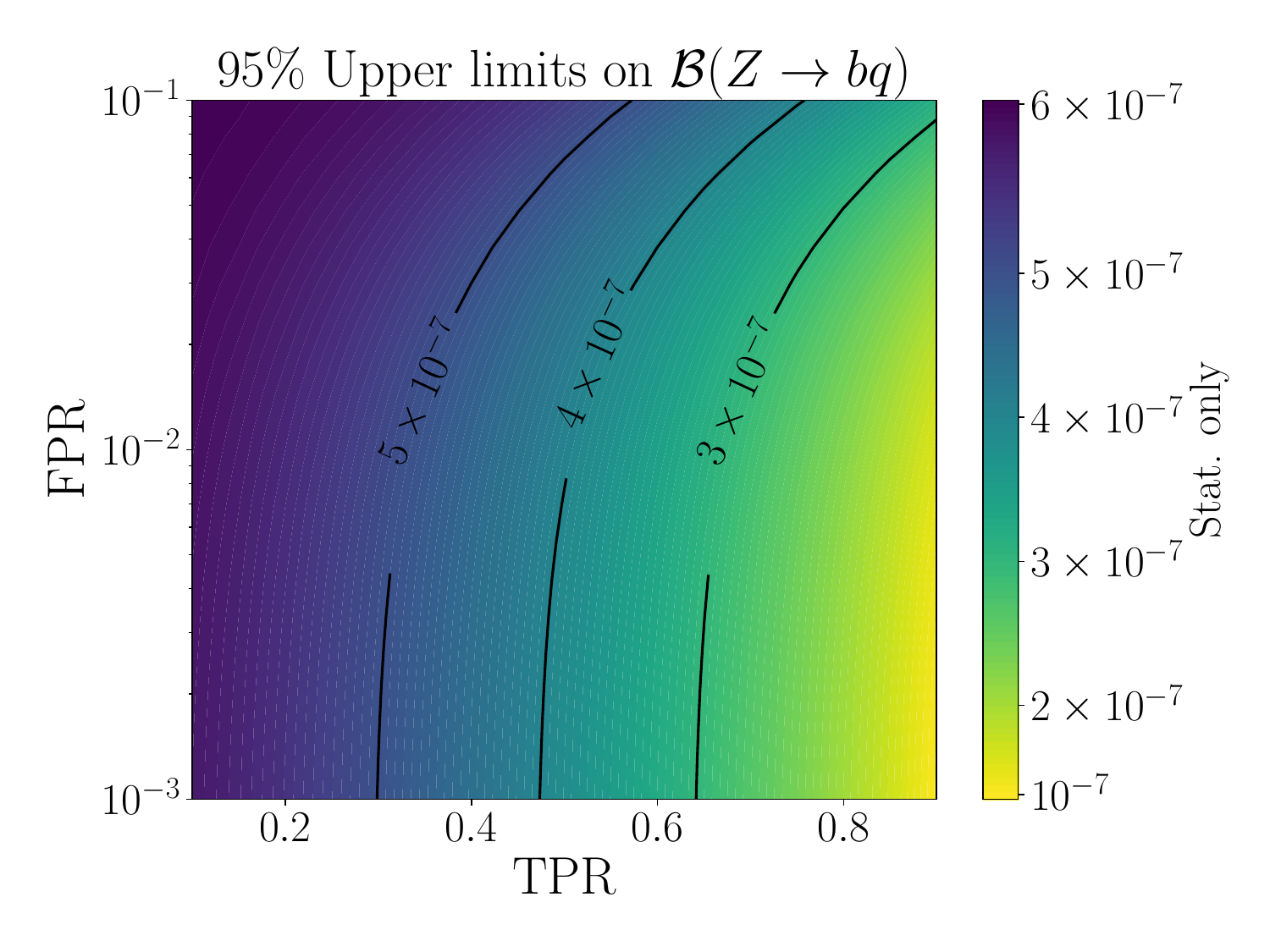}~
\includegraphics[width=0.33\linewidth]{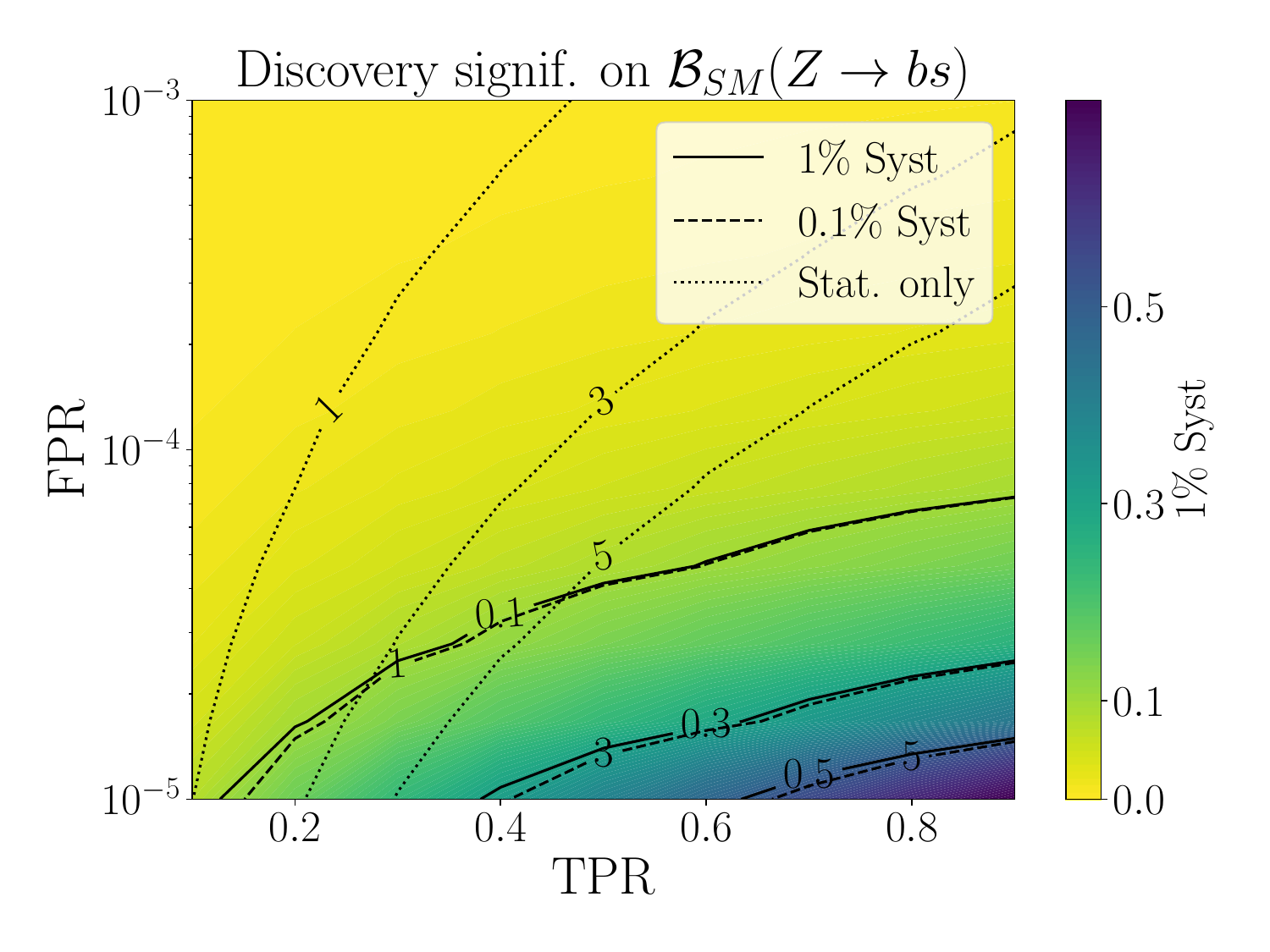}~
\includegraphics[width=0.33\linewidth]{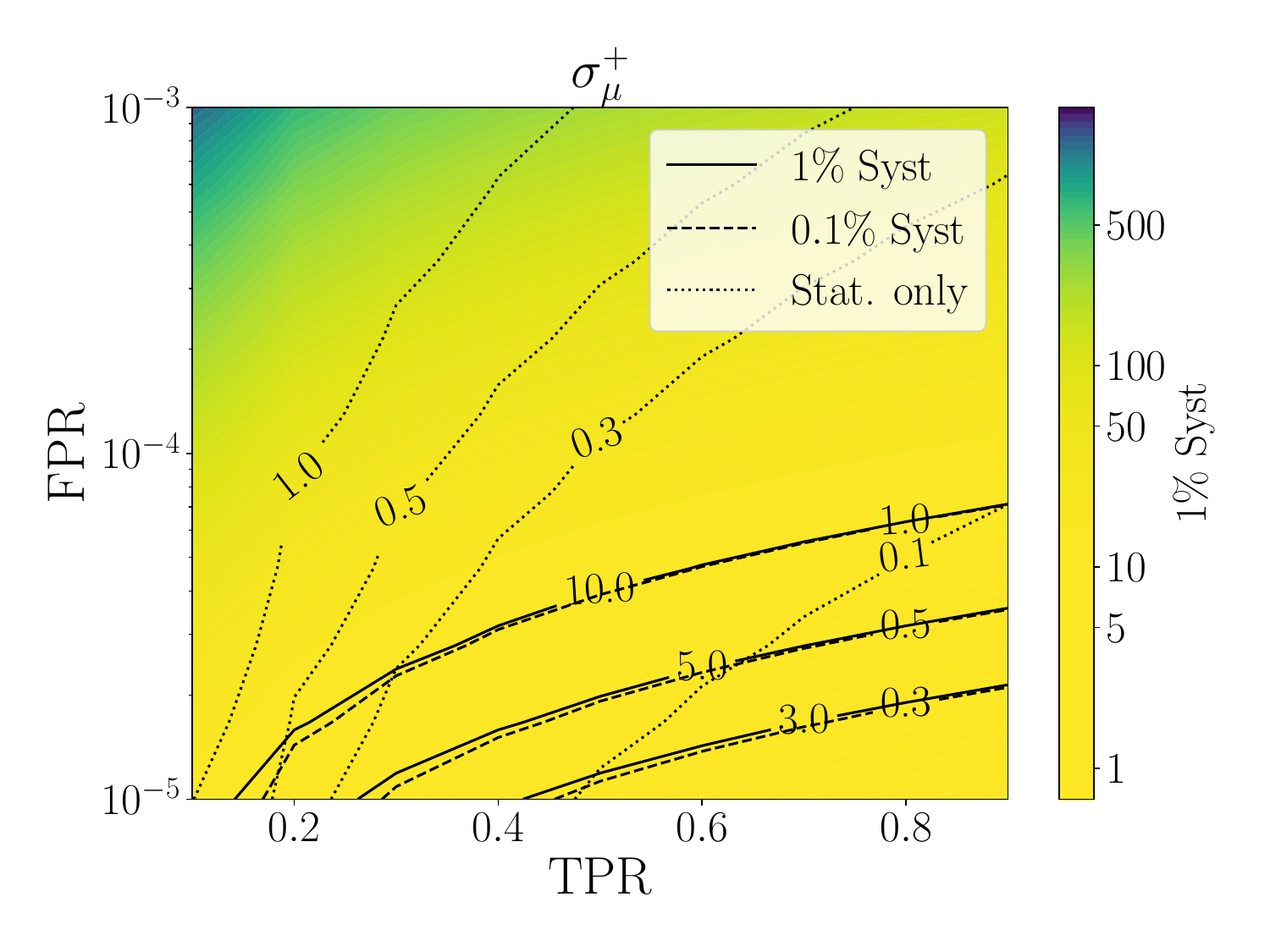}
  \vspace{-0.3cm}
  \end{center}
\caption{
{\bf Left:} The expected $95\%$ CL upper bounds on $\mathcal{B}(Z\to bq)$ as a function of the $b$-tagger efficiencies, assuming no systematic uncertainties.
{\bf Middle:} The expected discovery significance for the SM ${\cal B}(Z\to bs)$ as a function of TPR and FPR. Solid (dashed, dotted) lines and colors are for the default systematic uncertainties of $1\%$ ($0.1\%, 0\%$). {\bf Right:} The expected uncertainty $\sigma_\mu^+$ as a function of TPR and FPR. Solid lines and colors are with default systematic uncertainties while dashed lines correspond to reduced tagging uncertainties and dotted lines to no systematic uncertainties, see text for details. }\label{fig:result_tpr_fpr:Zbs}
\end{figure}

\begin{table*}[t]
\renewcommand{\arraystretch}{1}
\centering
\begin{tabular}{cccc}
\hline\hline
 (TPR,\,FPR,\,${\Delta \epsilon^{\alpha}_{\beta}}/{\epsilon^{\alpha}_{\beta}}$) & $\sigma^{+}_{\mu}$ for $\mu_{\text{true}} = 1$  & Discov. signif. (in $\sigma$)&  $\mathcal{B}(Z\to bs)$ ($95\%$ CL)\\ \hline 
 (0.4, $10^{-4}$,~~\,\,$1\%$) & $0.40$(stat.)$+32$(syst.) & 0.032 & $1.8\times 10^{-6}$ \\ 
 (0.4, $10^{-4}$, $0.1\%$) & $0.40$(stat.)$+3.2$(syst.) & 0.32 & $1.8\times 10^{-7}$ \\ 
(0.2, $10^{-5}$, ~~$1\%$) & $0.36$(stat.)$+6.3$ (syst.) & 0.16 & $4.2\times 10^{-7}$  \\ 
(0.2, $10^{-5}$, $0.1\%$) & $0.36$(stat.)$+0.63$ (syst.) & 1.4 & $4.2\times 10^{-8}$  \\ \hline\hline
\end{tabular}
\caption{Examples of possible tagger choices with their uncertainties and the corresponding measurement uncertainty, discovery significance and $95\%$ upper limits for $Z\to bs$. 
}
\label{tab:comparison:Zbs}
\end{table*}

As for the Higgs decays, we first consider the statistical reach on $\mathcal{B}(Z\to bq)=\mathcal{B}(Z\to bd)+\mathcal{B}(Z\to bs)$, i.e., summing over the $Z\to bd$ and $Z\to bs$ decay modes, thus only using the $b$-tagger. The resulting bounds as functions of TPR and FPR are shown in Fig.~\ref{fig:result_tpr_fpr:Zbs} (left). We see that even ignoring systematic uncertainties, the projected sensitivity is well above the SM $\mathcal{B}(Z\to bq)$ ratio. 

Once we introduce the $s$-tagger, we can explore the discovery potential of $Z\to bs$ decays and the respective confidence interval. We first consider a simplified set-up which allows for a two-dimensional scan, and parameterize the taggers as a set of true positive rates (TPR) $\epsilon^{b}_{b}=\epsilon^{s}_{s}$ and a set of false positive rates (FPR) $\epsilon^{b}_{udsc}=\epsilon^{s}_{udcb}$. The uncertainties on the flavour conserving branching ratios were taken from the Particle Data Group~\cite{ParticleDataGroup:2022pth}, while the uncertainties on the number of $Z$ bosons and the acceptances were taken from Ref.~\cite{FCC:2018evy}. These are shown in Table~\ref{tab:syst}.

Table~\ref{tab:comparison:Zbs} shows the upper $1\sigma$ error, $\sigma_\mu^+$, on the measurement of $\mathcal{B}(Z\to bs)_{\rm SM}$ (2nd column), the discovery significance (3d column), and the expected $95\%$ upper limits (4th column) for two working points, each with either the default $1\%$ systematic and the significantly reduced $0.1\%$ tagger uncertainties. The two working points are chosen as (TPR,FPR)=$(0.4,10^{-4})$ and (TPR,FPR)=$(0.2,10^{-5})$. While the first is a reasonable choice for a future tagger performace, the latter represents a very aggressive projection which is clearly hard to achieve.
As shown in Table~\ref{tab:comparison:Zbs}, the aggressive choice of tagger efficiencies, coupled with a low systematic uncertainty, is the only scenario where FCC-ee can reach the SM value of ${\cal B}(Z\to bs)$, although with a low discovery significance and with systematic dominated uncertainty.   

The results discussed above are generalized in Fig.~\ref{fig:result_tpr_fpr:Zbs} (middle), where we show the expected  discovery significance as a function of FPR and TPR, with the solid (dashed) lines denoting the case of  $1\%$ ($0.1\%$) systematic errors on the taggers, while the dotted lines assume only statistical errors. Fig.~\ref{fig:result_tpr_fpr:Zbs} (right) shows, similarly, the expected  upper error on $\mu$ for each of these cases.  Neglecting systematics we observe that an uncertainty on the SM $Z\to bs$ branching ratio below $30\%$ is achieved already with the conservative choice of efficiencies, (TPR,FPR)=($0.5$,$10^{-4}$), which is not altogether impossible to implement with current state of the art taggers. However, this performance is highly degraded by the introduction of systematic uncertainties, i.e., the measurement is expected to be completely systematic dominated.

From this preliminary study, we can conclude that one will not be able to measure the $Z\to bs$ decay rate with enough precision to impact the searches for beyond standard model physics.

\subsection{The $Z\to cu$ decays}
\label{eq:sec:Zcu}

\begin{figure}[t]
  \vspace{-0.3cm}
\begin{center}
\includegraphics[width=0.33\linewidth]{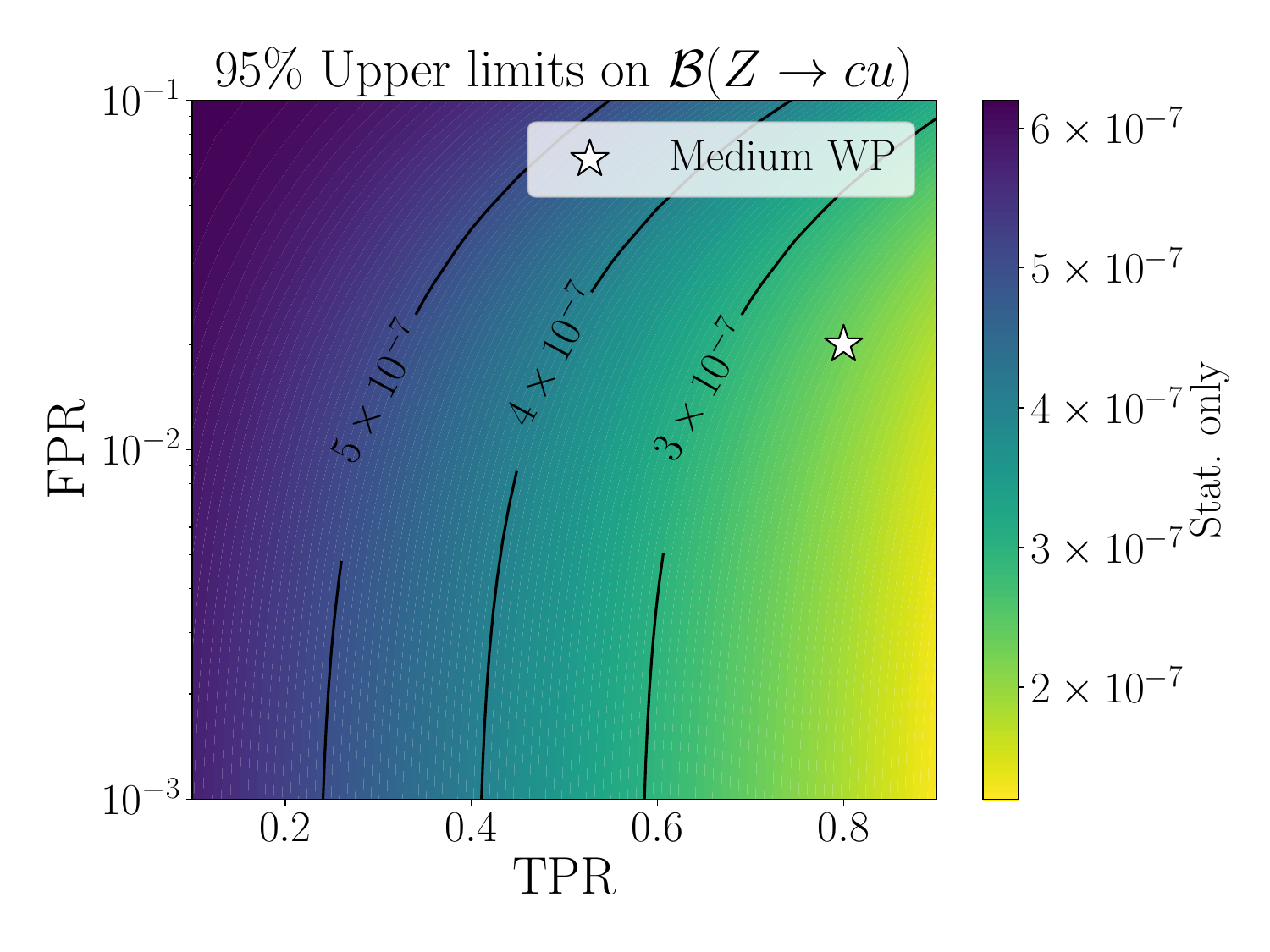}
\includegraphics[width=0.33\linewidth]{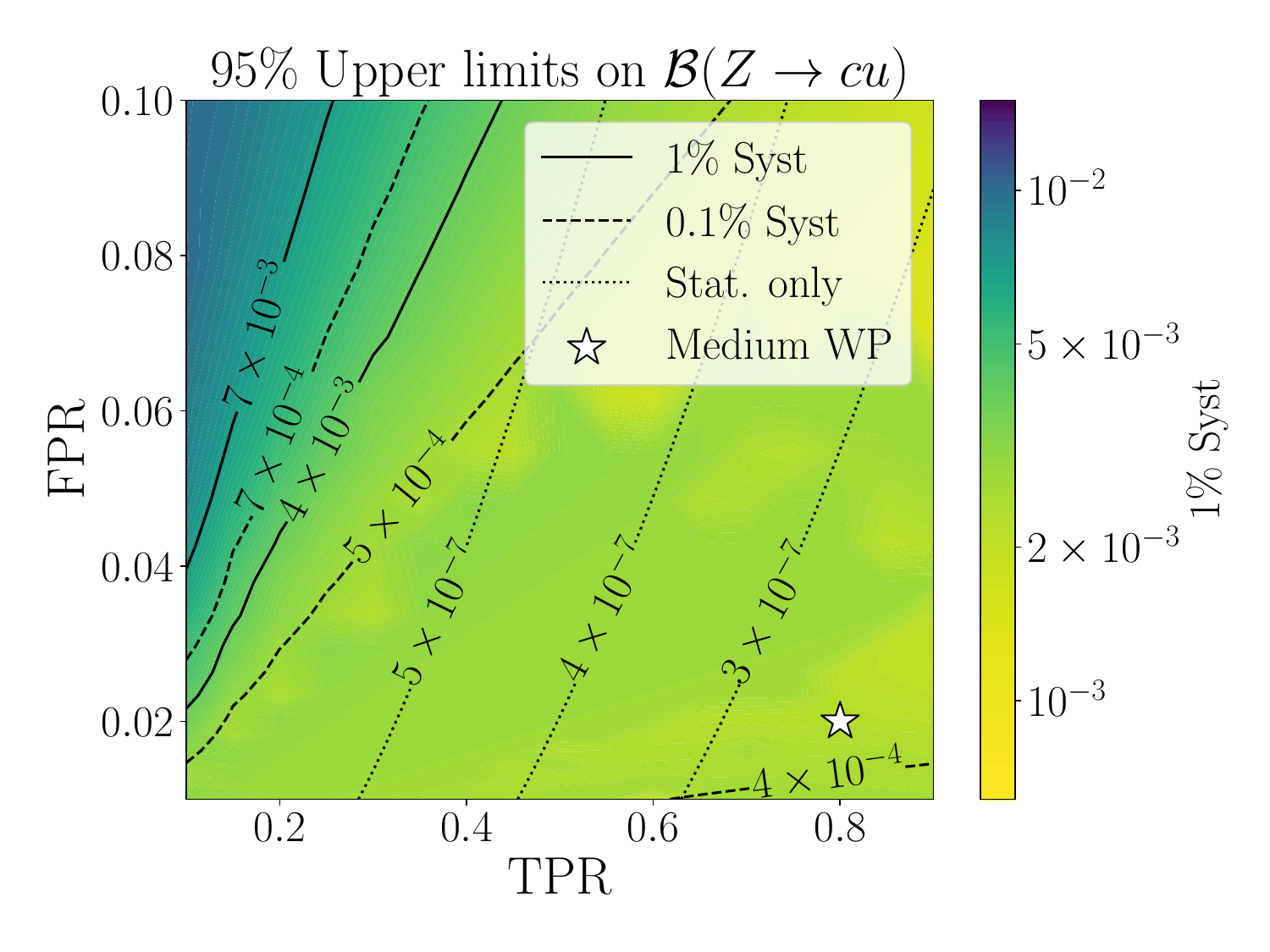}
  \vspace{-0.3cm}
  \end{center}
\caption{
The expected $95\%$ CL upper bounds on $\mathcal{B}(Z\to cu)$ as a function of the $c$-tagger efficiencies. {\bf Left:} Solid lines and colors are for no systematic uncertainties. {\bf Right:} Solid (dashed, dotted) lines and colors are for the default systematic uncertainties of $1\%$ ($0.1\%, 0\%$), while the temperature map assumes $1\%$ systematics.}\label{fig:result_Zcq}
\end{figure}

Similarly to the $h\to cu$ decays, we can obtain the expected $95\%$ CL upper limits on $\mathcal{B}(Z\to cu)$ when applying just the $c$-tagger. In Fig.~\ref{fig:result_Zcq} we show the projected upper limits as functions of charm tagger TPR and FPR, assuming again the three cases of 1\%, 0.1\% and no systematic uncertainties on the tagger efficiency. For a given value of (TPR, FPR) the upper bounds on $\mathcal{B}(Z\to cu)$, when assuming negligible systematics, are almost identical to the ones for $\mathcal{B}(Z\to bq)$, cf. Fig.~\ref{fig:result_tpr_fpr:Zbs} (top), obtaining an upper limit of $2.3\times 10^{-7}$ for the medium WP (TPR,FPR) = $(0.80,0.02)$. This is not surprising, given that in both cases only a single tagger is used as the discriminator, while the $Z\to bb$ and  $Z\to cc$ backgrounds are almost identical in size. Once the systematic effects are taken into account, these completely dominate the performance, saturating the achievable reach (note that for this reason the FPR range displayed in the left and right panels in Fig.~\ref{fig:result_Zcq} differ significantly). For the medium WP (TPR,FPR) = $(0.80,0.02)$ and $1\%$ ($0.1\%$) systematic uncertainties we obtain an expected $95\%$ CL upper limit on $\mathcal{B}(Z\to cu)$ of $2.3\times 10^{-3}$ ($4.0\times 10^{-4}$).

We  reiterate that the above results rely on just the $c$-tagger. As done for $h\to cu$, we could implement a $u$-tagger to study its impact. However, in this case the background is larger as both the $Z\to uu, dd$ decays are important (even if better than a random tagger, the reported tagger in Ref.~\cite{utagger} yields similar $\epsilon^{u}_{u}$ and $\epsilon^{u}_{d}$). This implies that the background from $Z\to uu+dd$ would be essentially twice the one given by $Z\to ss$ in the $Z\to bs$ case. If already for $Z\to bs$ the background contamination is a limiting factor that forces us to consider extreme efficiencies, and systematic uncertainties render a precise measurement very difficult, in this case we are searching for a smaller signal with a larger background and more experimental taggers. We thus deem the application of both a $c$- and $u$-tagger more unrealistic than for $h\to cu$. As mentioned for $h\to cu$, in principle one may also try to apply instead an $s$-tagger as an additional discriminator between signal and background. As in $h\to cu$, preliminary studies of applying an $s$-tagger as a light-jet tagger have shown no increase in the statistical power.  We leave a more complete study of applying a two-tagger analysis for $Z\to cu$ for future work.

\section{Updated calculations of the $Z/h$ flavor changing decay widths in SM}
\label{sec:UpdateSMpredictions}
The SM predictions for ${\cal B}(h\to bs,bd,cu)$ were presented in \cite{Benitez-Guzman:2015ana, Aranda:2020tqw}, and for ${\cal B}(Z\to bs,bd)$ in \cite{PhysRevD.22.214, PhysRevD.27.570, PhysRevD.27.579}. Here, we repeat the calculations and update the numerical predictions for the SM value of ${\cal B}(Z/h\to qq')\equiv{\cal B}(Z/h\to q\bar q') + {\cal B}(Z/h\to \bar q q')$, with the final results collected in  Table~\ref{tab:SMpredict}. The numerical inputs are summarized in Table~\ref{tab:NumInputs}. The SM RG evolution of the parameters is performed using the three-loop $\beta$-functions~\cite{Mihaila:2012fm, Chetyrkin:2012rz}, in order to obtain the $t$ and $b$ quark masses and $\alpha, \alpha_s$ at $\mu=m_h, m_Z$. For the values of the CKM matrix elements we use the results of the global fit from the CKMfitter collaboration~\cite{Charles:2004jd} (the Moriond 2021 update).

\begin{table}[t]
\renewcommand{\arraystretch}{1.3}
\centering
\begin{tabular}{cccccc}
\hline\hline
param. & value & param. & value & param & value \\
\hline
 $|V_{tb}|$ & $0.999142^{+0.000018}_{-0.000023}$ & $|V_{ts}|$ & $0.04065^{+0.00040}_{-0.00055}$ & $|V_{td}|$ & $0.008519^{+0.000075}_{-0.000146}$ \\
 $m_t(m_Z)$ & $171.512\pm0.329$\,GeV  & $m_t(m_h)$ & $167.036\pm0.315$\,GeV \\
 $m_b(m_Z)$ & $2.871\pm0.024$\,GeV  & $m_b(m_h)$ & $2.796\pm0.024$\,GeV \\
 $m_Z$ & $91.1876\pm0.0021$\,GeV  & $m_W$ & $80.377\pm0.012$\,GeV & $m_h$ & $125.25\pm0.17$\,GeV\\
 $\alpha^{-1}(m_Z)$ & $127.955\pm0.009$ & $\alpha^{-1}(m_h)$ & $127.506\pm0.009$ & $s_W^2(m_Z)$ & $0.23122\pm0.00004$  \\
 $\alpha_s(m_Z)$ & $0.1179\pm0.0009$ & $\alpha_s(m_h)$ & $0.1126\pm0.0008$ & $\alpha_s(m_t)$ & $0.1076\pm0.0007$\\
\hline\hline
\end{tabular}
\caption{The numerical inputs used for the SM prediction of $Z/h\to b\bar q$ decay widths. The $m_t$ and $m_b$ masses are given in the $\overline{\text{MS}}$ scheme for two values of $\mu$.}
\label{tab:NumInputs}
\end{table}

\begin{figure}[t]
\centering
\includegraphics[width=0.26\linewidth]{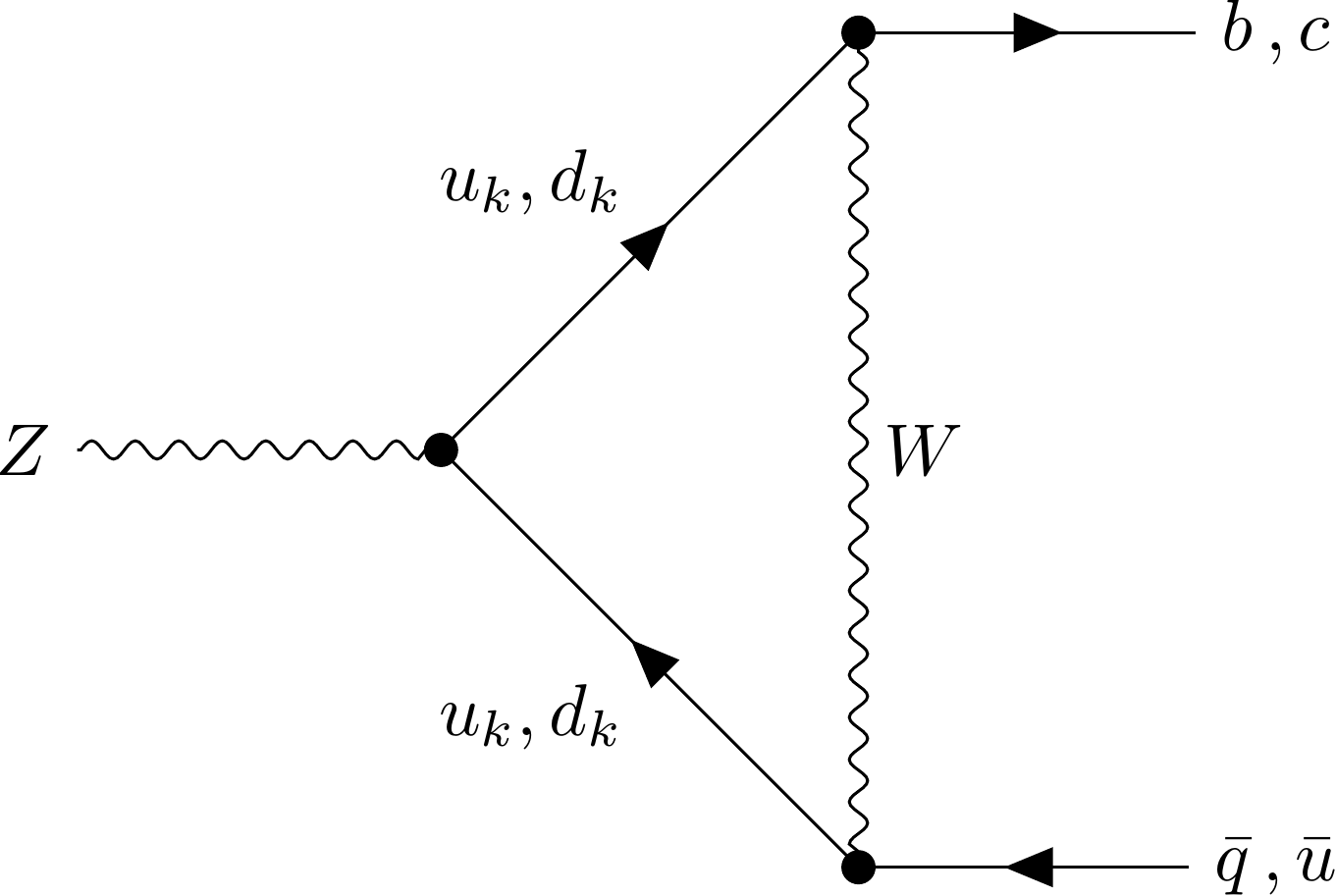}\hspace{2cm}
\includegraphics[width=0.26\linewidth]{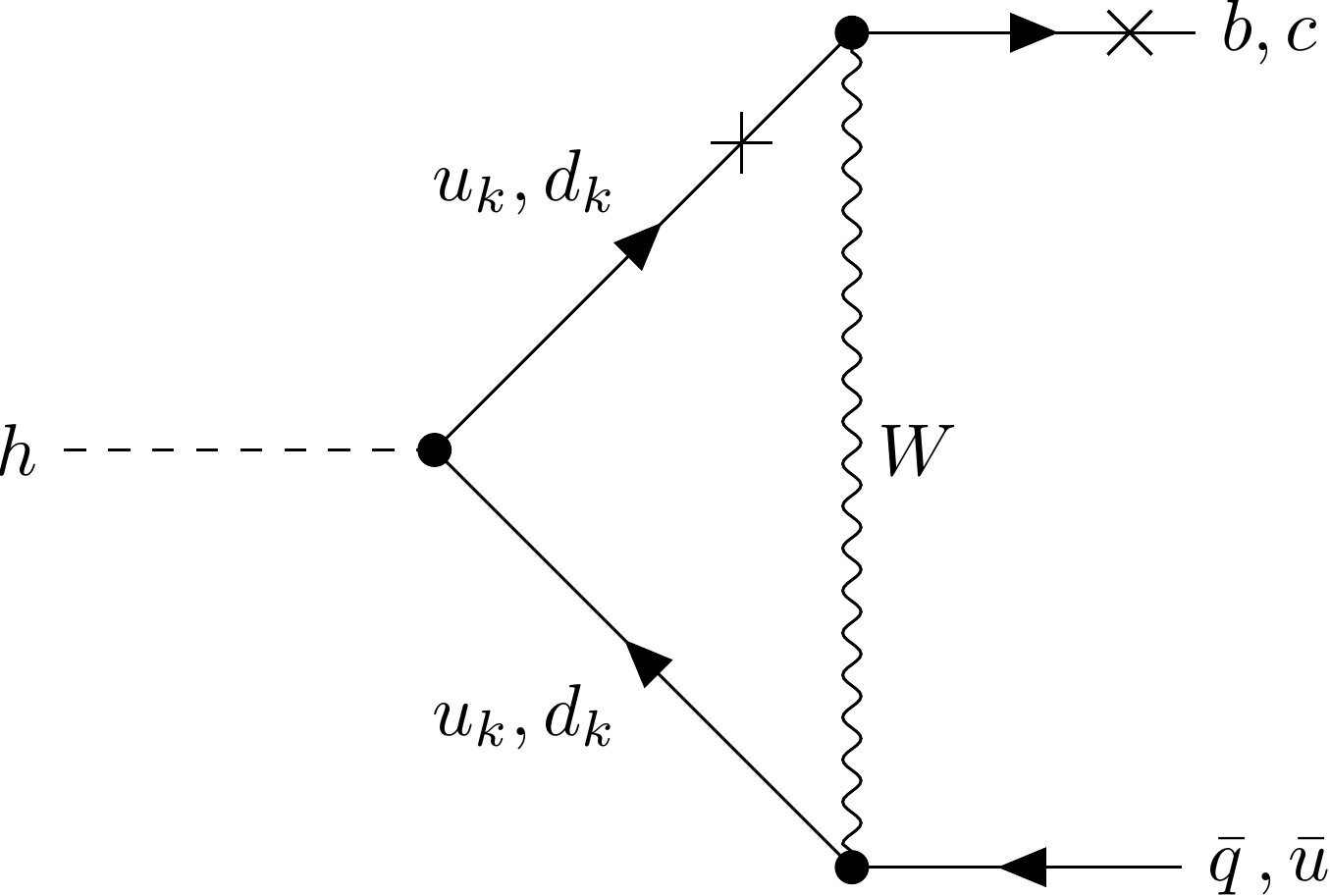}
\caption{Representative one-loop diagrams for the $Z\to q\bar q'$ (left) and $h\to q\bar q'$ (right) decays. The crosses indicate mass insertions.} 
\label{fig:ZtoBS:oneloopdiagram}
\end{figure}

The partial decay widths are, to the order we are working,  given by 
\beq\label{eq:SUPP:GammaZtoBS}
\Gamma(h/Z\to b\bar q) = N_C \frac{|\bar{\cal M}(h/Z\to q\bar q')|^2}{16\pi m_{h/Z}} \,,
\eeq
where $m_{h(Z)}$ is the mass of the Higgs ($Z$) boson,  $N_C = 3$ the number of  colors, and $|\bar{\cal M}(h/Z\to q\bar q')|^2$ the spin-averaged squared decay amplitude. 
In the SM, the $h/Z\to q\bar q'$ transitions occur at one-loop, through an up-type quark and $W$ boson exchange for $qq' = bs,bd$, while $qq' = cu$ requires a down-type quark. A representative diagram for each decay is shown in Fig.~\ref{fig:ZtoBS:oneloopdiagram}, where $u_k = u, c, t$ and $d_k = d, s, b$ for $k=1,2,3$. 

We first focus on the $Z\to bq $ decays. Due to the GIM mechanism, the decay amplitude is proportional to ${\cal M}\propto V_{kb}V_{kq}^* m_{u_k}^2$ and is thus dominated by the top quark contribution. Counting top mass as $m_t\sim m_Z$, gives the following naive dimensional analysis estimate for the decay amplitude,
\beq\label{eq:SUPP:NaiveAmplitude:Z}
{\cal M}_{\rm NDA} (Z\to b\bar q) \sim g^3 m_Z \frac{V_{tb}V_{tq}^*}{(4\pi)^2}\,,
\eeq
where $g$ is the $SU(2)_L$ gauge coupling, $s_W = \sin(\theta_W)$ and $c_W = \cos(\theta_W)$ are the sine and cosine of the weak mixing angle respectively, while $V_{ij}$ are the CKM matrix elements entering the $u_i - d_j-W$ vertex. The corresponding estimates for the partial decay widths are 
\beq
\Gamma_{\rm NDA}(Z\to b\bar s) \simeq 9.2\times10^{-9}~{\rm GeV}\,, \qquad \Gamma_{\rm NDA}(Z\to b\bar d) \simeq 4.0\times10^{-10}~{\rm GeV}\,.
\eeq
These turn out to be good approximations to the exact result, given below. 

We perform the calculation of the one-loop decay amplitude by first generating the one loop  $Z\to b\bar q$ diagrams using {\tt FeynArts}~\cite{Hahn:2000kx}, and then evaluating the amplitudes using {\tt FeynCalc}~\cite{Mertig:1990an,Shtabovenko:2016sxi,Shtabovenko:2020gxv}, including the reduction of the loop integrals to the Passarino-Veltmann (PaVe) functions. Finally, we use {\tt LoopTools}~\cite{Hahn:1998yk} to perform the numerical evaluation of the full amplitude. As an additional check, we have also evaluated the PaVe functions analytically, using the {\tt Package-X}~\cite{Patel:2016fam}. Isolating the $1/\epsilon$ divergent term, this is of the form
\beq
{\cal M}_{\rm div} \propto \sum_{k=u,c,t} V_{kb} V_{kq}^* \lp \bar u_b \slashed\epsilon_Z P_L u_q  \rp\,,
\eeq
where $P_L = (1-\gamma_5)/2$, while $u_b$, $u_q$ and $\epsilon_Z$ are the $b$- and $q$-quark spinors, and the $Z$ boson polarization vector, respectively. From the unitarity of the CKM matrix we then obtain ${\cal M}_{\rm div} = 0$, giving an independent check on our calculation. In general, all the $m_k$-independent terms in the amplitude give vanishing contributions due to the CKM unitarity; we check independently that each of these pieces cancel. Using the inputs from Table \ref{tab:NumInputs}, gives for the $Z\to b\bar q$ partial decay widths, 
\beq\label{eq:SUPP:WidthNumericalResult:OneLoop}
\begin{split}
\Gamma(Z\to b\bar s) &= \lp 5.2 \pm 0.1 \rp\times10^{-8}~{\rm GeV}\,,\qquad \Gamma(Z\to b\bar d) = \lp 2.3 \pm 0.1 \rp\times10^{-9}~{\rm GeV}\,.
\end{split}
\eeq
The quoted theoretical uncertainties reflect only the uncertainties on the inputs, and are dominated by the errors on $|V_{ts}|$ and $|V_{td}|$ CKM elements, which amount in both cases to a relative uncertainty, at the 1$\sigma$ level, of $\sim1\%$, where for simplicity we have symmetrized the uncertainty interval reported in Table~\ref{tab:NumInputs}. The latter translates to a $\sim2\%$ uncertainty on the $Z\to b\bar s$ and $Z\to b\bar d$ widths. Other inputs are known with a sub-percent  precision and thus we neglect their contribution to the total error budget. 

\begin{figure}[t]
\centering
\includegraphics[width=0.25\linewidth]{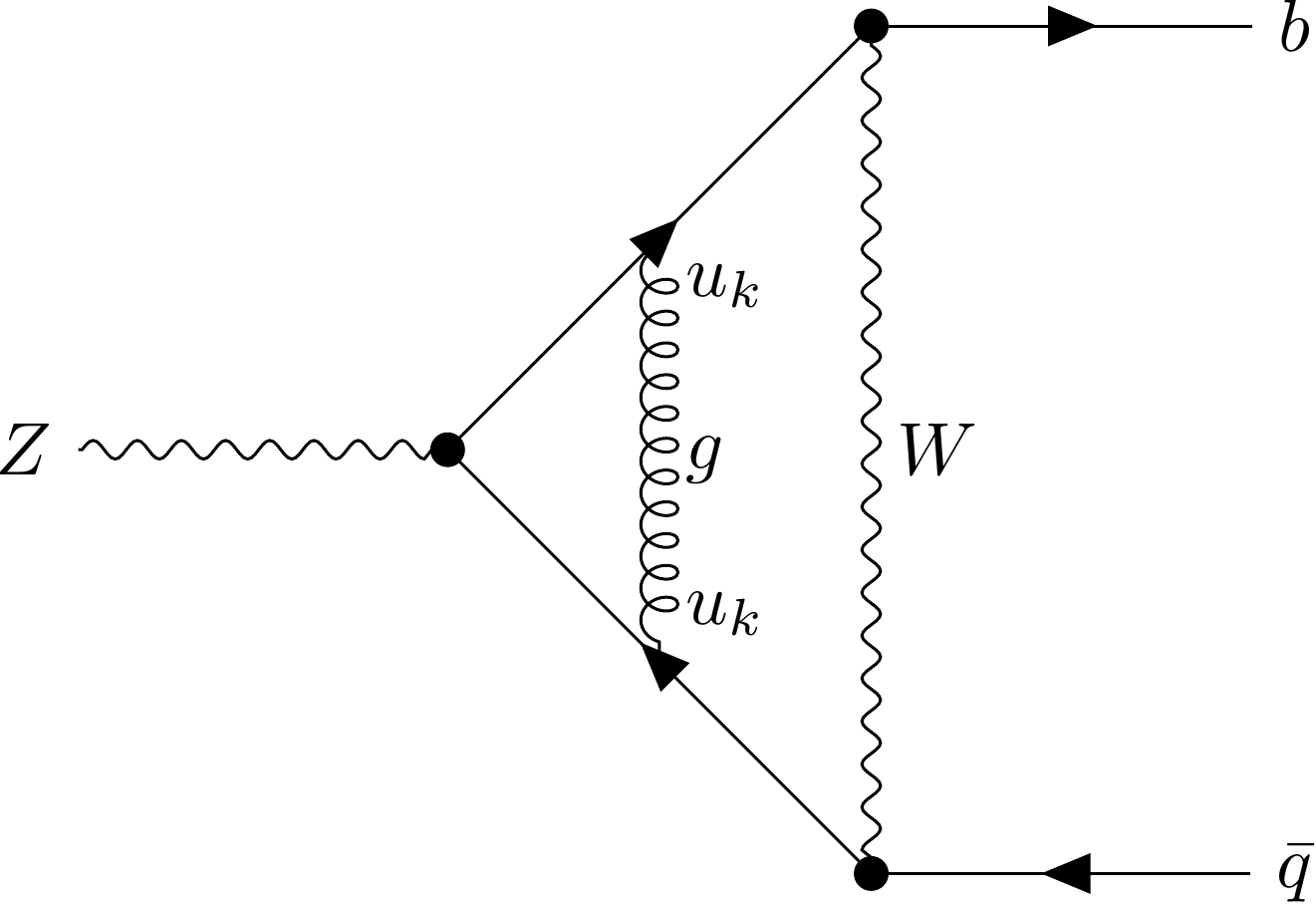}\hspace{1cm}
\includegraphics[width=0.25\linewidth]{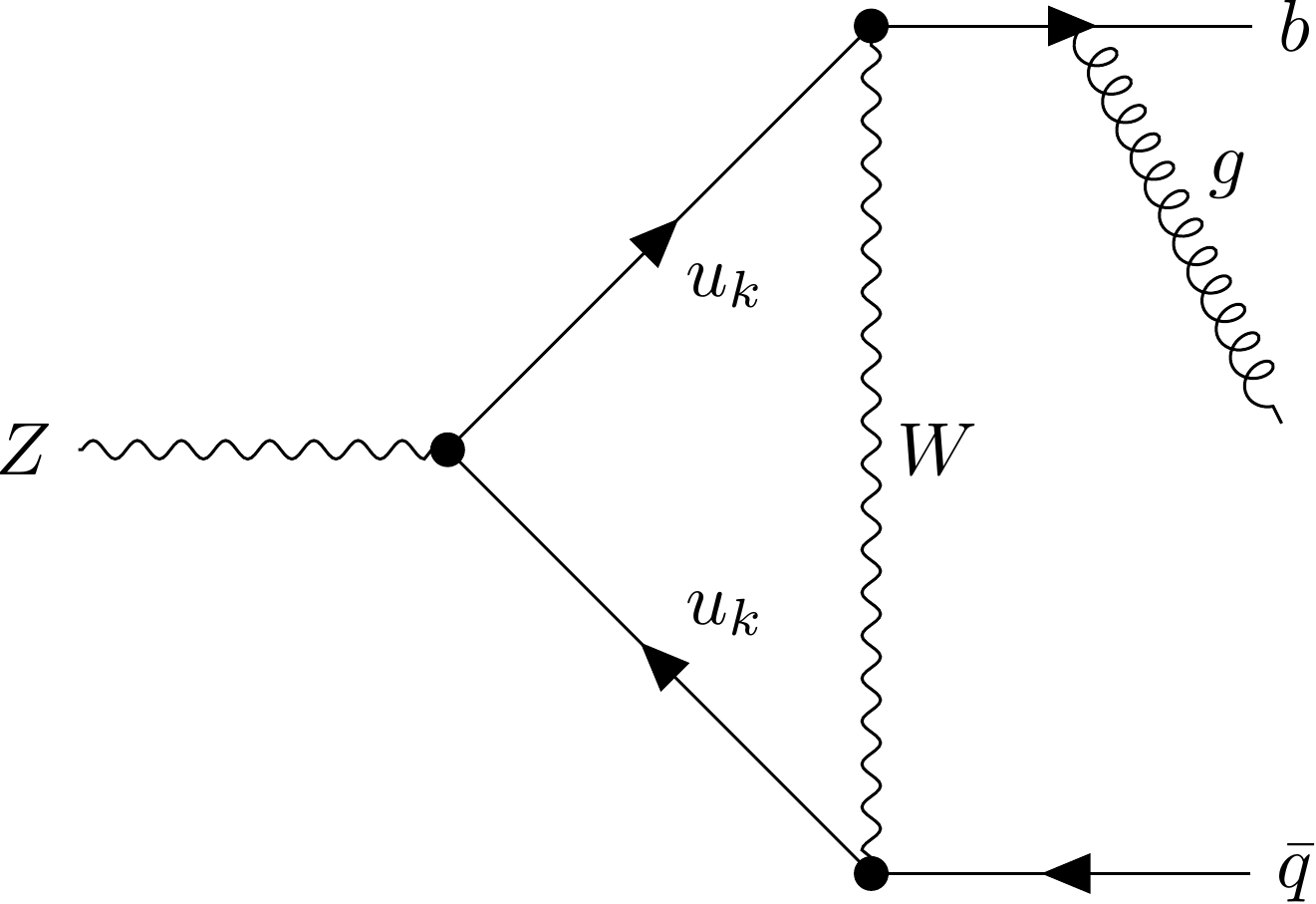}\hspace{1cm}
\includegraphics[width=0.25\linewidth]{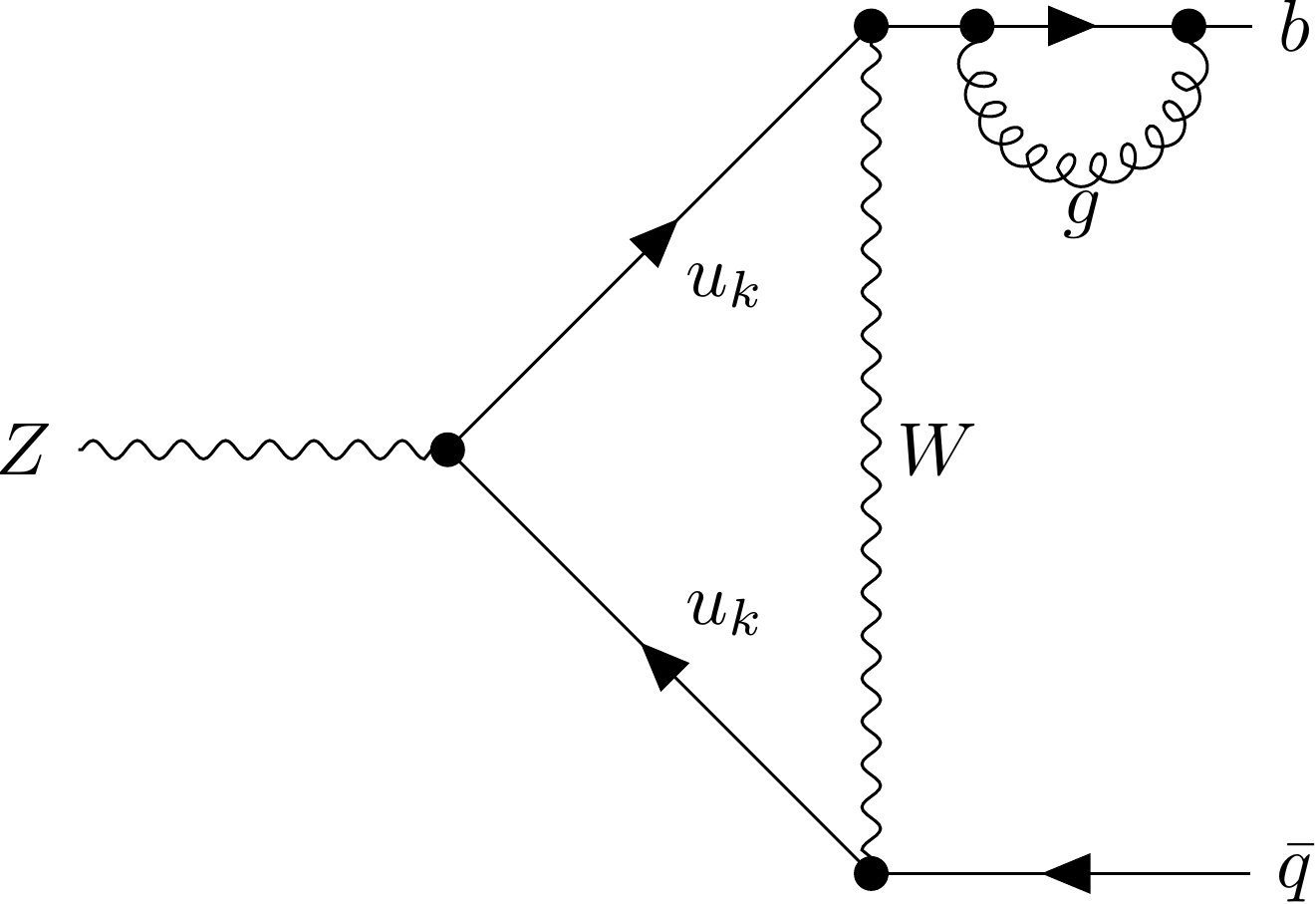}
\caption{Leading two-loop QCD corrections to the $Z\to b \bar q$ process. 
}
\label{fig:ZtoBS:twoloopdiagram}
\end{figure}

The second source of theoretical uncertainty are the higher order QCD corrections, which we estimate by using the partial two-loop calculation of mixed QCD-EW diagrams from~\cite{Chetyrkin:1990kr,Novikov:1995vu,Novikov:1999af}, and include them in the total uncertainty budget. In general, one can write 
\beq
\label{eq:Gammaalpha_s}
\Gamma_{\alpha_s}(Z\to b\bar q) = \frac{m_Z}{12\pi} \Big[ g_{Vbq}^2 \lp 1 \pm \lp R_V + 2 c_V \rp \rp + g_{Abq}^2 \lp 1 \pm \lp R_A + 2 c_A \rp \rp \Big] + \dots\,,
\eeq
where the ellipses denote the ${\cal O}(\alpha_s^2)$ corrections that we have neglected. The vector and axial couplings, $g_{V(A)bq}$, can be extracted from the full one-loop calculation. The radiators $R_{V(A)}$ represent the corrections from the sum of virtual and real (emitted) gluons attached to the external quark legs, see the middle and right diagrams in Fig.~\ref{fig:ZtoBS:twoloopdiagram}. To the order we are working these corrections are, in the $\overline{{\rm MS}}$ scheme, equal to $R_V = R_A \simeq \alpha_s(m_Z)/\pi\sim3.7\times10^{-2}$~\cite{Chetyrkin:1990kr,Novikov:1995vu}. 
The coefficients $c_{V(A)}$ in \eqref{eq:Gammaalpha_s} indicate the numerically leading two-loop gluon correction, see the left diagram in Fig.~\ref{fig:ZtoBS:twoloopdiagram}. Using the results from Ref.~\cite{Novikov:1999af} we get (again in the $\overline{{\rm MS}}$ scheme), $c_V = c_A =  -\alpha_s(m_t)\sim-0.1$. 
The factor of 2 in \eqref{eq:Gammaalpha_s} takes into account that this is a correction on the coupling, while the decay width is proportional to
\beq
g_{V(A)bq}^2 \to \left[ g_{Abq} (1 + c_{V(A)}) \right]^2 \simeq g_{V(A)bq}^2 \lp 1 + 2 c_{V(A)} + {\cal O}(\alpha_s^2) \rp\,.
\eeq
The sum of the two contributions gives $\lp R_{A,V} + 2 c_{A,V} \rp \simeq 0.17$. This shift in the predicted decay width is much larger than the uncertainty on the CKM elements. Our estimate for the relative error on the theoretical prediction is therefore $\sim 17\%$, leading to
\beq\label{eq:SUPP:WidthNumericalResult:OneLoopAndQCD}
\begin{split}
\Gamma(Z\to b\bar s) &= \lp 5.2 \pm 0.9 \rp\times10^{-8}~{\rm GeV}\,,\qquad \Gamma(Z\to b\bar d) = \lp 2.3 \pm 0.4 \rp\times10^{-9}~{\rm GeV}\,.
\end{split}
\eeq

Using the value of the total $Z$ width~\cite{Workman:2022ynf}, $\Gamma_Z = 2.4952 \pm 0.0023$ GeV, this then translates to the following predictions for the  ${\cal B}(Z\to bq)\equiv{\cal B}(Z\to b \bar q) + {\cal B}(Z\to \bar b q)$ branching ratios,
\beq\label{eq:SUPP:BRNumericalResult:OneLoopAndQCD}
\begin{split}
{\cal B}(Z\to bs) &= \lp 4.2 \pm 0.7 \rp\times10^{-8}\,, \qquad {\cal B}(Z\to bd) = \lp 1.8 \pm 0.3 \rp\times10^{-9}\,.
\end{split}
\eeq

We can repeat the above procedure for the calculation of the remaining decay widths.  A representative one loop diagram generating the $h\to b\bar q$ transition is shown in Fig.~\ref{fig:ZtoBS:oneloopdiagram} (right). To obtain the correct chirality, we need at least two mass insertions, indicated with a cross in Fig.~\ref{fig:ZtoBS:oneloopdiagram}, one of which will be on the external quark legs. Thus we expect the dominant contribution to the $h\to b\bar q$ amplitude to be suppressed by an additional factor of bottom Yukawa,  $y_b$, compared to the result for the $Z\to b\bar q$ decay, Eq.~\eqref{eq:SUPP:NaiveAmplitude:Z}. The NDA estimate for the $h\to b \bar q$ decay amplitude is thus
\beq\label{eq:SUPP:HtoBQ:NDA}
{\cal M}_{\rm NDA} (h\to b\bar q) \sim g^2 y_t^2 y_b m_h \frac{V_{tb}V_{tq}^*}{(4\pi)^2}\,,
\eeq
and the corresponding NDA estimates for the partial decay widths
\beq
\Gamma_{\rm NDA}(h\to b\bar s) \simeq 2\times10^{-11}~{\rm GeV}\,, \qquad \Gamma_{\rm NDA}(h\to b\bar d) \simeq 8.7\times10^{-13}~{\rm GeV}\,.
\eeq

Using the same {\tt FeynArts}+ {\tt FeynCalc}+{\tt LoopTools} pipeline as for the $Z\to b q$ decays above, along with the numerical inputs  at $\mu=m_h$ in Table~\ref{tab:NumInputs}, gives 
\beq\label{eq:SUPP:WidthNumericalResult:OneLoopAndQCD:Higgs}
\begin{split}
\Gamma(h\to b\bar s) &= \lp 1.8 \pm 0.3 \rp\times10^{-10}~{\rm GeV}\,, \qquad \Gamma(h\to b\bar d) = \lp 7.9 \pm 1.3 \rp\times10^{-12}~{\rm GeV}\,.
\end{split}
\eeq
where as a rough guidance we assigned the same $\sim 17\%$ uncertainty due to the missing higher order QCD corrections, which we assigned above for the $Z\to b q$ decay. 
Dividing by the SM prediction for the Higgs width, $\Gamma_h = 4.12\pm0.06$ MeV~\cite{LHCHiggsCrossSectionWorkingGroup:2013rie}\footnote{The latest updates can also be found on the LHCWG TWiki page \url{https://twiki.cern.ch/twiki/bin/view/LHCPhysics/LHCHWG}}, then gives the SM predictions for the  ${\cal B}(Z\to bq)\equiv{\cal B}(Z\to b \bar q) + {\cal B}(Z\to \bar b q)$ branching ratios 

\beq\label{eq:SUPP:BRNumericalResult:OneLoopAndQCD:Higgs}
{\cal B}(h\to bs) = \lp 8.9 \pm 1.5 \rp\times10^{-8}\,, \qquad {\cal B}(h\to bd) = \lp 3.8 \pm 0.6 \rp\times10^{-9}\,.
\eeq

Finally, we give predictions for the $Z/h\to cu$ decays. The one-loop decay amplitudes involve down-type quarks in the loop, and are thus proportional to ${\cal M}\propto\sum_k V_{ck}V_{uk}^* m_{d_k}$, where the sum runs over $k=d,s,b$. The NDA of the amplitudes read
\beq
\begin{split}
{\cal M}_{\rm NDA}(Z\to c\bar u) &\sim g^3 m_Z \sum_{k=d,s,b} \lp\frac{m_k}{m_Z}\rp^2 \frac{V_{ck} V_{uk}^*}{(4\pi)^2}\,, \\
{\cal M}_{\rm NDA}(h\to c\bar u) &\sim g^2 m_h y_c \sum_{k=d,s,b} y_k^2 \frac{V_{ck} V_{uk}^*}{(4\pi)^2}\,,
\end{split}
\eeq
where in the last expression we neglected terms proportional to $y_u$. The GIM mechanism now leads  to numerically even more suppressed decays, with the $Z/h\to c\bar u$ decay amplitudes about ${\cal O}(10^6)$ times smaller than the $Z/h\to b\bar s$ decay amplitudes in Eqs.~\eqref{eq:SUPP:NaiveAmplitude:Z} and \eqref{eq:SUPP:HtoBQ:NDA}.
The full numerical evaluation leads to
\beq
{\cal B}(Z\to cu) = \lp 1.4 \pm 0.2 \rp \times 10^{-18}\,, \quad {\cal B}(h\to cu) = \lp 2.7 \pm 0.5 \rp \times 10^{-20}\,,
\eeq
where the errors reflect our rough estimate of higher order QCD corrections, due to which we assign a $\sim 17\%$ uncertainty on the result.
Note that the ${\cal B}(h\to cu)$ prediction quoted above differs from the one reported in Ref.~\cite{Benitez-Guzman:2015ana} by orders of magnitude, most likely due to an incorrect normalization of the Passarino-Veltmann integrals, while it is consistent with the prediction in Ref.~\cite{Aranda:2020tqw}. As a consistency check  we obtained our result in two partially independent ways; after generating the diagrams and the corresponding amplitudes, the loop integrals were either computed numerically using {\tt LoopTools} or first computed analytically with {\tt Package-X} and then evaluated numerically, with the two results agreeing with each other.

\section{Additional details on the BSM models}
\label{sec:details:BSM}
Here we give further details on the indirect constraints on the FCNC couplings of the $Z$ boson and the Higgs to quarks, Eq.~\eqref{eq:ZbsHbsLagrangianPheno}. We show three different examples of NP effects. In section \ref{sec:indirect:Z} we show the constraints on the $Z-bs$, $Z-bd$ and $Z-cu$ couplings from low energy observables, assuming that these are the dominant NP effects. The same results for the $h-bs$ and $h-cu$ couplings, for which the only low energy constraints are due to $B_s-\bar B_s$ and $D-\bar D$ mixing, respectively, were already shown in the main text, cf. Fig. \ref{fig:newPhysics_bottom_charm}. Here we complete this list with constraints on $h-bd$ couplings, dominated by $B_d-\bar B_d$ mixing observables.  We also show constraints for two UV complete NP models, in section \ref{sec:VLQ} for the SM extended by a set of vector-like quarks, and in section \ref{sec:2HDM} for the type III two Higgs doublet model (2HDM) with a particular flavor violating structure of Yukawas. 

\begin{figure}[t]
\begin{center}
\includegraphics[width=0.3\linewidth]{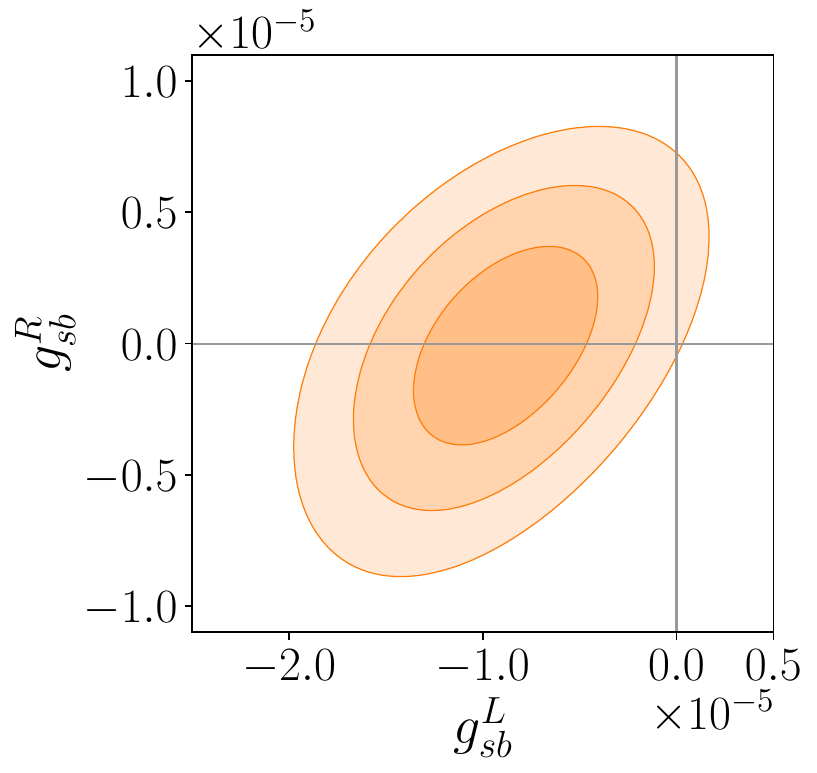}\hspace{1.5cm}
\includegraphics[width=0.28\linewidth]{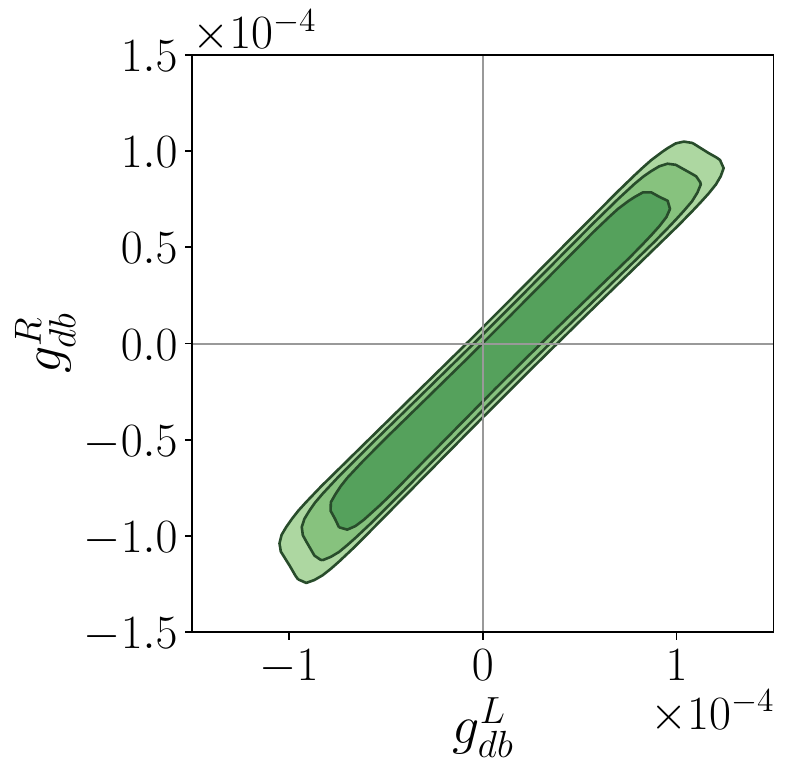}~
\end{center}
\caption{\textbf{Left}: The $1\sigma, 2\sigma, 3\sigma$ limits (from dark to light colors) on left-handed and right-handed FCNC couplings of $Z$ to $b$ and $s$ quarks, $g_{bs}^L, g_{bs}^R$, from current low-energy experiments, see Eq. \eqref{eq:ZbsHbsLagrangianPheno} in the main text. \textbf{Right}: The $1\sigma, 2\sigma, 3\sigma$ limits on couplings of $Z$ to $bd$.}\label{fig:SUPP:newphysicszbs}
\end{figure}

\subsection{Indirect bounds on FCNC $Z$ couplings}
\label{sec:indirect:Z}
The  $bs$ couplings of the $Z$ boson, $g_{sb}^{L,R}$, Eq.~\eqref{eq:ZbsHbsLagrangianPheno}, result in a shift  in a number of low-energy observables, such as the branching ratio of $B_s \to \mu^+ \mu^-$, as well as angular observables in  $B \to K^{(*)} \ell^+ \ell^-$ decays, etc. Integrating out the $Z$ gives a tree level contribution to the $\Delta B = \Delta S =  1$ weak effective Hamiltonian,

\begin{equation}\label{eq:WETHamiltonian}
-\mathcal{H}_{\rm WET} = \frac{4 G_F}{\sqrt{2}}\, \frac{\alpha_{\rm em}}{4 \pi}\,V_{tb}^\ast V_{ts} \sum_\ell \bigg(C_9 \mathcal{O}_9 + C_9^\prime \mathcal{O}_9^\prime + C_{10} \mathcal{O}_{10} + C_{10}^\prime \mathcal{O}_{10}^\prime + C_{\nu} \mathcal{O}_{\nu} + C_{\nu}^{\prime} \mathcal{O}_{\nu}^{\prime} +  \dots \bigg)\,,
\end{equation}
where 
\begin{equation}
\mathcal{O}_9^{(\prime)} = \big(\bar{s} \gamma_\mu b_{L(R)}\big)\big(\bar{\ell} \gamma^\mu \ell\big), \qquad \mathcal{O}_{10}^{(\prime)} = \big(\bar{s} \gamma_\mu b_{L(R)}\big)\big(\bar{\ell} \gamma^\mu \gamma_5 \ell\big)\,, \qquad \mathcal{O}_{\nu}^{(\prime)} = \big(\bar{s} \gamma_\mu b_{L(R)}\big) \big(\bar{\nu}_\ell \gamma^\mu(1 - \gamma_5) \nu_\ell\big)\,,
\end{equation}
and to the $\Delta B = 2$ four-quark operators:
\begin{equation}\label{eq:DelF2Hamiltonian}
    -\mathcal{H}_{\Delta F = 2} = C_{VL} (\bar{s} \gamma_\mu b_L)^2 + C_{VR} (\bar{s} \gamma_\mu b_R)^2 + C_{VLR}(\bar{s} \gamma_\mu b_L)(\bar{s} \gamma_\mu b_R) \,.
\end{equation}
The ellipses in \eqref{eq:WETHamiltonian} indicate operators that are less relevant for this analysis. The operators in~\eqref{eq:DelF2Hamiltonian} are kept dimensionful to align with the definitions in \texttt{flavio}. The Wilson coefficients are a sum of the SM, $C_i^{\rm SM}$, and the NP contributions, $\delta C_i$,

\begin{equation}
C_i = C_i^{\rm SM} + \delta C_i\,.
\end{equation}
The vector ($g_{Z \ell \ell, V}$) and axial ($g_{Z \ell \ell,A}$) couplings of the $Z$ to charged leptons, as well as to neutrinos ($g_{Z \nu \nu}$) are well measured, so that to good approximation in our analysis they can be taken to be equal to their SM values,  $g_{Z \ell \ell, V}= g\big(T_\ell^3 - 2\, Q_\ell s_W^2\big)/(2 c_W)$, $g_{Z \ell \ell,A}= - gT_\ell^3/(2 c_W) $, $g_{Z \nu \nu} = gT_\nu^3/(2 c_W)$, where $g = 2m_W/v$ is the weak coupling constant, $T_{\ell,\nu}^3$ the weak isospin eigenvalue, $Q_\ell$ the charge of the lepton, and $s_W$ ($c_W$) the sine (cosine) of the weak mixing angle. Integrating out the $Z$ at tree level gives, 

\begin{align}
\label{eq:deltaC9}
&\delta C_{9, \ell \ell}^{(\prime)} = \mathcal{N} g_{sb}^{L(R)} g_{Z \ell \ell, V} \simeq 6.04 \times 10^{3} g_{sb}^{L(R)}\,, \\
\label{eq:deltaC9prime}
&\delta C_{10, \ell \ell}^{(\prime)} = \mathcal{N} g_{sb}^{L(R)} g_{Z \ell \ell, A} \simeq -5.67 \times 10^{4} g_{sb}^{L(R)} \,, \\
&\delta C_{\nu}^{(\prime)} = \mathcal{N}g_{sb}^{L(R)} g_{Z \nu \nu} \simeq -5.67 \times 10^{4} g_{sb}^{L(R)}\,, 
\end{align}
where $\mathcal{N} = \sqrt{2} \pi/\big(G_F \alpha_{\rm em} V_{tb} V_{ts}^\ast m_Z^2\big)$, and
\begin{equation}\label{eq:SUPP:4-quark-operators}
C_{VL} = \frac{(g_{sb}^L)^2}{2 m_Z^2}\,, \qquad C_{VR} = \frac{(g_{sb}^R)^2}{2 m_Z^2} \,, \qquad C_{VLR} = \frac{g_{sb}^L g_{sb}^R}{m_Z^2} \,.
\end{equation}

We perform a global fit to low-energy observables, comparing the measurements to their theoretical predictions (with and without the presence of NP) by importing first  the matching conditions  \eqref{eq:deltaC9} and \eqref{eq:deltaC9prime} to \texttt{flavio} \cite{Straub:2018kue},  and then utilizing \texttt{smelli} \cite{Aebischer:2018iyb} to construct the global likelihood. 
\noindent The RGE running of the operators down to the scale $\mu \sim m_b$ is performed via the \texttt{wilson} package \cite{Aebischer:2018bkb}. The most stringent constraints on $g_{sb}^L, g_{sb}^R$ couplings come from the $b \to s \ell^+ \ell^-$ transitions.  The global fit gives the limits on $g_{sb}^L$, $g_{sb}^R$ of the order $10^{-5}$, see Fig.~\ref{fig:SUPP:newphysicszbs} (left), where nonzero negative values of $g_{sb}^L$ are preferred by the current experimental results (at the $3 \sigma $ level). Note however that this improvement of the global fit is far from optimal, as the global minimum in the space of $(C_{9 \ell \ell}^{(\prime)}, C_{10 \ell \ell}^{(\prime)})$ lies outside the planes defined by $g_{sb}^L$ $(g_{sb}^R)$. In particular, the relation $\delta C_{9 \ell \ell} \simeq - \delta C_{10 \ell \ell}$, which would give a much better description of current data, is not attainable just by having nonzero $Z-bs$ couplings (for the current status of different NP flavor scenarios in the global $b \to s \ell^+ \ell^-$ fits see e.g. Ref.~\cite{Alguero:2023jeh}). Similarly, we performed the global fit to the $g_{db}^L$, $g_{db}^R$ couplings, shown in Fig.~\ref{fig:SUPP:newphysicszbs} (right), which leads to ${\cal O}(10^{-4})$ limits. In this case the most stringent bound on the difference $g_{db}^L - g_{db}^R$ comes from the branching ratio $B_d^0 \to \mu^+ \mu^-$, which depends on the difference of axial couplings, $\mathcal{B} (B_d^0 \to \mu^+ \mu^-) \sim |C_{10 \mu \mu}^{db} - (C_{10 \mu \mu}^{db})^\prime|^2$. The orthogonal direction is limited by requiring that the NP contributions in $B-\bar{B}$ mixing, induced by the operators of the form \eqref{eq:SUPP:4-quark-operators} (with the obvious replacement $s \to d$), is not too large. In this case the data are consistent with the SM point, $g_{db}^{L,R} = 0$, at the $1 \sigma$ level.

One can carry out a similar analysis for the case of nonzero $Z -cu$ couplings. Since the relevant observables are not yet implemented in \texttt{flavio}, we derived the bounds from a combined fit to the weak decays $D^0 \to \mu^+ \mu^-$ \cite{LHCb:2022uzt}, $D^0 \to e^+ e^-$ \cite{Belle:2010ouj}, $D^0 \to \pi^0 \nu \bar{\nu}$ \cite{BESIII:2021slf} and $D^+ \to \pi^+ \mu^+ \mu^-$ \cite{Bause:2019vpr,LHCb:2013hxr}. Note that the bounds on SMEFT operators from the analyses of high-$p_T$ events \cite{Descotes-Genon:2023pen,Allwicher:2022mcg,Fuentes-Martin:2020lea} are not directly applicable here, since the assumption of point interactions is not valid for the much lighter $Z$.

\begin{figure}
    \centering
    \includegraphics[width = 0.6\linewidth]{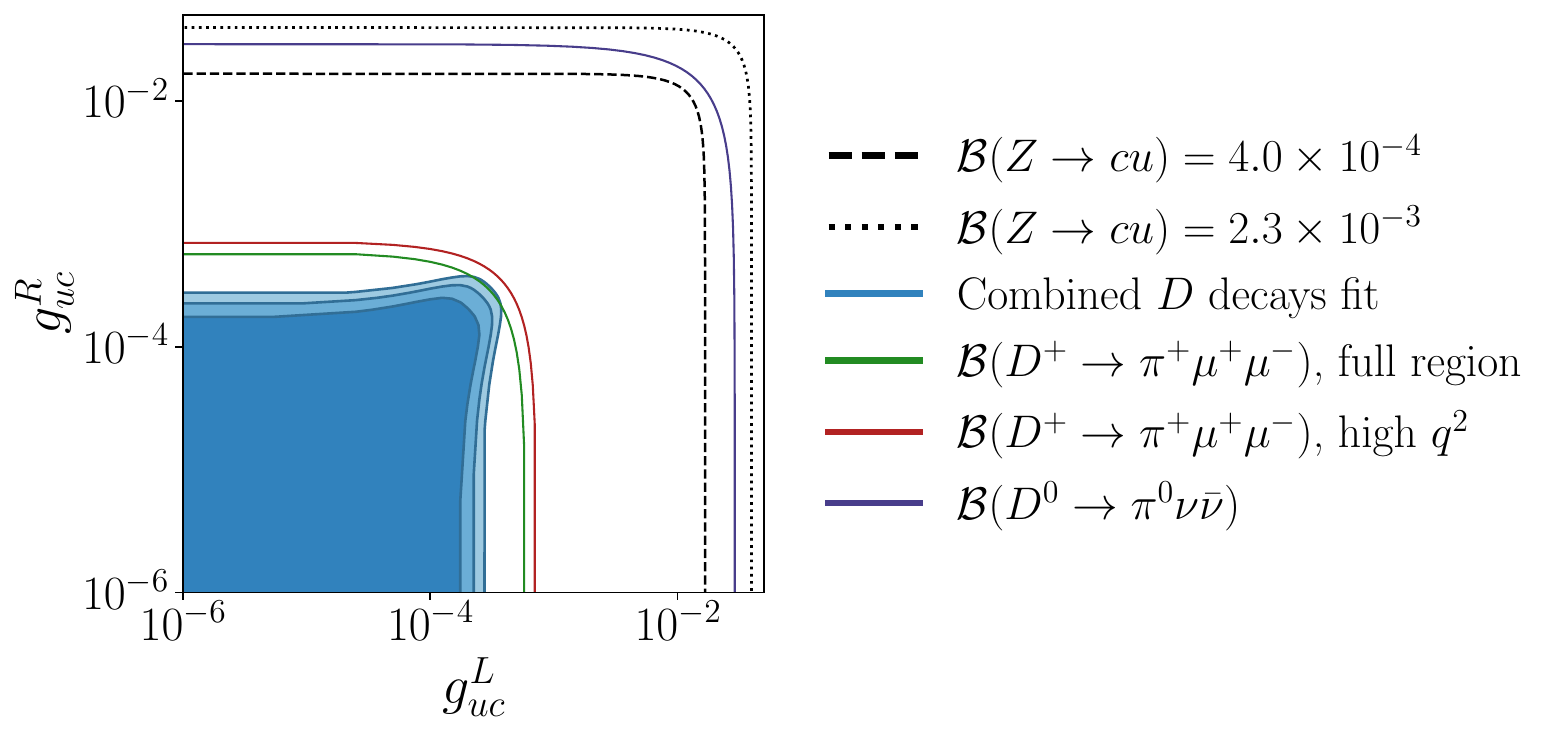}
    \caption{The $1\sigma, 2\sigma, 3\sigma$ limits (from dark to light blue)  on $g_{uc}^{L}$, $g_{uc}^R$ couplings of the $Z$, Eq.~\eqref{eq:ZbsHbsLagrangianPheno}, from rare $D$ meson decay data. The lines denote the $3 \sigma$ limits that were obtained from various constraints, as indicated in the legend. The two projected 95\% C.L. upper limits at the FCC-ee for the medium WP with $1\%$ ($0.1\%$) systematic uncertainties are shown as dotted (dashed) lines, cf. Sec. \ref{eq:sec:Zcu}.}\label{fig:SUPP:ZcuLimits}
\end{figure}

We summarize the results derived from charm decays in Fig.~\ref{fig:SUPP:ZcuLimits}. Due to parity constraints, the  $D^0 \to \ell^+ \ell^-$ decays are sensitive to the difference of the couplings $g_{uc}^L - g_{uc}^R$, whereas $D^0 \to \pi^0 \nu \bar{\nu}$ and $D^+ \to \pi^+ \mu^+ \mu^-$ are sensitive to their sum. Note that as before, we can assume that the couplings are lepton flavor universal as required by  LEP measurements. The maximum allowed value of couplings for either quark chirality is then given by $|g_{uc}^{L,R}| \lesssim 5.7 \times 10^{-4}$.

\subsection{Vector-like quarks}
\label{sec:VLQ}

Next, we turn to a concrete NP model, where we add to the SM a single generation of vector-like singlet down-type quarks, $(D_L, D_R)$, singlets under $SU(2)_L$ and with hypercharge $-1/3$. These have Yukawa couplings to the SM quarks, see, e.g., \cite{Fajfer:2013wca},

\begin{equation}\label{eq:SUPP:LagrDownTypeSingletZ}
    -\mathcal{L}_{\rm int}  \supset y_d^{ij} \bar{q}_L^i H d_R^j + y_u^{ij}\bar{q}_L^i \tilde{H} u_R^j+ y_D^i \bar{q}_L^i H D_R + M_D \bar{D}_L D_R + \mathrm{h.c.} \,,
\end{equation}
where $y_{d,u}^{ij} $ are the SM Yukawa couplings, $y_D^i$ are the Yukawa couplings to the vector-like quarks, and $M_D$ is the vector-like quark mass. 
After EWSB,  the down-type SM quarks and the vectorlike-quarks mix. Diagonalization of the mass matrix leads to flavor-changing couplings of down quarks with the $Z$ and the Higgs boson,

\beq
    \mathcal{L}_{\mathrm{VLQ}}^{D}  \supset  \frac{g}{2 c_W}X^d_{ij}\big(\bar{d}^i \gamma^\mu P_L d^j\big)Z_\mu + X^d_{ij} \frac{m_j}{v} \big(\bar{d}^i P_R d^j\big)h + \mathrm{h.c.}\label{eq:SUPP:LagrDownTypeSingletH}\,,
\eeq
where $X^d_{ij}$ in general has both nonzero diagonal and off-diagonal entries. Focusing on the $bs$ couplings, this gives for the couplings in the effective Lagrangian \eqref{eq:ZbsHbsLagrangianPheno},
\beq\label{eq:SUPP:VLQmatchingCond}
g^L_{sb}=\frac{g}{2c_W} \big(X_{sb}^d + X_{bs}^{d \ast} \big), \quad g^R_{sb}=0, \quad y_{sb}= X_{sb}^d m_b/v, \quad y_{bs}= X_{bs}^d m_s/v\,.
\eeq

\begin{figure}
    \centering
    \includegraphics[width = 0.3\linewidth]{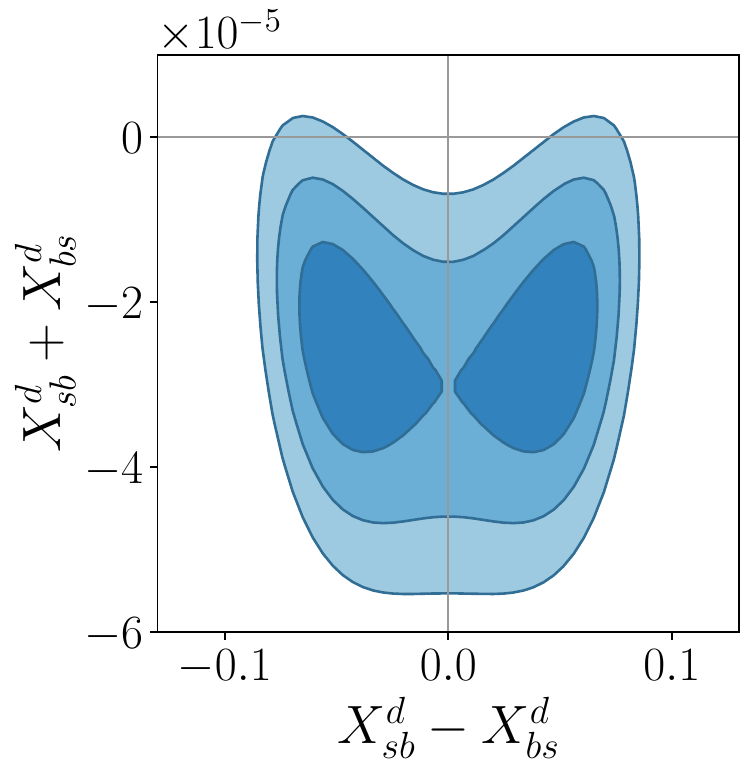}\hspace{1.5cm}
    \includegraphics[width = 0.3\linewidth]{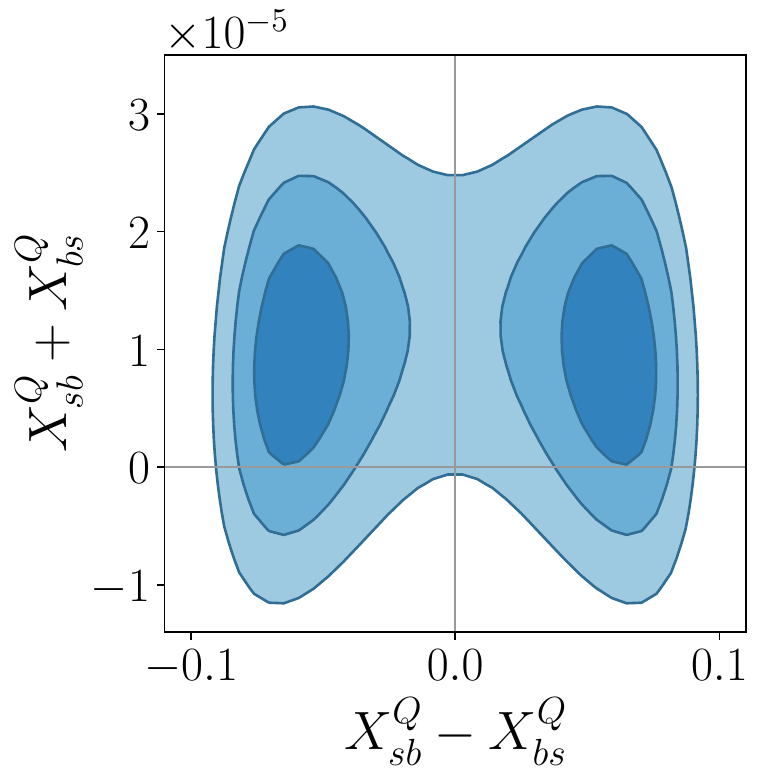}
    \caption{The parameter space, Eqs.~\eqref{eq:SUPP:LagrDownTypeSingletH} and \eqref{eq:VLQ:Q},  of the SM extended by either a down-type singlet vector-like quark (left) or  a doublet vector-like quark (right), which is allowed at  the $1\sigma, 2\sigma, 3\sigma$ level (from dark to light blue). For simplicity, we assume $X_{bs}^{d,Q}$ and $X_{sb}^{d,Q}$ to be real.}
    \label{fig:SUPP:VLQs}
\end{figure}

Figure~\ref{fig:SUPP:VLQs} (left) displays the results of the global fit to the low energy observables using {\tt flavio} and {\tt smelli}, and assuming real values of the couplings for simplicity. The negative values of the sum $X^d_{sb} + X^d_{bs}$ are favored by the fit, since this particular combination enters $\delta C_{9, \ell \ell}$ and $\delta C_{10, \ell \ell}$ Wilson coefficients. That is, the NP contributions to these two Wilson coefficients are in the direction set by $g_{bs}^L$, for which negative values are preferred, see Fig.~\ref{fig:SUPP:newphysicszbs}. The orthogonal direction of the parameter space is constrained by the $B_s - \bar{B}_s$ mixing, which receives contributions from both the tree level Higgs and $Z$ exchanges.

Alternatively, one could extend the SM by a single generation of doublet vector-like quarks $(Q_L, Q_R)$ with hypercharge $1/6$. These can have Yukawa couplings of $Q_L$ with  both the right-handed SM up- and down-quarks. The Lagrangian reads:
\begin{equation}\label{eq:SUPP:doubletVLQLagr}
    -\mathcal{L}_Q = y_d^{ij} \bar{q}_L^i H d_R^j + y_u^{ij}\bar{q}_L^i \tilde{H} u_R^j + y_D^i \bar{Q}_L H d_R^i + y_U^i \bar{Q}_L \tilde{H}u_R^i + M_Q \bar{Q}_L Q_R + \mathrm{h.c.} \,.
\end{equation}
After mass diagonalization these then generate right-handed flavor-changing neutral currents in both up and down sectors. Focusing on the down quark sector, the Lagrangian~\eqref{eq:SUPP:LagrDownTypeSingletH} is replaced by 

\begin{equation}
\label{eq:VLQ:Q}
    \mathcal{L}_{\mathrm{VLQ}}^{Q}  \supset  \frac{g}{2 c_W}X^Q_{ij}\big(\bar{d}^i \gamma^\mu P_R d^j\big)Z_\mu + X^Q_{ij} \frac{m_j}{v} \big(\bar{d}^i P_R d^j\big)h + \mathrm{h.c.}\,,
\end{equation}
while the couplings in the effective Lagrangian \eqref{eq:ZbsHbsLagrangianPheno} are now given by, 
\beq\label{eq:sup:gRL:VLQ}
g^R_{sb}=\frac{g}{2c_W} (X_{sb}^Q + X_{bs}^{Q \ast}), \quad g^L_{sb} = 0, \quad y_{sb}=X_{sb}^Q m_b/v, \quad y_{bs}=X_{bs}^Q m_s/v\,.
\eeq

The combination $X_{sb}^{Q} + X_{bs}^{Q \ast}$ is tightly constrained from the fits to the low energy $b \to s \ell\ell$ data. Extending the SM by a doublet VLQ  leads to flavor-violating right-handed currents in the direction of $g_{sb}^R$, whose values are closer to the SM expectation than the left-handed $g_{sb}^L$, explaining the relative position of allowed regions in the vertical direction in the left and right panels in  Fig. \ref{fig:SUPP:VLQs}.  The orthogonal linear combination $X_{sb}^Q - X_{bs}^{Q \ast}$ is limited by the $B_s - \bar{B}_s$ mixing. Both $X_{sb}^Q$ and $X_{bs}^{Q \ast}$ generate effective contributions to $h \to bs$; $X_{sb}^Q$ ($X_{bs}^{\ast Q}$) induces the effective operator $C_2$ ($C_2^\prime$), and these both enter in $C_4$. Since $X_{sb}^Q$ contribution is enhanced by an $m_b/m_s$ factor compared $X_{bs}^Q$, the bounds imposed by the $B_s - \bar{B}_s$ mixing on tree level Higgs exchanges are relevant mostly for $X_{sb}^Q $ and are saturated at approximately $\sim \mathcal{O}(0.1)$, cf. Fig.~\ref{fig:SUPP:VLQs} (similar discussion applies to $X_{sb,bs}^d$).

\subsection{Type III two Higgs doublet model}
\label{sec:2HDM}

The second example we consider is the type-III two-Higgs-doublet model \cite{Branco:2011iw}. Denoting the two Higgs fields as $H_{1,2}$, the most general form of the quark Yukawa couplings is given by 
\beq\label{eq:SUPP:2HDMLagrangian}
    \mathcal{L}_{\rm{2HDM}} \supset - \frac{\sqrt{2}m_i}{v}\delta_{ij}\bar{q}_L^i H_1 d_R^j - \sqrt{2}\,Y^d_{ij}\,\bar{q}_L^i H_2 d_R^j 
    - \frac{\sqrt{2}m_i}{v}\delta_{ij}\bar{q}_L^{\prime i} \tilde{H}_1 u_R^j - \sqrt{2}\,Y_{ij}^u\,\bar{q}_L^{\prime i} \tilde{H}_2 u_R^j\,,
\eeq
where $q_L^i$ ($q_L^{\prime i}$) are the left-handed quark doublet fields written in the down (up) quark mass basis.
We work in the Higgs basis, where only $H_1$ has a nonzero vev, so that the two Higgs doublets are given by, 
\begin{equation}
	H_1 = \begin{pmatrix}G^+\\[0.2em] \frac{1}{\sqrt{2}}\big(v+h_1+iG^0\big) \end{pmatrix},\quad
	H_2 = \begin{pmatrix}H^+ \\[0.2em] \frac{1}{\sqrt{2}} \big(h_2+iA\big)\end{pmatrix}.
	\label{eq:SUPP:HiggsDoublets}
\end{equation}
Here, $G^0$ and  $G^+$ are the Goldstone bosons, and $A$ the CP-odd heavy Higgs. The CP-even Higgs mass eigen-states $h, H$ are an admixture of $h_1$ and $h_2$,
\begin{equation}
	\begin{pmatrix}h_1\\h_2\end{pmatrix} = 
	\begin{pmatrix}
	c_\alpha &s_\alpha\\
	-s_\alpha &c_\alpha
	\end{pmatrix} \begin{pmatrix}h\\H\end{pmatrix}\,.
 \label{eq:SUPP:higgsMixingMatrix}
\end{equation}
Here $h$ denotes the SM-like Higgs, and we abbreviated $c_\alpha\equiv \cos\alpha$, $s_\alpha\equiv \sin\alpha$. 

We work in the limit where $H$ and $A$ are much heavier than the $h$. Diagonalizing the mass matrix gives for the light Higgs coupling to quarks, after electroweak symmetry breaking, 

\begin{equation}\label{eq:SUPP:HiggsTerm2HDM}
\mathcal{L}_{\rm Yukawa } \supset - \Big(\frac{m_{i}}{v}\delta_{ij} c_\alpha - Y^d_{ij} s_\alpha\Big)\bar{d}_{Li} d_{Rj} h + \mathrm{h.c.} + \dots \,,
\end{equation}
with ellipses denoting the couplings of the SM-like Higgs to the up-quark sector, and similarly for the Heavy scalars. In the numerical analysis we will assume that $Y_{ii}^{d,u}=0$, which leads to a simple rescaling of the theory predictions for Higgs phenomenology -- all the SM Higgs production rates and decay widths get multiplied by $c_\alpha^2$.

The tree level $h$, $H$ and $A$ exchanges induce meson mixing. Integrating out the heavy scalars gives the WET Lagrangian, 
\beq
{\cal L}_{\rm WET}\supset C_{2} (\bar{s}_R b_L)^2 + C_{2}^\prime (\bar{s}_L b_R)^2 + C_{4} (\bar{s}_L b_R)(\bar{s}_R b_L),
\eeq
where we focus on the $bs$ couplings (the WET Lagrangian induced by $cu$ couplings is obtained by replacing $b\to c, s\to u$). The Wilson coefficients of the scalar operators are given by \cite{Crivellin:2017upt}

\begin{align}
    C_{2} &= - \frac{\big(Y^{d \ast}_{bs}\big)^2}{2} \Bigg( \frac{s_\alpha^2}{m_h^2} + \frac{c_\alpha^2}{m_H^2} - \frac{1}{m_A^2} \Bigg), \\
    C_{2}^{\prime} &= - \frac{\big(Y^d_{sb}\big)^2}{2}\Bigg( \frac{s_\alpha^2}{m_h^2} + \frac{c_\alpha^2}{m_H^2} - \frac{1}{m_A^2} \Bigg),  \\
    C_{4} &= - \big(Y^{d \ast}_{bs} Y^d_{sb}\big) \Bigg( \frac{s_\alpha^2}{m_h^2} + \frac{c_\alpha^2}{m_H^2} + \frac{1}{m_A^2} \Bigg) \,. 
\end{align}
We use these expressions in \texttt{flavio} in order to obtain bounds on the 2HDM parameter space by comparing the theoretical predictions with the current bounds on $B_s - \bar{B}_s$ mixing. 

\begin{figure}
    \centering
    \includegraphics[width = 0.3\linewidth]{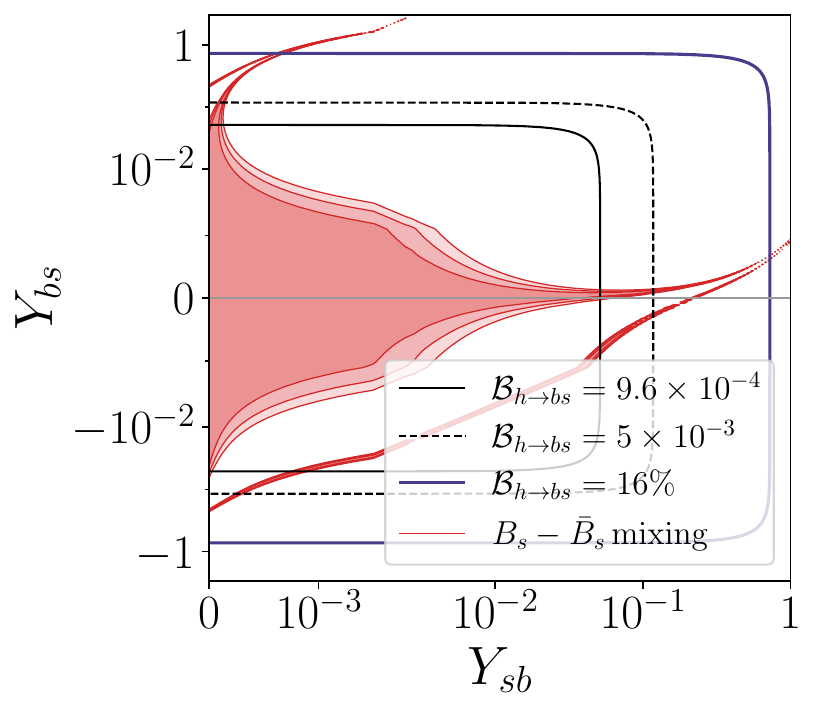}\hspace{1.5cm}
    \includegraphics[width = 0.3\linewidth]{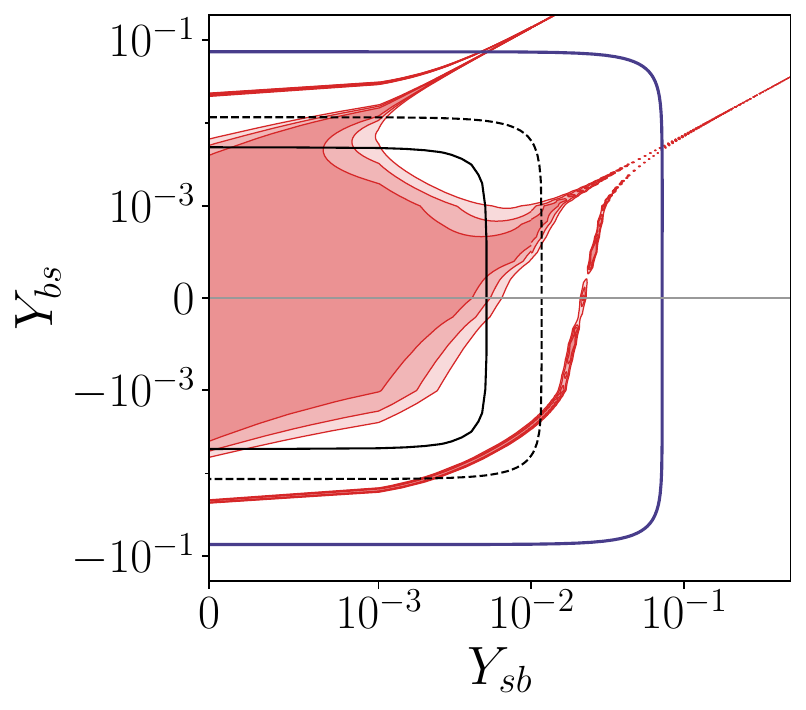}
    \caption{Constraints on the parameter space of the off-diagonal Yukawas $Y_{sb}$ and $Y_{bs}$ in the full 2HDM model, for values of the mixing angle $\sin \alpha = 1 \times 10^{-2}$ (left) and $\sin \alpha = 1 \times 10^{-1}$ (right). The masses of the heavier Higgses are assumed to be $m_H = m_A = 1 \, \mathrm{TeV}$. The legend in the left panel holds also for the right panel. 
    \label{fig:SUPP:2HDM_full_sinAlpha}
    }
\end{figure}

\begin{figure}
    \centering
    \includegraphics[width = 0.3\linewidth]{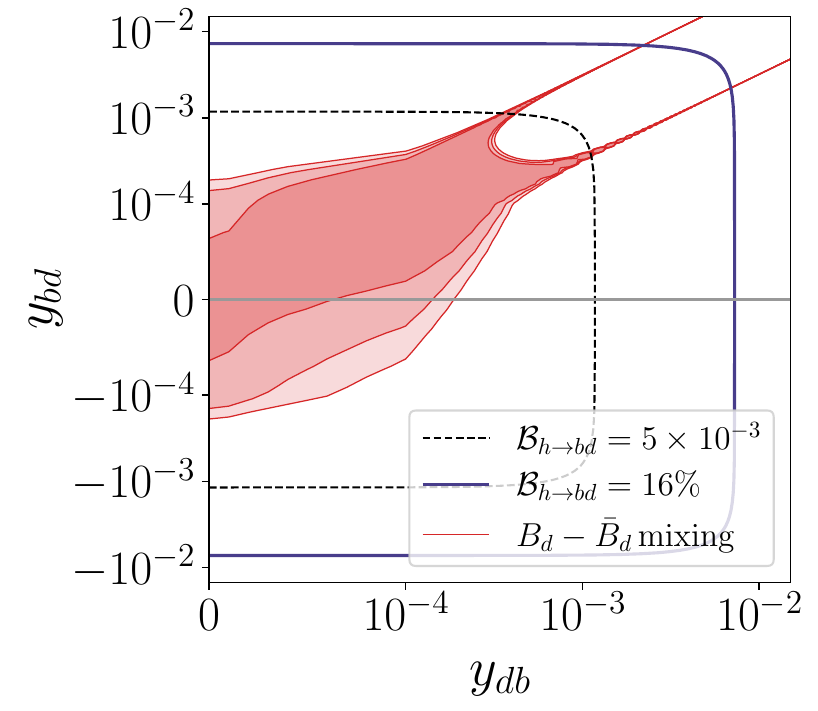}
    \includegraphics[width = 0.3\linewidth]{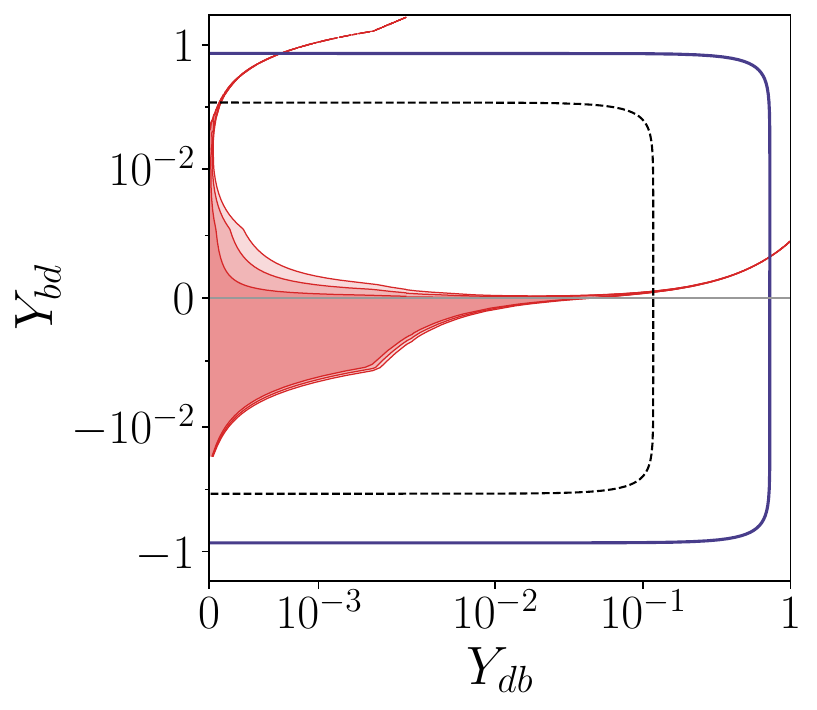}
    \includegraphics[width = 0.3\linewidth]{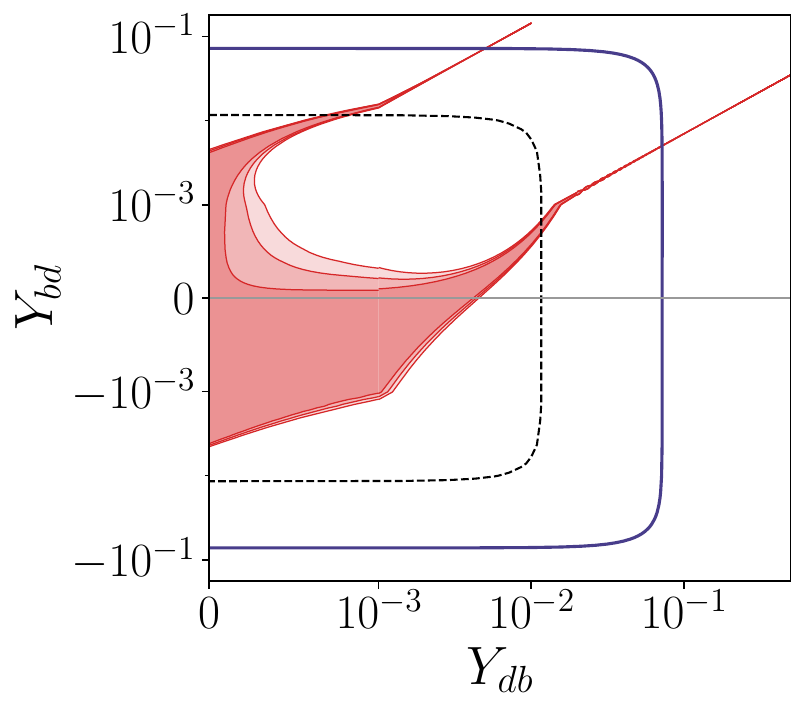}
    \caption{\textbf{Left:} Constraints on the parameter space of the off-diagonal Yukawas $y_{db}$ and $y_{bd}$ in the limit of the 2HDM model where the light Higgs exchange dominates. \textbf{Center:} Constraints on the parameter space of the off-diagonal Yukawas $Y_{db}$ and $Y_{bd}$ in the full 2HDM model, for value of the mixing angle $\sin \alpha = 1 \times 10^{-2}$. \textbf{Right:} Same as the central plot, but for $\sin \alpha = 1 \times 10^{-1}$. The legend in the left panels applies to all three.}
    \label{fig:SUPP:higgs_bd}
\end{figure}

\begin{figure}
    \centering
    \includegraphics[width = 0.3\linewidth]{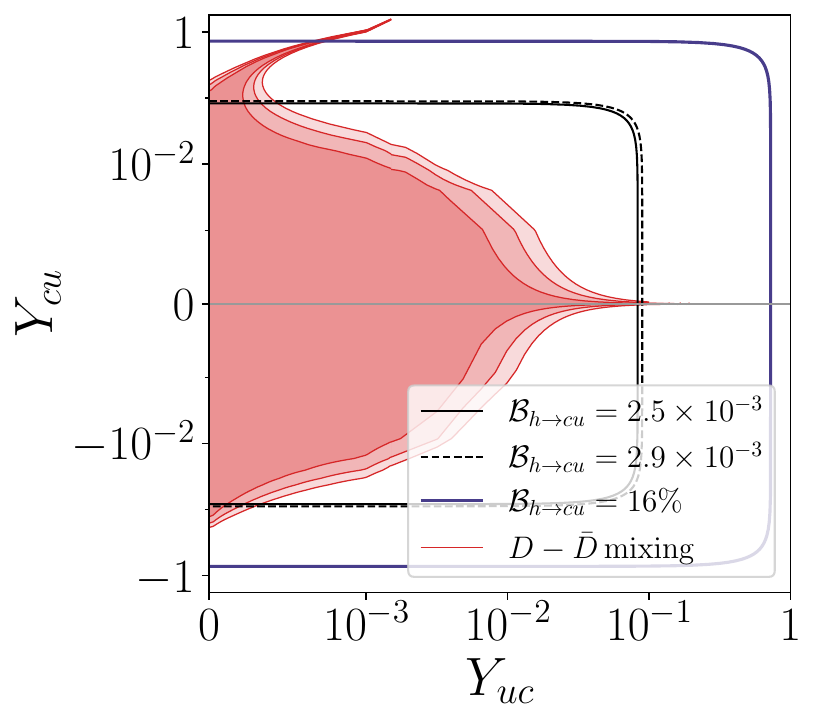}\hspace{1.5cm}
    \includegraphics[width = 0.3\linewidth]{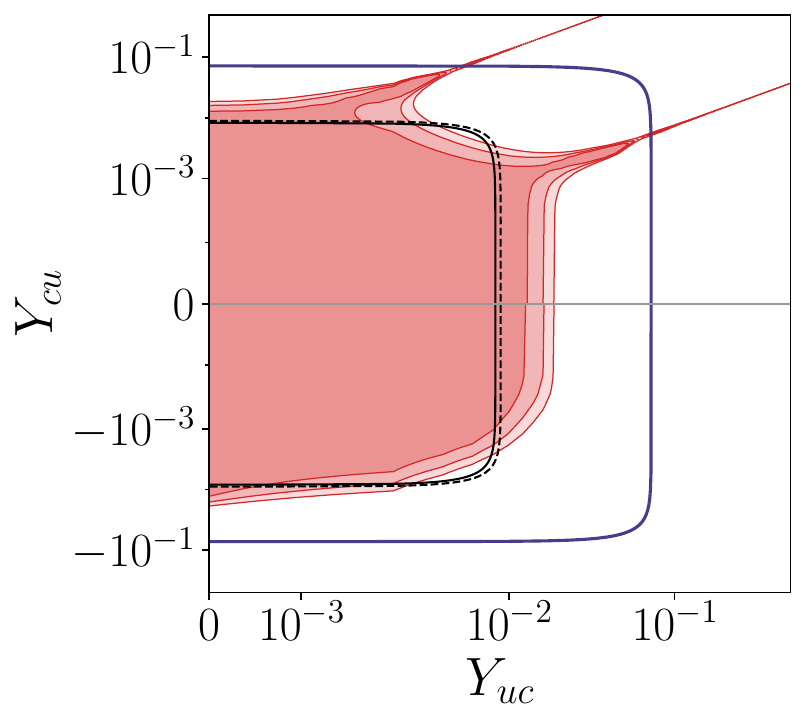}
    \caption{Constraints on the parameter space of the off-diagonal Yukawas $Y_{uc}$ and $Y_{cu}$ in the type III 2HDM model, for values of the mixing angle $\sin \alpha = 1 \times 10^{-2}$ (left) and $\sin \alpha = 1 \times 10^{-1}$ (right). The masses of the heavier Higgses are assumed to be $m_H = m_A = 1 \, \mathrm{TeV}$. The legend in the left panel applies also to the right panel.}
    \label{fig:SUPP:2HDM_full_sinAlpha_cu}
\end{figure}

In the main text we showed the bounds on the Higgs flavor violating Yukawas, cf. Eq. \eqref{eq:ZbsHbsLagrangianPheno}. This corresponds to the 2HDM parameter space in the limit where the $H$ and $A$ contributions are numerically small, and can be neglected. In this limit the NP contributions to meson mixing are completely determined by the Higgs flavor violating Yukawas
\beq
y_{bs, sb}= Y_{bs,sb}^d s_\alpha, \qquad y_{cu, uc}= Y_{cu,uc}^u s_\alpha.
\eeq
The corresponding bounds are shown in Fig.~\ref{fig:newPhysics_bottom_charm} and in the left plot of Fig.~\ref{fig:SUPP:higgs_bd} for the $bd$ case.  In  Figs.~\ref{fig:SUPP:2HDM_full_sinAlpha}, \ref{fig:SUPP:higgs_bd} (center and right) and \ref{fig:SUPP:2HDM_full_sinAlpha_cu} we also show two examples of bounds on the 2HDMIII Yukawa couplings, $Y_{bs, sb}$, $Y_{bd, db}$ and $Y_{cu,uc}$, from $B_s - \bar{B}_s$, $B_d - \bar{B}_d$ and $D - \bar{D}$ mixings respectively, in the regime where contributions from the heavy Higgses cannot be neglected. As an illustrative example we set $m_H=m_A=1 \,\mathrm{TeV}$, and $\sin \alpha = 10^{-2}$ (left panels) or $\sin \alpha = 10^{-1}$ (right panels). While the detailed shapes of the allowed regions in the space of flavor violating Yukawas changes for each of these benchmark points the main message remains unchanged: the FCC-ee can probe through $h\to bs, cu$ decays large regions of flavor changing parameter space that is not accessible via indirect probes.

Lastly, we comment briefly on existing bounds coming from the branching ratio of Higgs decays to an undetermined final state, ${\cal B}(h\to {\rm undet.})< 0.16$ at 95\% CL~\cite{ATLAS:2021vrm,CMS:2022dwd}, by assuming that this channel is saturated by flavor-changing decays. In the 2HDMIII, the decay rate is given by
\begin{equation}\label{eq:SUPP:HiggsDecayTobs}
\Gamma (h \to qq') \simeq \frac{3 s_\alpha^2 m_h}{8 \pi}(|Y_{qq'}|^2 + |Y_{q'q}|^2)\,,
\end{equation}
which in turn results in the branching ratio
\begin{equation}\label{eq:SUPP:HiggsBRTobs}
{\cal B}(h \to qq') = \frac{\Gamma(h \to qq')}{\Gamma(h \to qq') + c_\alpha^2 \, \Gamma_h^{\rm SM}}\,,
\end{equation}
where $\Gamma_h^{\rm SM} = 4.12$ MeV is the SM prediction for the Higgs total width. Typically we have $\sin\alpha\ll1$ and thus $\cos^2\alpha\sim1$. In this limit we obtain the bound quoted in the main text
\beq
 (|y_{qq'}|^2 + |y_{q'q}|^2)^{1/2} < 7.3\times 10^{-3}\,.
\eeq
This result applies to the $h\to bs$, $h\to bd$ and $h\to cu$ decay channels equally, as the expressions in eqs.~\eqref{eq:SUPP:HiggsDecayTobs} and \eqref{eq:SUPP:HiggsBRTobs} do not depend on final state parameters. 
Furthermore, we have checked that for such relatively small Yukawa couplings the Higgs boson production cross-section at the LHC is minimally affected, and the obtained bounds are thus internally consistent. They are shown as purple lines in Figs.~\ref{fig:newPhysics_bottom_charm}, \ref{fig:SUPP:2HDM_full_sinAlpha}, \ref{fig:SUPP:higgs_bd} and \ref{fig:SUPP:2HDM_full_sinAlpha_cu}.

\end{document}